\documentclass[prd,superscriptaddress,nofootinbib,secnumarabic,preprintnumbers]{revtex4-2}[10pt]
\pdfoutput=1
\usepackage{latexsym}
\usepackage{mathrsfs, amsmath, amssymb}
\usepackage{diagbox, slashed, makecell}
\usepackage{hyperref, url}
\usepackage{color}
\usepackage{graphicx, subcaption}
\usepackage{appendix}
\usepackage{verbatim}
\usepackage{multirow}
\usepackage{ucs}
\usepackage{indentfirst, layout, geometry}
\usepackage{enumitem}

\newcommand{\mr}{\mathrm}
\newcommand{\mc}{\mathcal}
\newcommand{\sLQ}{${}^\mr{s}\mr{LQ}$}
\graphicspath{{figures/}}

\begin{document}

\begin{sloppypar}

\title{Scalar leptoquark contributions to the $gg\rightarrow Zh$ process}

\author{Shi-Ping He}
\email[Corresponding author: ]{heshiping@tyut.edu.cn}
\affiliation{College of Physics and Optoelectronic Engineering, Taiyuan University of Technology, Taiyuan 030024, P. R. China}

\date{\today}

\begin{abstract}
In this manuscript, we study the general scalar leptoquark contributions to the $gg\rightarrow Zh$ process at one-loop level. We find that the contributions only show up for the off-diagonal Higgs and $Z$ boson interactions with scalar leptoquarks, which can lead to new tensor structures different from the standard model. In the one-field and two-field models, the contributions vanish exactly; thus, the contributions are possible at least in the three-field models. We exemplify the contributions in the $S_1+\widetilde{R}_2+S_3$ model, which can appear in the presence of scalar leptoquark mass splittings and $CP$ violation. In view of the heavy scalar leptoquark mass suppression, the leading contributions are typically small and difficult to observe at hadron colliders. However, this work offers a quantitative and systematic investigation of scalar leptoquark contributions, also making it valuable for the new physics studies in other models.
\end{abstract}

\maketitle

\tableofcontents
\clearpage

\section{Introduction}

Searching for new physics beyond the standard model (SM) is one of the crucial topics in elementary particle physics. In the new physics theories and models, the extra particles can be colorless and colored. For the colorless scalars and leptons, they are widely considered. At hadron colliders, new colored particles can lead to rich phenomenology because of the color nature of the initial states governed by quantum chromodynamics (QCD). One typical class of colored fermions are vector-like quark (VLQ) \cite{Aguilar-Saavedra:2013qpa}, which naturally occur in the composite Higgs models and grand unified theories. Meanwhile, the colored bosons are ubiquitous in many new physics models, for example, leptoquark (LQ) \cite{Dorsner:2016wpm}, scalar top partner \cite{Martin:1997ns}, color sextet scalar \cite{Chen:2008hh, Dorsner:2012pp}, and color octet scalar \cite{Manohar:2006ga, Dorsner:2012pp}. The LQ couples to quark and lepton simultaneously; thus, they are able to address many flavour problems in the quark and lepton sector. Usually, they are classified as scalar LQ (\sLQ) and vector LQ (${}^\mr{v}\mr{LQ}$) \cite{Blumlein:1992ej, Dorsner:2016wpm},\footnote{We add the upper script to distinguish it from the VLQ.} depending on the spin. In this work, we mainly concentrate on the \sLQ{}, of which the phenomenology has been investigated extensively. Currently, the pair production of direct search sets the \sLQ{} lower mass bound to be 2 TeV \cite{ATLAS:2024huc, CMS:2024bnj}. Their oblique corrections are discussed in Refs. \cite{Dorsner:2019itg, Crivellin:2020ukd, He:2022zjz, Albergaria:2023pgl}. The constraints from Higgs signal strength are also considered in the di-photon \cite{Chang:2012ta, Zhang:2019jwp, Crivellin:2020ukd}, di-quark \cite{Bhaskar:2020kdr}, and $\gamma Z$ decay \cite{Crivellin:2020ukd, Boto:2023bpg} channels. In Ref. \cite{Gherardi:2020qhc}, the authors studied the contributions in $B$ physics. Through the \sLQ{} Yukawa interactions, it can also be a solution to the $(g-2)_{\mu}$ anomaly \cite{Dorsner:2019itg, He:2021yck, He:2022glw, He:2024cut}.

However, none of the proposed new physics models have been experimentally confirmed to date. The Large Hadron Collider (LHC) operating at a center-of-mass energy of $\sqrt{s}=13\mr{TeV}$, has achieved an integrated luminosity of nearly $140\mr{fb}^{-1}$ and is projected to reach $3000\mr{fb}^{-1}$ in future runs. Therefore, precision measurements can play a vital role in searching for new physics. Since the Higgs boson's discovery over a decade ago \cite{ATLAS:2012yve, CMS:2012qbp}, the recorded data have pushed the Higgs physics to a high precision frontier. Currently, the main production and decay channels have already been measured at the percent level \cite{ATLAS:2024fkg, CMS:2022dwd}. Nevertheless, several rare production and decay channels are still not observed yet. Generally, these loop-induced processes are sensitive to the effects of new physics. In particular, gluon initiated productions can serve as a probe of new colored particles. For example, the di-Higgs and $gg\rightarrow Zh$ processes are two of the rare production channels. For the di-Higgs production, we constrain the flavour changing neutral Yukawa couplings in the VLQ models \cite{He:2020fqj}. The effects of colored scalars have also been extensively discussed in di-Higgs production, for example, scalar top partner \cite{Belyaev:1999mx, BarrientosBendezu:2001di, Dawson:2015oha, Batell:2015koa, Huang:2017nnw}, color octet scalar \cite{Kribs:2012kz}, and \sLQ{} \cite{Enkhbat:2013oba, DaRold:2021pgn, Bhaskar:2022ygp, DaRold:2023hst}. 

Compared to the $gg\rightarrow hh$ process, the $gg\rightarrow Zh$ channel can provide extra information through $Z$ polarization and is free of the triple Higgs coupling. Compared to the $h\rightarrow \gamma Z$ process, the $gg\rightarrow Zh$ channel can offer rich information through the kinematic distributions. Consequently, the $gg\rightarrow Zh$ channel is an important and compelling rare process. For the $gg\rightarrow Zh$ production, it was considered in the SM many years ago \cite{Kniehl:1990iva}. Recently, the QCD corrections are studied in several papers \cite{Brein:2003wg, Altenkamp:2012sx, Davies:2020drs, Chen:2020gae, Alasfar:2021ppe, Wang:2021rxu, Degrassi:2022mro}. Furthermore, it has also been investigated in various beyond the SM frameworks, including the VLQ model \cite{Harlander:2018yio}, two Higgs doublet model (2HDM) \cite{Harlander:2018yio}, and SM effective field theory (SMEFT) \cite{Rossia:2023hen}. In Refs. \cite{Kniehl:2011aa, Harlander:2018yio}, it is stated that the squark contributions vanish in supersymmetric models. However, the contributions of \sLQ{} remain to be investigated systematically, which is the dedication of this manuscript.

This paper is organized as follows. In Sec. \ref{sec:model}, we exhibit the \sLQ{} models including the one-field models and general models with mixing. In particular, we present the related Higgs and $Z$ boson interactions in the two-field and three-field models. In Sec. \ref{sec:contributions}, we investigate the general tensor structures relevant to the $gg\rightarrow Zh$ process. Furthermore, we derive the one-loop analytic contributions in the \sLQ{} models. In Sec. \ref{sec:contri:expansion}, we perform heavy internal mass expansion of the general \sLQ{} contributions and obtain the specific results in the three-field model $S_1+\widetilde{R}_2+S_3$. Moreover, this process is discussed from the SMEFT operators. The Sec. \ref{sec:summary} is devoted to the summary and conclusions. In App. \ref{app:model:2LQ:specific}, we enumerate specific two scalar leptoquark extended models as examples. In App. \ref{app:interactions:G0LQLQ}, we demonstrate absence of the diagonal $G^0-\mr{LQ}-\mr{LQ}$ type interactions. In App. \ref{app:loop}, we present the definitions and expansions of the loop integrals. In App. \ref{app:amp}, we show the computation details of one-loop amplitudes. In App. \ref{app:results}, we exhibit the explicit results of the form factors $G_{1,2,3}$.
\section{Model setup}\label{sec:model}
For the loop-induced $Z$ and Higgs associated production, both the SM and \sLQ{} interactions are involved. After electroweak symmetry breaking (EWSB), the related bosonic interactions in SM can be written as 
\begin{align}
&\frac{g_2Z_{\mu}}{2c_W}[h(\partial^{\mu}G^0)-(\partial^{\mu}h)G^0]+\frac{g_2m_Z}{2c_W}hZ_{\mu}Z^{\mu}.
\end{align}
Here, $G^0$ is the Goldstone of $Z$ boson. The related fermionic interactions in SM can be written as
\begin{align}
&-\frac{m_f}{v}h\bar{f}f+\frac{g_2Z_{\mu}}{c_W}[(I_3^f-Q_fs_W^2)\overline{f_L}\gamma^{\mu}f_L-Q_fs_W^2\overline{f_R}\gamma^{\mu}f_R]+g_s\bar{f}(\frac{\lambda^a}{2}G_{\mu}^a)f.
\end{align}
Here, the third component of weak isospin $I_3^f$ is $1/2$ ($-1/2$) for left-handed up-type (down-type) quark.

Typically speaking, there are five type of \sLQ{}s \cite{Dorsner:2016wpm}, which carry a conserved quantum number $F\equiv3B+L$. Here, the $B$ and $L$ are the baryon and lepton numbers. In Tab. \ref{tab:SLQ:rep}, we list their representations and labels.\footnote{For the $\bar{S}_1$ under the representation of $(\bar{3},1,-2/3)$, it only couples to a right-handed neutrino.} We take the conventions from Ref. \cite{Dorsner:2016wpm}, then the singlet and triplet \sLQ{}s correspond to complex conjugate fields of those in Ref. \cite{Crivellin:2021ejk}.
\begin{table}[!htb]
\begin{center}
\begin{tabular}{c|c|c}
\hline
\makecell{$SU(3)_C\times SU(2)_L\times U(1)_Y$ \\ representation} & Label & $F$ \\
\hline
\rule{0pt}{12pt} $(\bar{3},3,1/3)$ & $S_3\equiv(S_3^{4/3},S_3^{1/3},S_3^{-2/3})$ & $-2$ \\
\hline
$(3,2,7/6)$ & $R_2\equiv(R_2^{5/3},R_2^{2/3})$ & 0 \\
\hline
\rule{0pt}{12pt} $(3,2,1/6)$ & $\widetilde{R}_2\equiv(\widetilde{R}_2^{2/3},\widetilde{R}_2^{-1/3})$ & 0 \\
\hline
\rule{0pt}{12pt} $(\bar{3},1,4/3)$ & $\widetilde{S}_1$ & $-2$ \\
\hline
\rule{0pt}{12pt} $(\bar{3},1,1/3)$ & $S_1$ & $-2$ \\
\hline
\end{tabular}
\caption{The \sLQ{} representations.} \label{tab:SLQ:rep}
\end{center}
\end{table}
To summarize, we adopt $\Phi$ to mark a general $(2I+1)$ dimensional scalar multiplet. For $I=0,1/2,1$, it leads to the singlet, doublet, and triplet, respectively. In the following, we will investigate the \sLQ{} interactions. Hereafter, the covariant derivative on \sLQ{} is defined as $D_{\mu}\Phi\equiv(\partial_{\mu}-ig_1YB_{\mu}-ig_2t^aW_{\mu}^a-ig_sT^aG_{\mu}^a)\Phi$. The $t^a$ and $T^a=\lambda^a/2$ are the generators of $SU_L(2)$ and $SU_C(3)$, respectively. Here, the $\lambda^a(a=1,2,...,8)$ label the Gell-Mann matrices.
\subsection{One scalar leptoquark extended models}\label{sec:model:1LQ}
\subsubsection{Gauge interactions of a single scalar leptoquark}\label{sec:model:1LQ:gauge}

For the scalar multiplet $\Phi$, the $\Phi_{I_3}$ labels the $I_3$ component of electroweak scalar multiplet $\Phi$. We have hidden the color indices of $\Phi$ for simplicity. For the electroweak singlet of $I=0$, the $t^aW_{\mu}^a$ is absent. For the electroweak doublet of $I=1/2$, the representation $t^a$ is just $\sigma^a/2$ with $\sigma^a$ being the Pauli matrices. For the electroweak triplet of $I=1$, we have $(t^a)_{bc}=i\epsilon^{abc}$.\footnote{For the scalar triplet, it can also be parameterized in the matrix form \cite{He:2022zjz}.}

After EWSB, the relevant gauge interactions can be written as
\begin{align}
&\frac{ig_2Z_{\mu}}{c_W}\sum_{-I\le I_3\le I}(I_3-Q_{I_3}s_W^2)[\Phi_{I_3}^{\dag}(\partial^{\mu}\Phi_{I_3})-(\partial^{\mu}\Phi_{I_3}^{\dag})\Phi_{I_3}]\nonumber\\
&+ig_s\sum_{-I\le I_3\le I}[\Phi_{I_3}^{\dag}(\frac{\lambda^a}{2}G_{\mu}^a)(\partial^{\mu}\Phi_{I_3})-(\partial^{\mu}\Phi_{I_3}^{\dag})(\frac{\lambda^a}{2}G_{\mu}^a)\Phi_{I_3}]\nonumber\\
&+\sum_{-I\le I_3\le I}\Phi_{I_3}^{\dag}[\frac{g_2Z_{\mu}}{c_W}(I_3-Q_{I_3}s_W^2)+g_s(\frac{\lambda^a}{2}G_{\mu}^a)]^2\Phi_{I_3}.
\end{align}
In the above, the first line indicates the $Z-\mr{LQ}-\mr{LQ}$ type interactions, the second line gives the $g-\mr{LQ}-\mr{LQ}$ type interactions, and the third line shows the $g-g-\mr{LQ}-\mr{LQ}$ and $g-Z-\mr{LQ}-\mr{LQ}$ type interactions.

\subsubsection{Scalar sector interactions of a single scalar leptoquark}\label{sec:model:1LQ:scalar}
In this section, we derive the relevant scalar interactions. For the $gg\rightarrow Zh$ process, the involved interactions are physical Higgs type $h-\mr{LQ}-\mr{LQ}$ and Goldstone type $G^0-\mr{LQ}-\mr{LQ}$. For a general electroweak multiplet $\Phi$, the relevant gauge invariant scalar interactions can be parameterized as\footnote{For the doublet, there are also interactions of $(\Phi^{\dag}H)(H^{\dag}\Phi)$ and $(H^Ti\sigma_2\Phi)^{\dag}(H^Ti\sigma_2\Phi)$ \cite{Crivellin:2021ejk}. However, they are not independent interactions and can be converted into the standard forms through the relations $(\sigma^a)_{ij}(\sigma^a)_{kl}=2\delta_{il}\delta_{kj}-\delta_{ij}\delta_{kl}$ and $\epsilon_{ik}\epsilon_{jl}=\delta_{ij}\delta_{kl}-\delta_{il}\delta_{kj}$.}
\begin{align}
&-\mu_{\Phi}^2(\Phi^{\dag}\Phi)-\lambda_{1}(H^{\dag}H)(\Phi^{\dag}\Phi)-\lambda_{2}(H^{\dag}i\sigma^aH)(\Phi^{\dag}it^a\Phi).
\end{align}

After EWSB, the mass spectrums can be written as \cite{Lavoura:1993nq, Earl:2013jsa}
\begin{align}
&m_{\Phi_{I_3}}^2=\mu_{\Phi}^2+\frac{1}{2}(\lambda_{1}+I_3\cdot\lambda_{2})v^2=\frac{1}{2}(m_{\Phi_{I}}^2+m_{\Phi_{-I}}^2)+\frac{I_3}{2I}(m_{\Phi_{I}}^2-m_{\Phi_{-I}}^2)\quad(I_3=-I,...,I-1,I).
\end{align}

The related physical Higgs interactions are
\begin{align}
-vh\sum_{-I\le I_3\le I}(\lambda_{1}+I_3\cdot\lambda_{2})(\Phi_{I_3})^{\dag}\Phi_{I_3}.
\end{align}

For the single representation \sLQ{} models, there is no $G^0-\mr{LQ}-\mr{LQ}$ type interaction.
\subsection{Scalar leptoquark interactions with mixing: general formalism}
In the models extended by more than two fields of $S_3$, $R_2$, $\widetilde{R}_2$, $\widetilde{S}_1$, $S_1$, there can be mixing between the same electrically charged \sLQ. After EWSB, the general mass terms, physical Higgs and Goldstone interactions can be parameterized as
\begin{align}
-\sum_{i,j}M_{0,ij}^2(\phi_i)^{\dag}\phi_j-vh\sum_{i,j}\Gamma_{0,ij}^h(\phi_i)^{\dag}\phi_j-ivG^0\sum_{i,j}\Gamma_{0,ij}^{G^0}(\phi_i)^{\dag}\phi_j.
\end{align}
Typically, the $M_{0}^2$ and $\Gamma_{0}^h$ are Hermitian matrices, while the $\Gamma_{0}^{G^0}$ is anti-Hermitian. Then, we need to rotate them into mass eigenstates through the transformation of
\begin{align}\label{eqn:model:general:U}
\left[\begin{array}{c}\phi_1\\\phi_2\\\vdots\\\phi_n\end{array}\right]\rightarrow
	U\left[\begin{array}{c}\phi_1\\\phi_2\\\vdots\\\phi_n\end{array}\right].
\end{align}
Here, $U$ is a $n\times n$ unitary matrix, which can be parameterized in terms of rotation angle and phase parameters. The mass matrix $M_{0}^2$ is diagonalized as $M^2\equiv\mr{diag}\{m_{\phi_1}^2,m_{\phi_2}^2,\dots,m_{\phi_n}^2\}$ by the unitary transformation, which implies
\begin{align}\label{eqn:model:general:M2}
&U^{\dag}M_{0}^2U=M^2\quad\Longleftrightarrow\quad M_{0}^2=UM^2U^{\dag}.
\end{align}
After diagonalization, we denote the $h$ and $G^0$ coupling matrices in mass eigenstates as $\Gamma^h$ and $\Gamma^{G^0}$, which are Hermitian and anti-Hermitian, respectively. They are expressed as
\begin{align}\label{eqn:model:general:gamma}
&\Gamma^h=U^{\dag}\Gamma_{0}^hU,\qquad \Gamma^{G^0}=U^{\dag}\Gamma_{0}^{G^0}U.
\end{align}
Equivalently, the component forms are expressed as
\begin{align}\label{eqn:model:general:gammaCom}
&\Gamma^h_{ij}=\sum_{m,n}U_{mi}^{\ast}\Gamma_{0,mn}^hU_{nj},\qquad \Gamma^{G^0}_{ij}=\sum_{m,n}U_{mi}^{\ast}\Gamma_{0,mn}^{G^0}U_{nj}.
\end{align}

Taking into account the mixing between different \sLQ, the $G^0-\mr{LQ}-\mr{LQ}$ type interaction can appear \cite{Crivellin:2021ejk}. However, there is no diagonal interaction $G^0(\phi_i)^{\dag}\phi_i$. Only off-diagonal interactions $G^0(\phi_i)^{\dag}\phi_j$ ($i\neq j$) is possible. \footnote{In Ref. \cite{Branco:2011iw}, the similar $G^0S_iS_j$ interaction is proportional to $(m_i^2-m_j^2)$, which also causes the disappearance of diagonal interaction $G^0S_iS_i$.} In App. \ref{app:interactions:G0LQLQ}, we present the detailed demonstrations for the two-\sLQ{} and three-\sLQ{} models.

The $g-\mr{LQ}-\mr{LQ}$ and $g-g-\mr{LQ}-\mr{LQ}$ type interactions are still diagonal because of $SU_C(3)$ gauge symmetry, which can be parameterized as
\begin{align}
&ig_s\sum_{i}[\phi_i^{\dag}(\frac{\lambda^a}{2}G_{\mu}^a)(\partial^{\mu}\phi_i)-(\partial^{\mu}\phi_i^{\dag})(\frac{\lambda^a}{2}G_{\mu}^a)\phi_i]+g_s^2\sum_{i}\phi_i^{\dag}(\frac{\lambda^a}{2}G_{\mu}^a)^2\phi_i.
\end{align}
After EWSB, the general $Z-\mr{LQ}-\mr{LQ}$ and $g-Z-\mr{LQ}-\mr{LQ}$ type interactions can be parameterized as
\begin{align}
&\frac{ig_2Z_{\mu}}{c_W}\sum_{i,j}g_{0,ij}^Z[\phi_i^{\dag}(\partial^{\mu}\phi_j)-(\partial^{\mu}\phi_i^{\dag})\phi_j]+\frac{g_2Z_{\mu}}{c_W}g_sG_{\mu}^a\sum_{i,j}g_{0,ij}^{gZ}\phi_{i}^{\dag}\lambda^a\phi_j.
\end{align}
Here, we have $g_{0,ij}^Z=g_{0,ij}^{gZ}=I_{3,\phi_i}\delta_{ij}-Q_{\phi_i}s_W^2\delta_{ij}$. After diagonalization, we denote the $Z-\mr{LQ}-\mr{LQ}$ and $g-Z-\mr{LQ}-\mr{LQ}$ coupling matrices in mass eigenstates as $g^Z$ and $g^{gZ}$, which are Hermitian. They are expressed as
\begin{align}\label{eqn:model:general:gZ}
&g^Z=g^{gZ}\equiv g_I^Z-Qs_W^2,\qquad g_I^Z=U^{\dag}\mr{diag}\{I_{3,\phi_1},I_{3,\phi_2},\cdots,I_{3,\phi_2}\}U.
\end{align}
Equivalently, the component forms are expressed as
\begin{align}\label{eqn:model:general:gZCom}
&g_{I,ij}^Z=\sum_m I_{3,\phi_m}U_{mi}^{\ast}U_{mj}.
\end{align}
Because of $U(1)$ electromagnetic gauge symmetry, the mixings merely occur between \sLQ{s} with the same electric charge. Thus, the $Q_{\phi_i}s_W^2\delta_{ij}$ part is automatically diagonal in mass eigenstates. For the off-diagonal interactions, we only need to consider the $g_I^Z$ part.

Combing Eq. \eqref{eqn:model:general:gammaCom} and Eq. \eqref{eqn:model:general:gZCom} together, we have the following results directly relevant to the $gg\rightarrow Zh$ process:
\begin{align}\label{eqn:model:general:gZgammaCom}
&\mr{Im}(g_{I,ij}^Z\Gamma^h_{ji})=\mr{Im}(\sum_{m'}I_{3,\phi_{m'}}U_{m'i}^{\ast}U_{m'j}\cdot\sum_{m,n}U_{mj}^{\ast}\Gamma_{0,mn}^hU_{ni}).
\end{align}
Because $g_{I}^Z$ and $\Gamma^h$ are both Hermitian matrices, only the $i\neq j$ case is non-zero. Unlike the traditional two Higgs doublet model, the \sLQ{s} interactions with Higgs are strongly constrained by the $SU_C(3)$ color symmetry. The $\Gamma_{0,mn}^h$ and $M_{0,mn}^2$ are linearly correlated if $m\neq n$; thus, we define a ratio $\kappa_{mn}\equiv\Gamma_{0,mn}^h/M_{0,mn}^2$ for simplicity. Then, we find that $\kappa_{mn}=1/v^2$ ($m\neq n$) for the cubic type interactions $H\,\mr{LQ}^2$ and $\kappa_{mn}=2/v^2$ ($m\neq n$) for the quartic type interactions $H^2\mr{LQ}^2$ in gauge eigenstates. Adopting the transformation of mass matrix in \eqref{eqn:model:general:M2} and the real number nature of $\kappa_{mn}$, we can obtain the following results:
\begin{align}\label{eqn:model:general:gZgammaCom2}
&\mr{Im}(g_{I,ij}^Z\Gamma^h_{ji})=\sum_{m'}\sum_{m}I_{3,\phi_{m'}}\Gamma_{0,mm}^h\mr{Im}(U_{m'i}^{\ast}U_{m'j}U_{mj}^{\ast}U_{mi})\nonumber\\
&+\sum_{m'}\sum_{m\neq n}\sum_{\alpha}I_{3,\phi_{m'}}\kappa_{mn}m_{\phi_{\alpha}}^2\mr{Im}(U_{m'i}^{\ast}U_{m'j}U_{mj}^{\ast}U_{m\alpha}U_{n\alpha}^{\ast}U_{ni}).
\end{align}
In the above, we have utilized the facts that $\Gamma_{0,mm}^h$ and $\kappa_{mn}$ ($m\neq n$) are real. We can see that there are two terms in the $\mr{Im}(g_{I,ij}^Z\Gamma^h_{ji})$. The first term actually includes the fourth order invariant $J^{(4)}\equiv\mr{Im}(U_{m'i}^{\ast}U_{m'j}U_{mj}^{\ast}U_{mi})$, just as the Jarlskog invariant $J\equiv\mr{Im}(U_{\alpha i}U_{\beta j}U_{\alpha j}^{\ast}U_{\beta i}^{\ast})$ defined in Ref. \cite{Jarlskog:1985ht}. The second term includes the sixth order invariant $J^{(6)}\equiv\mr{Im}(U_{m'i}^{\ast}U_{m'j}U_{mj}^{\ast}U_{m\alpha}U_{n\alpha}^{\ast}U_{ni})$, which also appears in Ref. \cite{Wu:1985ea}. \footnote{The Jarlskog-like invariants for scalars have been studied in Refs. \cite{Mendez:1991gp, Lavoura:1994fv, Botella:1994cs}.}

When there are two \sLQ{} generations, we can check $J^{(4)}=J^{(6)}=0$. Let us take $i=1,j=2$ for demonstrations:\\
The fourth order invariant is turned as $J^{(4)}\equiv\mr{Im}(U_{m'1}^{\ast}U_{m'2}U_{m2}^{\ast}U_{m1})$. If $m'=m$, we have $J^{(4)}=0$ obviously; if $m'=1,m=2$, we have $J^{(4)}=\mr{Im}(U_{11}^{\ast}U_{12}U_{22}^{\ast}U_{21})=0$ because of the orthogonal relation $U_{22}^{\ast}U_{21}+U_{12}^{\ast}U_{11}=0$; if $m'=2,m=1$, we have $J^{(4)}=0$ similarly.\\
As for the sixth order invariant, it is turned as $J^{(6)}=\mr{Im}(U_{m'1}^{\ast}U_{m'2}U_{m2}^{\ast}U_{m\alpha}U_{n\alpha}^{\ast}U_{n1})$ with $m\neq n$. For $m=1,n=2$, we have $J^{(6)}=\mr{Im}(U_{m'1}^{\ast}U_{m'2}U_{12}^{\ast}U_{1\alpha}U_{2\alpha}^{\ast}U_{21})$. If $m'=1,\alpha=1$, it is $J^{(6)}=\mr{Im}(U_{11}^{\ast}U_{12}U_{12}^{\ast}U_{11}U_{21}^{\ast}U_{21})=0$; if $m'=1,\alpha=2$, it is $J^{(6)}=\mr{Im}(U_{11}^{\ast}U_{12}U_{12}^{\ast}U_{12}U_{22}^{\ast}U_{21})=|U_{12}|^2\mr{Im}(U_{11}^{\ast}U_{12}U_{22}^{\ast}U_{21})=0$; if $m'=2,\alpha=1$, it is $J^{(6)}=\mr{Im}(U_{21}^{\ast}U_{22}U_{12}^{\ast}U_{11}U_{21}^{\ast}U_{21})=|U_{21}|^2\mr{Im}(U_{21}^{\ast}U_{22}U_{12}^{\ast}U_{11})=0$; if $m'=2,\alpha=2$, it is $J^{(6)}=\mr{Im}(U_{21}^{\ast}U_{22}U_{12}^{\ast}U_{12}U_{22}^{\ast}U_{21})=0$. For $m=2,n=1$, we can also verify $J^{(6)}=0$. This agrees with the general results in two scalar leptoquark extended models shown in Sec. \ref{sec:model:2LQ}.

In Refs. \cite{Bernabeu:1986fc, Branco:1999fs}, the authors propose the condition of $CP$ invariance applicable to an arbitrary number of fermion generations. We can perform analogous procedures for the \sLQ{} generations. The $CP$ violation here originates from the interactions between Higgs and \sLQ{}; thus, a quantity for $CP$ violation can be constructed from these interactions. One key point is that such quantity should be weak bais invariant, and we know that $\mr{Tr}[(M_{0}^2)^r(\Gamma_{0}^h)^s]=\mr{Tr}[(M^2)^r(\Gamma^h)^s]$ is invariant under the weak basis transformations, where $r$ and $s$ are non-negative integers. On the other hand, $CP$ transformation $K$ acts as
\begin{align}
K^{\dag}M^2K=(M^2)^{\ast}=(M^2)^{T},\qquad K^{\dag}\Gamma^hK=(\Gamma^h)^{\ast}=(\Gamma^h)^T.
\end{align}
In the above, the Hermiticity of $M^2$ and $\Gamma^h$ is applied. Then, we can obtain
\begin{align}
K^{\dag}[M^2,\Gamma^h]^rK=[(M^2)^{\ast},(\Gamma^h)^{\ast}]^r=([\Gamma^h,M^2]^r)^T,
\end{align}
which leads to the identity $\mr{Tr}([M^2,\Gamma^h]^r)=0$ for odd $r$. We can diagonalize the $\Gamma^h$ through a unitary matrix $V$, namely $\Gamma^h=V\Gamma_{diag}^hV^{\dag}$. \footnote{The $V$ matrix should not be confused with the $U$ matrix.} When there are three \sLQ{} generations, the following quantity can serve as a physical measure for $CP$ violation:
\begin{align}\label{eqn:model:general:CP3gen}
&\mr{Tr}([M^2,\Gamma^h]^3)=6i~\mr{Im}\{\mr{Tr}[(M^2)^2(\Gamma^h)^2M^2\Gamma^h]\}=6i~\mr{Im}\{\mr{Tr}[(M^2)^2V(\Gamma_{diag}^h)^2V^{\dag}M^2V\Gamma_{diag}^hV^{\dag}]\}\nonumber\\
&=6i\sum_{\alpha,\beta,i,j}m_{\phi_{\alpha}}^4m_{\phi_{\beta}}^2(\Gamma_{diag,ii}^h)^2\Gamma_{diag,jj}^h\mr{Im}(V_{\alpha j}^{\ast}V_{\alpha i}V_{\beta i}^{\ast}V_{\beta j})\nonumber\\
&=6i(m_{\phi_1}^2-m_{\phi_2}^2)(m_{\phi_2}^2-m_{\phi_3}^2)(m_{\phi_1}^2-m_{\phi_3}^2)\cdot\nonumber\\
	&(\Gamma_{diag,11}^h-\Gamma_{diag,22}^h)(\Gamma_{diag,22}^h-\Gamma_{diag,33}^h)(\Gamma_{diag,11}^h-\Gamma_{diag,33}^h)\mr{Im}(V_{13}^{\ast}V_{12}V_{22}^{\ast}V_{23}).
\end{align}
It means that the $CP$ violation will be not observable if any two of the three \sLQ{} masses are degenerate, which is validated explicitly in the $S_1+\widetilde{R}_2+S_3$ model in Sec. \ref{sec:contri:expansion:S1R2tS3}.
\subsection{Two scalar leptoquark extended models}\label{sec:model:2LQ}
For the case of two \sLQ{s}, the unitary matrix contains one rotation angle and one phase parameter, which can be parameterized as
\begin{align}\label{eqn:model:2LQ:U}
&U\equiv\left[\begin{array}{cc}\cos\theta&\sin\theta e^{i\delta}\\-\sin\theta e^{-i\delta}&\cos\theta\end{array}\right].
\end{align}
Inserting Eq. \eqref{eqn:model:2LQ:U} into Eq. \eqref{eqn:model:general:M2}, we have the relations
\begin{align}\label{eqn:model:2LQ:M0sq}
M_{0}^2=\left[\begin{array}{cc}m_{\phi_1}^2\cos^2\theta+m_{\phi_2}^2\sin^2\theta&(m_{\phi_2}^2-m_{\phi_1}^2)\sin\theta\cos\theta e^{i\delta}\\(m_{\phi_2}^2-m_{\phi_1}^2)\sin\theta\cos\theta e^{-i\delta}&m_{\phi_1}^2\sin^2\theta+m_{\phi_2}^2\cos^2\theta\end{array}\right].
\end{align}

Inserting Eq. \eqref{eqn:model:2LQ:U} into Eq. \eqref{eqn:model:general:gamma}, the $h$ coupling matrix in mass eigenstates is computed as
\begin{align}
\Gamma^h=\left[\begin{array}{cc}\makecell{\Gamma_{0,11}^h\cos^2\theta+\Gamma_{0,22}^h\sin^2\theta\\[0.5ex]-\sin\theta\cos\theta[\Gamma_{0,12}^he^{-i\delta}+(\Gamma_{0,12}^h)^{\ast}e^{i\delta}]}
	&\makecell{e^{i\delta}[\Gamma_{0,12}^h\cos^2\theta e^{-i\delta}\\[0.5ex]-(\Gamma_{0,12}^h)^{\ast}\sin^2\theta e^{i\delta}+\sin\theta\cos\theta(\Gamma_{0,11}^h-\Gamma_{0,22}^h)]}\\[4ex]
	\makecell{e^{-i\delta}[(\Gamma_{0,12}^h)^{\ast}\cos^2\theta e^{i\delta}\\[0.5ex]-\Gamma_{0,12}^h\sin^2\theta e^{-i\delta}+\sin\theta\cos\theta(\Gamma_{0,11}^h-\Gamma_{0,22}^h)]}
	&\makecell{\Gamma_{0,11}^h\sin^2\theta+\Gamma_{0,22}^h\cos^2\theta\\[0.5ex]+\sin\theta\cos\theta[\Gamma_{0,12}^he^{-i\delta}+(\Gamma_{0,12}^h)^{\ast}e^{i\delta}]}\end{array}\right].
\end{align}
Note that $\Gamma_{11}^h$ and $\Gamma_{22}^h$ are real. For the $\Gamma_{12}^h$ and $\Gamma_{21}^h$, we have the following relations:
\begin{align}
&\Gamma_{12}^h=e^{i\delta}[\Gamma_{0,12}^h\cos^2\theta e^{-i\delta}-(\Gamma_{0,12}^h)^{\ast}\sin^2\theta e^{i\delta}+\sin\theta\cos\theta(\Gamma_{0,11}^h-\Gamma_{0,22}^h)]\nonumber\\
&=e^{i\delta}[\frac{\Gamma_{0,12}^h}{M_{0,12}^2}(m_{\phi_2}^2-m_{\phi_1}^2)\sin\theta\cos^3\theta-\frac{(\Gamma_{0,12}^h)^{\ast}}{(M_{0,12}^2)^{\ast}}(m_{\phi_2}^2-m_{\phi_1}^2)\sin^3\theta\cos\theta+\sin\theta\cos\theta(\Gamma_{0,11}^h-\Gamma_{0,22}^h)],\nonumber\\
&\Gamma_{21}^h=(\Gamma_{12}^h)^{\ast}.
\end{align}

Inserting Eq. \eqref{eqn:model:2LQ:U} into Eq. \eqref{eqn:model:general:gZ}, the $g_I^Z$ coupling matrix in mass eigenstates is computed as
\begin{align}
g_I^Z=\left[\begin{array}{cc}I_{3,\phi_1}\cos^2\theta+I_{3,\phi_2}\sin^2\theta & (I_{3,\phi_1}-I_{3,\phi_2})\sin\theta\cos\theta e^{i\delta}\\
	(I_{3,\phi_1}-I_{3,\phi_2})\sin\theta\cos\theta e^{-i\delta} & I_{3,\phi_1}\sin^2\theta+I_{3,\phi_2}\cos^2\theta\end{array}\right].
\end{align}

Hence, the $\mr{Im}(g_{12}^{Z}\Gamma_{21}^h)$ is computed as
\begin{align}
&\mr{Im}(g_{12}^{Z}\Gamma_{21}^h)=\mr{Im}(g_{I,12}^{Z}\Gamma_{21}^h)\nonumber\\
&=(I_{3,\phi_1}-I_{3,\phi_2})(m_{\phi_2}^2-m_{\phi_1}^2)\sin^2\theta\cos^2\theta\cdot\mr{Im}[\frac{(\Gamma_{0,12}^h)^{\ast}}{(M_{0,12}^2)^{\ast}}\cos^2\theta-\frac{\Gamma_{0,12}^h}{M_{0,12}^2}\sin^2\theta]\nonumber\\
&=(I_{3,\phi_2}-I_{3,\phi_1})(m_{\phi_2}^2-m_{\phi_1}^2)\sin^2\theta\cos^2\theta\cdot\mr{Im}(\frac{\Gamma_{0,12}^h}{M_{0,12}^2}).
\end{align}

We have known that the $SU_C(3)$ color symmetry restricts the \sLQ{} interactions with the Higgs boson to be cubic type $H\,\mr{LQ}^2$ and quartic type $H^2\mr{LQ}^2$. Consequently, the $\mr{Im}(g_{12}^{Z}\Gamma_{21}^h)$ is zero because the ratio $\Gamma_{0,12}^h/M_{0,12}^2$ is real, which is also presented explicitly in App. \ref{app:model:2LQ:specific} with specific two scalar leptoquark extended model examples. As reasoned in following Sec. \ref{sec:contri:LQ:explicit}, it leads to the fact that the two-field models do not contribute to the $gg\rightarrow Zh$ process exactly. \footnote{This is not a surprise, because we know that physical $CP$ violation can only appear when there are at least three generations in the SM.}

\subsection{Three scalar leptoquark $S_1+\widetilde{R}_2+S_3$ model}\label{sec:model:S1R2tS3}
For this model, there can be mixing interactions of
\begin{align}
-A_{1\widetilde{2}}(\widetilde{R}_2^{\dag}H)S_1^{\dag}+Y_{13}H^{\dag}(S_3^a\sigma_a)^{\dag}HS_1+A_{\widetilde{2}3}\widetilde{R}_2^{\dag}(S_3^a\sigma_a)^{\dag}H+\mathrm{h.c.}.
\end{align}
After EWSB, there are mixings between the $2/3$ and $-1/3$ electrically charged \sLQ{} individually. For the \sLQ{} with electric charge $2/3$, the $2\times2$ mixing terms are presented in the $\widetilde{R}_2+S_3$ model of App. \ref{app:model:2LQ:specific}. For the \sLQ{} with electric charge $-1/3$, the related mass terms and scalar interactions can be parameterized as
\begin{align}
&-\left[\begin{array}{c}S_1\quad(\widetilde{R}_2^{-1/3})^{\dag}\quad S_3^{1/3}\end{array}\right]M_{0}^2\left[\begin{array}{c}(S_1)^{\dag}\\[0.5ex]\widetilde{R}_2^{-1/3}\\[0.5ex](S_3^{1/3})^{\dag}\end{array}\right]
	-vh\left[\begin{array}{c}S_1\quad(\widetilde{R}_2^{-1/3})^{\dag}\quad S_3^{1/3}\end{array}\right]\Gamma_{0}^h\left[\begin{array}{c}(S_1)^{\dag}\\[0.5ex]\widetilde{R}_2^{-1/3}\\[0.5ex](S_3^{1/3})^{\dag}\end{array}\right]\nonumber\\
	&-ivG^0\left[\begin{array}{c}S_1\quad(\widetilde{R}_2^{-1/3})^{\dag}\quad S_3^{1/3}\end{array}\right]\Gamma_{0}^{G^0}\left[\begin{array}{c}(S_1)^{\dag}\\[0.5ex]\widetilde{R}_2^{-1/3}\\[0.5ex](S_3^{1/3})^{\dag}\end{array}\right].
\end{align}
In the above, the mass, $h$ coupling, and $G^0$ coupling matrices are given as
\begin{align}\label{eqn:model:S1R2tS3:M0sqGam0}
&M_{0}^2=\left[\begin{array}{ccc}\mu_{S_1}^2+\frac{1}{2}\lambda_{1S_1}v^2 & \frac{1}{\sqrt{2}}A_{1\widetilde{2}}^{\ast}v & \frac{1}{2}Y_{13}v^2 \\[1ex]
\frac{1}{\sqrt{2}}A_{1\widetilde{2}}v & \mu_{\widetilde{R}_2}^2+\frac{1}{2}(\lambda_{1\widetilde{R}_2}-\frac{1}{2}\lambda_{2\widetilde{R}_2})v^2 & \frac{1}{\sqrt{2}}A_{\widetilde{2}3}v \\[1ex]
\frac{1}{2}Y_{13}^{\ast}v^2 & \frac{1}{\sqrt{2}}A_{\widetilde{2}3}^{\ast}v & \mu_{S_3}^2+\frac{1}{2}\lambda_{1S_3}v^2 \end{array}\right],\nonumber\\
	&\Gamma_{0}^h=\left[\begin{array}{ccc}\lambda_{1S_1} & \frac{1}{\sqrt{2}v}A_{1\widetilde{2}}^{\ast} & Y_{13}\\[1ex]
\frac{1}{\sqrt{2}v}A_{1\widetilde{2}} & \lambda_{1\widetilde{R}_2}-\frac{1}{2}\lambda_{2\widetilde{R}_2} & \frac{1}{\sqrt{2}v}A_{\widetilde{2}3} \\[1ex]
Y_{13}^{\ast} & \frac{1}{\sqrt{2}v}A_{\widetilde{2}3}^{\ast} & \lambda_{1S_3} \end{array}\right],\qquad
	\Gamma_{0}^{G^0}=\left[\begin{array}{ccc} 0 & -\frac{1}{\sqrt{2}v}A_{1\widetilde{2}}^{\ast} & 0 \\[0.5ex]
\frac{1}{\sqrt{2}v}A_{1\widetilde{2}} & 0 & \frac{1}{\sqrt{2}v}A_{\widetilde{2}3} \\[0.5ex]
0 & -\frac{1}{\sqrt{2}v}A_{\widetilde{2}3}^{\ast} & 0 \end{array}\right].
\end{align}

Although we can diagonalize the mass matrix analytically through a  $3\times3$ unitary matrix defined in Eq. \eqref{eqn:model:general:U}, the expressions will be quite lengthy and tedious. Assuming $Y_{13}v^2\ll |A_{1\widetilde{2}}|v,|A_{\widetilde{2}3}|v\ll m_i^2$, the unitary matrix up to $\mc{O}(v^3)$ is computed as
\begin{align}\label{eqn:model:S1R2tS3:Uapp}
&U\approx\left[\begin{array}{ccc}1-\frac{|A_{1\widetilde{2}}|^2v^2}{4(m_{1\widetilde{2}}^2)^2} & -\frac{A_{1\widetilde{2}}^{\ast}v}{\sqrt{2}m_{1\widetilde{2}}^2}+U_{12}^{(3)} & \frac{(-Y_{13}m_{\widetilde{2}3}^2+A_{1\widetilde{2}}^{\ast}A_{\widetilde{2}3})v^2}{2m_{13}^2m_{\widetilde{2}3}^2}\\[1.5ex]
\frac{A_{1\widetilde{2}}v}{\sqrt{2}m_{1\widetilde{2}}^2}+U_{21}^{(3)} & 1-\frac{|A_{1\widetilde{2}}|^2v^2}{4(m_{1\widetilde{2}}^2)^2}-\frac{|A_{\widetilde{2}3}|^2v^2}{4(m_{\widetilde{2}3}^2)^2} & -\frac{A_{\widetilde{2}3}v}{\sqrt{2}m_{\widetilde{2}3}^2}+U_{23}^{(3)} \\[1.5ex]
\frac{(Y_{13}^{\ast}m_{1\widetilde{2}}^2+A_{1\widetilde{2}}A_{\widetilde{2}3}^{\ast})v^2}{2m_{13}^2m_{1\widetilde{2}}^2} & \frac{A_{\widetilde{2}3}^{\ast}v}{\sqrt{2}m_{\widetilde{2}3}^2}+U_{32}^{(3)} & 1-\frac{|A_{\widetilde{2}3}|^2v^2}{4(m_{\widetilde{2}3}^2)^2} \end{array}\right].
\end{align}
Here, we adopt the notations of $m_{1\widetilde{2}}^2\equiv m_{S_1}^2-m_{\widetilde{R}_2^{1/3}}^2$, $m_{13}^2\equiv m_{S_1}^2-m_{S_3^{1/3}}^2$, $m_{\widetilde{2}3}^2\equiv m_{\widetilde{R}_2^{1/3}}^2-m_{S_3^{1/3}}^2$. Note that our $U$ matrix agrees with $(W^{-1/3})^{\dag}$ given in Ref. \cite{Crivellin:2021ejk} up to $\mc{O}(v^2)$. The explicit formulae of $U_{12/21/23/32}^{(3)}$ are given as
\begin{align}\label{eqn:model:S1R2tS3:Uappv3}
&U_{12}^{(3)}=-\frac{A_{\widetilde{2}3}^{\ast}Y_{13}v^3}{2\sqrt{2}m_{1\widetilde{2}}^2m_{\widetilde{2}3}^2}-\frac{A_{1\widetilde{2}}^{\ast}|A_{1\widetilde{2}}|^2v^3}{4\sqrt{2}(m_{1\widetilde{2}}^2)^3},\nonumber\\
&U_{21}^{(3)}=\frac{A_{\widetilde{2}3}Y_{13}^{\ast}v^3}{2\sqrt{2}m_{1\widetilde{2}}^2m_{13}^2}+\frac{A_{1\widetilde{2}}|A_{1\widetilde{2}}|^2v^3}{4\sqrt{2}(m_{1\widetilde{2}}^2)^3}+\frac{A_{1\widetilde{2}}|A_{\widetilde{2}3}|^2v^3}{2\sqrt{2}(m_{1\widetilde{2}}^2)^2}(\frac{1}{m_{13}^2}-\frac{1}{m_{\widetilde{2}3}^2}),\nonumber\\
&U_{23}^{(3)}=\frac{A_{1\widetilde{2}}Y_{13}v^3}{2\sqrt{2}m_{13}^2m_{\widetilde{2}3}^2}-\frac{A_{\widetilde{2}3}|A_{\widetilde{2}3}|^2v^3}{4\sqrt{2}(m_{\widetilde{2}3}^2)^3}+\frac{A_{\widetilde{2}3}|A_{1\widetilde{2}}|^2v^3}{2\sqrt{2}(m_{\widetilde{2}3}^2)^2}(\frac{1}{m_{1\widetilde{2}}^2}-\frac{1}{m_{13}^2}),\nonumber\\
&U_{32}^{(3)}=-\frac{A_{1\widetilde{2}}^{\ast}Y_{13}^{\ast}v^3}{2\sqrt{2}m_{1\widetilde{2}}^2m_{\widetilde{2}3}^2}+\frac{A_{\widetilde{2}3}^{\ast}|A_{\widetilde{2}3}|^2v^3}{4\sqrt{2}(m_{\widetilde{2}3}^2)^3}.
\end{align}
As shown in the following, the $U_{12/21/23/32}^{(3)}$ are of crucial influences to the $gg\rightarrow Zh$ process.

Inserting $I_{3,(S_1)^{\dag}}=0,~I_{3,\widetilde{R}_2^{-1/3}}=-1/2,~I_{3,(S_3^{1/3})^{\dag}}=0$ into Eq. \eqref{eqn:model:general:gZCom}, the $g_{I,ij}^Z$ is computed as
\begin{align}
&g_{I,ij}^Z=-\frac{1}{2}U_{2i}^{\ast}U_{2j}.
\end{align}
Taking the expressions of $\Gamma_{0,ji}^h$ given in Eq. \eqref{eqn:model:S1R2tS3:M0sqGam0} further, we can obtain
\begin{align}
&\mr{Im}(g_{I,ij}^Z\Gamma^h_{ji})=-\frac{1}{2}\mr{Im}(U_{2i}^{\ast}U_{2j}\sum_{m,n}U_{mj}^{\ast}\Gamma_{0,mn}^hU_{ni})\nonumber\\
&=-\frac{1}{2}\Gamma_{0,11}^h\mr{Im}(U_{2i}^{\ast}U_{2j}U_{1j}^{\ast}U_{1i})-\frac{1}{2}\Gamma_{0,33}^h\mr{Im}(U_{2i}^{\ast}U_{2j}U_{3j}^{\ast}U_{3i})-\frac{1}{2\sqrt{2}v}|U_{2i}|^2\mr{Im}(U_{2j}U_{1j}^{\ast}A_{1\widetilde{2}}^{\ast})\nonumber\\
	&-\frac{1}{2\sqrt{2}v}|U_{2j}|^2\mr{Im}(U_{2i}^{\ast}U_{1i}A_{1\widetilde{2}})-\frac{1}{2\sqrt{2}v}|U_{2j}|^2\mr{Im}(U_{2i}^{\ast}U_{3i}A_{\widetilde{2}3})-\frac{1}{2\sqrt{2}v}|U_{2i}|^2\mr{Im}(U_{2j}U_{3j}^{\ast}A_{\widetilde{2}3}^{\ast})\nonumber\\
	&-\frac{1}{2}\mr{Im}(U_{2i}^{\ast}U_{2j}U_{1j}^{\ast}U_{3i}Y_{13})-\frac{1}{2}\mr{Im}(U_{2i}^{\ast}U_{2j}U_{3j}^{\ast}U_{1i}Y_{13}^{\ast}).
\end{align}
Note that there is no sum over the indices $i,j$ and we can take $i<j$.

Adopting the approximations of the unitary matrix exhibited in Eq. \eqref{eqn:model:S1R2tS3:Uapp} and Eq. \eqref{eqn:model:S1R2tS3:Uappv3}, we have the following results of $\mr{Im}(g_{I,ij}^Z\Gamma^h_{ji})$ up to
$\mc{O}(v^2)$:
\begin{align}\label{eqn:model:S1R2tS3:ImgZgammahapp}
&\mr{Im}(g_{I,12}^Z\Gamma^h_{21})\approx-\frac{v^2}{4m_{1\widetilde{2}}^2m_{\widetilde{2}3}^2}\mr{Im}(A_{1\widetilde{2}}^{\ast}A_{\widetilde{2}3}Y_{13}^{\ast}),\nonumber\\
&\mr{Im}(g_{I,13}^Z\Gamma^h_{31})\approx\frac{v^2}{4m_{1\widetilde{2}}^2m_{\widetilde{2}3}^2}\mr{Im}(A_{1\widetilde{2}}^{\ast}A_{\widetilde{2}3}Y_{13}^{\ast}),\nonumber\\
&\mr{Im}(g_{I,23}^Z\Gamma^h_{32})\approx-\frac{v^2}{4m_{1\widetilde{2}}^2m_{\widetilde{2}3}^2}\mr{Im}(A_{1\widetilde{2}}^{\ast}A_{\widetilde{2}3}Y_{13}^{\ast}).
\end{align}

\subsection{Three scalar leptoquark $R_2+\widetilde{R}_2+S_3$ model}\label{sec:model:R2R2tS3}

For this model, there can be mixing interactions of
\begin{align}
Y_{2\widetilde{2}}(R_2^{\dag}H)(H^Ti\sigma_2\widetilde{R}_2)+A_{\widetilde{2}3}\widetilde{R}_2^{\dag}(S_3^a\sigma_a)^{\dag}H+\mathrm{h.c.}.
\end{align}
After EWSB, there are mixings between the $2/3$ and $-1/3$ electrically charged \sLQ{} individually. For the \sLQ{} with electric charge $-1/3$, the $2\times2$ mixing terms are presented in the $\widetilde{R}_2+S_3$ model of App. \ref{app:model:2LQ:specific}. For the \sLQ{} with electric charge $2/3$, the related mass terms and scalar interactions can be parameterized as
\begin{align}
&-\left[\begin{array}{c}(R_2^{2/3})^{\dag}\quad(\widetilde{R}_2^{2/3})^{\dag}\quad S_3^{-2/3}\end{array}\right]M_{0}^2\left[\begin{array}{c}R_2^{2/3}\\[0.5ex]\widetilde{R}_2^{2/3}\\[0.5ex](S_3^{-2/3})^{\dag}\end{array}\right]
	-vh\left[\begin{array}{c}(R_2^{2/3})^{\dag}\quad(\widetilde{R}_2^{2/3})^{\dag}\quad S_3^{-2/3}\end{array}\right]\Gamma_{0}^h\left[\begin{array}{c}R_2^{2/3}\\[0.5ex]\widetilde{R}_2^{2/3}\\[0.5ex](S_3^{-2/3})^{\dag}\end{array}\right]\nonumber\\
	&-ivG^0\left[\begin{array}{c}(R_2^{2/3})^{\dag}\quad(\widetilde{R}_2^{2/3})^{\dag}\quad S_3^{-2/3}\end{array}\right]\Gamma_{0}^{G^0}\left[\begin{array}{c}R_2^{2/3}\\[0.5ex]\widetilde{R}_2^{2/3}\\[0.5ex](S_3^{-2/3})^{\dag}\end{array}\right].
\end{align}
In the above, the mass, $h$ coupling, and $G^0$ coupling matrices are given as
\begin{align}\label{eqn:model:R2R2tS3:M0sqGam0}
&M_{0}^2=\left[\begin{array}{ccc}\mu_{R_2}^2+\frac{1}{2}(\lambda_{1R_2}-\frac{1}{2}\lambda_{2R_2})v^2 & \frac{1}{2}Y_{2\widetilde{2}}v^2 & 0 \\[1ex]
\frac{1}{2}Y_{2\widetilde{2}}^{\ast}v^2 & \mu_{\widetilde{R}_2}^2+\frac{1}{2}(\lambda_{1\widetilde{R}_2}+\frac{1}{2}\lambda_{2\widetilde{R}_2})v^2 & -A_{\widetilde{2}3}v \\[1ex]
0 & -A_{\widetilde{2}3}^{\ast}v & \mu_{S_3}^2+\frac{1}{2}(\lambda_{1S_3}-\lambda_{2S_3})v^2 \end{array}\right],\nonumber\\
	&\Gamma_{0}^h=\left[\begin{array}{ccc}\lambda_{1R_2}-\frac{1}{2}\lambda_{2R_2} & Y_{2\widetilde{2}} & 0\\[1ex]
Y_{2\widetilde{2}}^{\ast} & \lambda_{1\widetilde{R}_2}+\frac{1}{2}\lambda_{2\widetilde{R}_2} & -\frac{1}{v}A_{\widetilde{2}3} \\[1ex]
0 & -\frac{1}{v}A_{\widetilde{2}3}^{\ast} & \lambda_{1S_3}-\lambda_{2S_3} \end{array}\right],\qquad
	\Gamma_{0}^{G^0}=\left[\begin{array}{ccc} 0 & Y_{2\widetilde{2}} & 0 \\[0.5ex]
-Y_{2\widetilde{2}}^{\ast} & 0 & -\frac{1}{v}A_{\widetilde{2}3} \\[0.5ex]
0 & \frac{1}{v}A_{\widetilde{2}3}^{\ast} & 0 \end{array}\right].
\end{align}

Assuming $Y_{13}v^2\ll |A_{1\widetilde{2}}|v,|A_{\widetilde{2}3}|v\ll m_i^2$, the unitary matrix up to $\mc{O}(v^3)$ is computed as
\begin{align}\label{eqn:model:R2R2tS3:Uapp}
&U\approx\left[\begin{array}{ccc}1 & -\frac{Y_{2\widetilde{2}}v^2}{2m_{2\widetilde{2}}^2} & U_{13}^{(3)} \\[1.5ex]
\frac{Y_{2\widetilde{2}}^{\ast}v^2}{2m_{2\widetilde{2}}^2} & 1-\frac{|A_{\widetilde{2}3}|^2v^2}{2(m_{\widetilde{2}3}^2)^2} & \frac{A_{\widetilde{2}3}v}{m_{\widetilde{2}3}^2}+U_{23}^{(3)} \\[1.5ex]
U_{31}^{(3)} & -\frac{A_{\widetilde{2}3}^{\ast}v}{m_{\widetilde{2}3}^2}+U_{32}^{(3)} & 1-\frac{|A_{\widetilde{2}3}|^2v^2}{2(m_{\widetilde{2}3}^2)^2} \end{array}\right].
\end{align}
Here, we adopt the notations of $m_{2\widetilde{2}}^2\equiv m_{R_2^{2/3}}^2-m_{\widetilde{R}_2^{2/3}}^2$, $m_{23}^2\equiv m_{R_2^{2/3}}^2-m_{S_3^{2/3}}^2$, $m_{\widetilde{2}3}^2\equiv m_{\widetilde{R}_2^{2/3}}^2-m_{S_3^{2/3}}^2$. Note that our $U$ matrix agrees with $(W^{+2/3})^{\dag}$ given in Ref. \cite{Crivellin:2021ejk} up to $\mc{O}(v^2)$. The explicit formulae of $U_{13/31/23/32}^{(3)}$ are given as
\begin{align}\label{eqn:model:R2R2tS3:Uappv3}
&U_{13}^{(3)}=-\frac{A_{\widetilde{2}3}Y_{2\widetilde{2}}v^3}{2m_{23}^2m_{\widetilde{2}3}^2},\qquad U_{31}^{(3)}=-\frac{A_{\widetilde{2}3}^{\ast}Y_{2\widetilde{2}}^{\ast}v^3}{2m_{2\widetilde{2}}^2m_{23}^2},\qquad U_{23}^{(3)}=\frac{A_{\widetilde{2}3}|A_{\widetilde{2}3}|^2v^3}{2(m_{\widetilde{2}3}^2)^3},\qquad U_{32}^{(3)}=-\frac{A_{\widetilde{2}3}^{\ast}|A_{\widetilde{2}3}|^2v^3}{2(m_{\widetilde{2}3}^2)^3}.
\end{align}

Inserting $I_{3,R_2^{2/3}}=-1/2,~I_{3,\widetilde{R}_2^{2/3}}=1/2,~I_{3,(S_3^{-2/3})^{\dag}}=1$ into Eq. \eqref{eqn:model:general:gZCom}, the $g_{I,ij}^Z$ is computed as
\begin{align}
&g_{I,ij}^Z=-\frac{1}{2}U_{1i}^{\ast}U_{1j}+\frac{1}{2}U_{2i}^{\ast}U_{2j}+U_{3i}^{\ast}U_{3j}.
\end{align}
Taking the expressions of $\Gamma_{0,ji}^h$ given in Eq. \eqref{eqn:model:R2R2tS3:M0sqGam0} further, we can obtain
\begin{align}
&\mr{Im}(g_{I,ij}^Z\Gamma^h_{ji})=-\frac{1}{2}\mr{Im}(U_{1i}^{\ast}U_{1j}\sum_{m,n}U_{mj}^{\ast}\Gamma_{0,mn}^hU_{ni})+\frac{1}{2}\mr{Im}(U_{2i}^{\ast}U_{2j}\sum_{m,n}U_{mj}^{\ast}\Gamma_{0,mn}^hU_{ni})\nonumber\\
	&+\mr{Im}(U_{3i}^{\ast}U_{3j}\sum_{m,n}U_{mj}^{\ast}\Gamma_{0,mn}^hU_{ni})\nonumber\\
&=-\frac{1}{2}\Gamma_{0,22}^h\mr{Im}(U_{1i}^{\ast}U_{1j}U_{2j}^{\ast}U_{2i})-\frac{1}{2}\Gamma_{0,33}^h\mr{Im}(U_{1i}^{\ast}U_{1j}U_{3j}^{\ast}U_{3i})-\frac{1}{2}|U_{1j}|^2\mr{Im}(U_{1i}^{\ast}U_{2i}Y_{2\widetilde{2}})\nonumber\\
	&-\frac{1}{2}|U_{1i}|^2\mr{Im}(U_{2j}^{\ast}U_{1j}Y_{2\widetilde{2}}^{\ast})+\frac{1}{2v}\mr{Im}(U_{1i}^{\ast}U_{1j}U_{2j}^{\ast}U_{3i}A_{\widetilde{2}3})+\frac{1}{2v}\mr{Im}(U_{1i}^{\ast}U_{1j}U_{3j}^{\ast}U_{2i}A_{\widetilde{2}3}^{\ast})\nonumber\\
	&+\frac{1}{2}\Gamma_{0,11}^h\mr{Im}(U_{2i}^{\ast}U_{2j}U_{1j}^{\ast}U_{1i})+\frac{1}{2}\Gamma_{0,33}^h\mr{Im}(U_{2i}^{\ast}U_{2j}U_{3j}^{\ast}U_{3i})+\frac{1}{2}|U_{2i}|^2\mr{Im}(U_{1j}^{\ast}U_{2j}Y_{2\widetilde{2}})\nonumber\\
	&+\frac{1}{2}|U_{2j}|^2\mr{Im}(U_{2i}^{\ast}U_{1i}Y_{2\widetilde{2}}^{\ast})-\frac{1}{2v}|U_{2j}|^2\mr{Im}(U_{2i}^{\ast}U_{3i}A_{\widetilde{2}3})-\frac{1}{2v}|U_{2i}|^2\mr{Im}(U_{3j}^{\ast}U_{2j}A_{\widetilde{2}3}^{\ast})\nonumber\\
	&+\Gamma_{0,11}^h\mr{Im}(U_{3i}^{\ast}U_{3j}U_{1j}^{\ast}U_{1i})+\Gamma_{0,22}^h\mr{Im}(U_{3i}^{\ast}U_{3j}U_{2j}^{\ast}U_{2i})+\mr{Im}(U_{3i}^{\ast}U_{3j}U_{1j}^{\ast}U_{2i}Y_{2\widetilde{2}})\nonumber\\
	&+\mr{Im}(U_{3i}^{\ast}U_{3j}U_{2j}^{\ast}U_{1i}Y_{2\widetilde{2}}^{\ast})-\frac{1}{v}|U_{3i}|^2\mr{Im}(U_{2j}^{\ast}U_{3j}A_{\widetilde{2}3})-\frac{1}{v}|U_{3j}|^2\mr{Im}(U_{3i}^{\ast}U_{2i}A_{\widetilde{2}3}^{\ast}).
\end{align}
Again, there is no sum over the indices $i,j$ and we can take $i<j$.

Adopting the approximations of the unitary matrix exhibited in Eq. \eqref{eqn:model:R2R2tS3:Uapp} and Eq. \eqref{eqn:model:R2R2tS3:Uappv3}, we have the following results of $\mr{Im}(g_{I,ij}^Z\Gamma^h_{ji})$ up to
$\mc{O}(v^2)$:
\begin{align}
&\mr{Im}(g_{I,12}^Z\Gamma^h_{21})\approx\mr{Im}(g_{I,13}^Z\Gamma^h_{31})\approx\mr{Im}(g_{I,23}^Z\Gamma^h_{32})\approx0.
\end{align}
Note the approximation is valid for $|Y_{ij}|v^2\ll |A_{ij}|v\ll m_i^2$. If $|Y_{ij}|v^2\sim|A_{ij}|v\ll m_i^2$, we need to take into account the $\mc{O}(v^4)$ corrections. \footnote{In Sec. \ref{sec:contri:expansion:S1R2tS3}, we estimate the leading order results under heavy internal mass expansion, which shows the negligible effects compared to the SM contributions. Therefore, there is no need to perform the hard diagonalization at $\mc{O}(v^4)$.}

\section{Contributions to the $gg\rightarrow Zh$ process}\label{sec:contributions}
\subsection{General amplitude structure}\label{sec:contributions:general}
For the $g(k_1)g(k_2)\rightarrow Z(p_1)h(p_2)$ process, momentum conservation requires $k_1+k_2=p_1+p_2$; then, we can choose $k_1,k_2,p_1$ as independent momenta. The Mandelstam variables are defined as $s=(k_1+k_2)^2=(p_1+p_2)^2,t=(k_1-p_1)^2=(k_2-p_2)^2,u=(k_1-p_2)^2=(k_2-p_1)^2$. The amplitude can be written as $\mathcal{M}=\mathcal{M}^{\mu\nu\rho}(k_1,k_2,p_1)\epsilon_{\mu}^{a}(k_1)\epsilon_{\nu}^{a}(k_2)\epsilon_{\rho}^{\ast}(p_1)$, in which $a$ is the color index of gluon. Generally, $\mathcal{M}^{\mu\nu\rho}(k_1,k_2,p_1)$ can be constructed from the basic tensor structures: momentum $\{k_2^{\mu}/p_1^{\mu},k_1^{\nu}/p_1^{\nu},k_1^{\rho}/k_2^{\rho}\}$,\footnote{The $\{k_1^{\mu},k_2^{\nu},p_1^{\rho}\}$ terms do not contribute because of the polarization conditions $k_1^{\mu}\epsilon_{\mu}^{a}(k_1)=0$, $k_2^{\nu}\epsilon_{\nu}^{a}(k_2)=0$, and $p_1^{\rho}\epsilon_{\rho}(p_1)=0$.} metric tensor $\{g^{\mu\nu},g^{\nu\rho},g^{\mu\rho}\}$, and anti-symmetric tensor $\{\{\epsilon^{\mu\nu\rho\times}\},\{\epsilon^{\mu\nu\times\times},\epsilon^{\nu\rho\times\times},\epsilon^{\mu\rho\times\times}\},\{\epsilon^{\mu\times\times\times},\epsilon^{\nu\times\times\times},\epsilon^{\rho\times\times\times}\}\}$. Here, the symbol ``$\times$" labels the contraction with $k_1,k_2,p_1$. For the physical amplitude, the Ward identity requires
\begin{align}\label{eqn:contri:LQ:general:Ward}
k_{1,\mu}\mathcal{M}^{\mu\nu\rho}(k_1,k_2,p_1)=k_{2,\nu}\mathcal{M}^{\mu\nu\rho}(k_1,k_2,p_1)=0.
\end{align}
Moreover, the Bose symmetry requires
\begin{align}\label{eqn:contri:LQ:general:boson}
\mathcal{M}^{\mu\nu\rho}(k_1,k_2,p_1)=\mathcal{M}^{\nu\mu\rho}(k_2,k_1,p_1).
\end{align}
Then, the possible amplitude structure can be constructed respecting the Ward identity in Eq. \eqref{eqn:contri:LQ:general:Ward} and Bose symmetry in Eq. \eqref{eqn:contri:LQ:general:boson}.

Considering the Levi-Civita tensor, the general $CP$-even amplitude is constructed as
\begin{align}\label{eqn:contri:LQ:general:structureEven}
&\mathcal{M}_{even}^{\mu\nu\rho}(k_1,k_2,p_1)=\Big\{F_1[(k_1\cdot k_2)\epsilon^{\mu\nu\rho k_2}-k_2^{\mu}\epsilon^{\nu\rho k_1k_2}]+F_2(p_1^{\mu}-\frac{p_1\cdot k_1}{k_1\cdot k_2}k_2^{\mu})\epsilon^{\nu\rho k_1k_2}\nonumber\\
	&+F_3(p_1^{\mu}-\frac{p_1\cdot k_1}{k_1\cdot k_2}k_2^{\mu})\epsilon^{\nu\rho k_2p_1}+(k_1,\mu)\leftrightarrow(k_2,\nu)\Big\}\nonumber\\
	&+F_4[-(k_1\cdot k_2)\epsilon^{\mu\nu\rho p_1}-g^{\mu\nu}\epsilon^{\rho k_1k_2p_1}+k_2^{\mu}\epsilon^{\nu\rho k_1p_1}-k_1^{\nu}\epsilon^{\mu\rho k_2p_1}].
\end{align}
Here, $F_{1,2,3,4}\equiv F_{1,2,3,4}(s,t,u)$ are the form factors. The contributions in SM are in the form of $\mathcal{M}_{even}^{\mu\nu\rho}(k_1,k_2,p_1)$ \cite{Kniehl:1990iva, Kniehl:1990zu}, in which $F_{4}(s,t,u)=-F_{4}(s,u,t)$.

Without considering the Levi-Civita tensor, the general $CP$-odd amplitude is constructed as
\begin{align}\label{eqn:contri:LQ:general:structureOdd}
&\mathcal{M}_{odd}^{\mu\nu\rho}(k_1,k_2,p_1)=G_1[k_2^{\mu}k_1^{\nu}-(k_1\cdot k_2)g^{\mu\nu}]k_1^{\rho}\nonumber\\
	&+G_2[k_2^{\mu}p_1^{\nu}k_1^{\rho}-(k_1\cdot k_2)g^{\mu\rho}p_1^{\nu}+(p_1\cdot k_2)g^{\mu\rho}k_1^{\nu}-(p_1\cdot k_2)g^{\mu\nu}k_1^{\rho}]\nonumber\\
	&+G_3[p_1^{\mu}p_1^{\nu}k_2^{\rho}-(p_1\cdot k_2)g^{\nu\rho}p_1^{\mu}-\frac{p_1\cdot k_1}{k_1\cdot k_2}k_2^{\mu}p_1^{\nu}k_2^{\rho}+\frac{(p_1\cdot k_1)(p_1\cdot k_2)}{k_1\cdot k_2}g^{\nu\rho}k_2^{\mu}]+(k_1,\mu)\leftrightarrow(k_2,\nu).
\end{align}
Note that $\mathcal{M}_{odd}^{\mu\nu\rho}$ contains three complete new tensor structures. It is not only induced by the \sLQ{} interactions in this paper, but also can appear in many other new physics models. Here, $G_{1,2,3}\equiv G_{1,2,3}(s,t,u)$ are the form factors. For simplicity, we can denote $G_i(s,u,t)$ as $\widetilde{G}_i\equiv\widetilde{G}_i(s,t,u)$.

For the squared amplitude, we average the initial spin and color degrees of freedom and sum over spin of final states. There are no interference terms between the $CP$-even and $CP$-odd tensor structures. For the $CP$-odd structure $\mathcal{M}_{odd}^{\mu\nu\rho}(k_1,k_2,p_1)$,
the squared amplitude is calculated as
\begin{align}
&256\overline{|\mc{M}|^2}=\frac{s^2[(m_Z^2-t)G_1+m_Z^2G_2]^2}{8m_Z^2}+\frac{s^2[(m_Z^2-u)\widetilde{G}_1+m_Z^2\widetilde{G}_2]^2}{8m_Z^2}+\frac{(tu-m_h^2m_Z^2)^2}{8m_Z^2}(G_2^2+\widetilde{G}_2^2)\nonumber\\
	&-\frac{s^2(m_Z^2-t)}{4}G_1\widetilde{G}_2-\frac{s^2(m_Z^2-u)}{4}\widetilde{G}_1G_2+(tu-m_h^2m_Z^2)(-\frac{s}{2}G_1G_3-\frac{s}{2}\widetilde{G}_1\widetilde{G}_3)\nonumber\\
	&+(tu-m_h^2m_Z^2)\big\{\frac{[(m_Z^2-u)G_3+s\widetilde{G}_2]^2}{2s}+\frac{[(m_Z^2-t)\widetilde{G}_3+sG_2]^2}{2s}\big\}\nonumber\\
	&+(tu-m_h^2m_Z^2-m_Z^2s)[\frac{s^2}{4m_Z^2}G_1\widetilde{G}_1+\frac{(m_Z^2-t)(m_Z^2-u)}{4m_Z^2}G_2\widetilde{G}_2-\frac{(tu-m_h^2m_Z^2)}{2s}G_3\widetilde{G}_3].
\end{align}
Note that there are no $G_2G_3$ and $G_1\widetilde{G}_3$ ($G_3\widetilde{G}_1$) terms because of the Lorentz contraction.
\subsection{Contributions in the scalar leptoquark models}
\subsubsection{Gauge independence of the contributions}
\begin{figure}[!h]
\centering
\subfloat[][]{\includegraphics[scale=0.45]{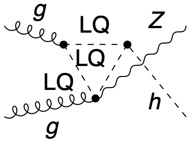}}\qquad
\subfloat[][]{\includegraphics[scale=0.45]{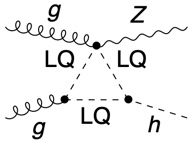}}\qquad
\subfloat[][]{\includegraphics[scale=0.45]{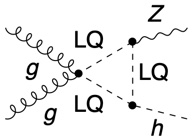}}
\\\rule[2mm]{12cm}{1pt}\\
\subfloat[][]{\includegraphics[scale=0.45]{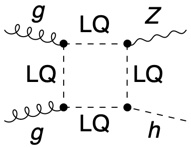}}\qquad
\subfloat[][]{\includegraphics[scale=0.45]{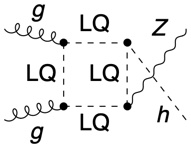}}\qquad
\subfloat[][]{\includegraphics[scale=0.45]{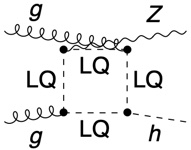}}
\captionsetup{justification=raggedright}
\caption{Triangle (a, b, c) and box (d, e, f) Feynman diagrams contributing to the $gg\rightarrow Zh$ process. Both the clockwise and counter-clockwise diagrams should be included for each diagram. The Feynman diagrams are generated by FeynArts \cite{Hahn:2000kx}.}
\label{fig:SLQcontri:Feynman}
\end{figure}

\begin{figure}[!h]\centering
\subfloat[][]{\framebox[1.1\width][s]{\includegraphics[scale=0.45]{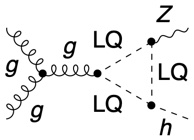}}}\qquad
\subfloat[][]{\framebox[1.1\width][s]{\includegraphics[scale=0.45]{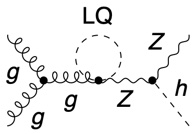}}}\qquad
\subfloat[][]{\framebox[1.1\width][s]{\includegraphics[scale=0.45]{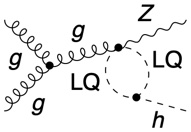}}}\\\vspace{2ex}
\subfloat[][]{\framebox[1.1\width][s]{\includegraphics[scale=0.45]{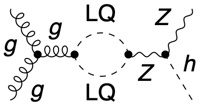}}}\hspace{20ex}
\subfloat[][]{\framebox[1.1\width][s]{\includegraphics[scale=0.45]{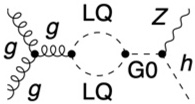}}}
\\[2ex]\rule[2mm]{12cm}{1pt}\\
\subfloat[][]{\includegraphics[scale=0.45]{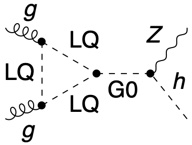}}\hspace{20ex}
\subfloat[][]{\includegraphics[scale=0.45]{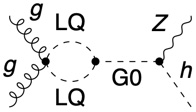}}
\captionsetup{justification=raggedright}
\caption{Feynman diagrams without contributing to the $gg\rightarrow Zh$ process. In the two rows of upper panel (a, b, c, d, e), the diagrams wrapped by the square frame vanish automatically because the color trace is zero. In the row of lower panel (f, g), the diagrams vanish automatically because the diagonal $G^0-\mr{LQ}-\mr{LQ}$ interactions are absent.}
\label{fig:SLQcontri:Feynman:vanish}
\end{figure}

\begin{figure}[!h]\centering
\subfloat[][]{\includegraphics[scale=0.45]{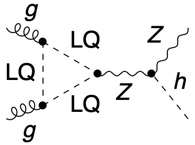}}
\\\rule[2mm]{12cm}{1pt}\\
\subfloat[][]{\includegraphics[scale=0.45]{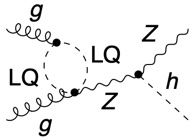}}\qquad
\subfloat[][]{\includegraphics[scale=0.45]{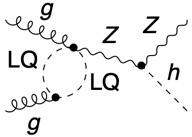}}\qquad
\subfloat[][]{\includegraphics[scale=0.45]{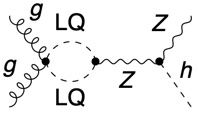}}
\captionsetup{justification=raggedright}
\caption{Feynman diagrams without contributing to the $gg\rightarrow Zh$ process. In the row of upper panel (a), both the clockwise and counter-clockwise triangle diagrams should be included, which make the zero total contribution. In the lower panel (b, c, d), the bubble-like diagrams vanish after loop integration.}
\label{fig:SLQcontri:Feynman:vanishInt}
\end{figure}

We classify all the related Feynman diagrams into three pieces, which are shown in Fig. \ref{fig:SLQcontri:Feynman}, Fig. \ref{fig:SLQcontri:Feynman:vanish}, and Fig. \ref {fig:SLQcontri:Feynman:vanishInt}. In fact, only the one-loop Feynman diagrams in Fig. \ref{fig:SLQcontri:Feynman} can contribute. Then, we investigate their contributions in the following Sec. \ref{sec:contri:LQ:explicit}. In Fig. \ref{fig:SLQcontri:Feynman}, there are no $Z$, Goldstone, and ghost propagators. Note that all the vertexes do not rely on the gauge parameter $\xi$.

For the diagrams (a, b, c, d, e) in Fig. \ref{fig:SLQcontri:Feynman:vanish}, the amplitudes vanish due to zero color trace. For the diagrams (e, f, g) in Fig. \ref{fig:SLQcontri:Feynman:vanish}, the amplitudes vanish due to absence of diagonal $G^0-\mr{LQ}-\mr{LQ}$ interactions, which is illustrated in App. \ref{app:interactions:G0LQLQ}.

For the diagram (a) in Fig. \ref{fig:SLQcontri:Feynman:vanishInt}, there are exact cancellation between the clockwise and counter-clockwise diagrams, which is demonstrated evidently in App. \ref{app:amp:vanish}. For the diagram (b, c, d) in Fig. \ref{fig:SLQcontri:Feynman:vanishInt}, the amplitudes vanish due to the identity $2B_1(k_1^2,m_i^2,m_i^2)+B_0(k_1^2,m_i^2,m_i^2)=0$, which is also demonstrated evidently in App. \ref{app:amp:vanish}.

Therefore, gauge independence of the total amplitude is guaranteed.

\subsubsection{Explicit amplitude formulae in the scalar leptoquark models}\label{sec:contri:LQ:explicit}
As aforementioned, only the one-loop Feynman diagrams in Fig. \ref{fig:SLQcontri:Feynman} can contribute. Here, we investigate their contributions. The computation details are presented in App. \ref{app:amp:contri}. Our results confirm the new tensor structures constructed in Eq. \eqref{eqn:contri:LQ:general:structureOdd}, and the form factor $G_{1,2,3}(s,t,u)$ are expressed in the form of
\begin{align}\label{eqn:SLQcontri:explicit:G123}
&G_{1,2,3}=-\sum_{i,j}\frac{g_2g_s^2\mr{Im}(g_{ij}^{Z}\Gamma_{ji}^h)v}{2\pi^2c_W}\sum_{m=1}^{10}\overline{G}_{1,2,3}^{(m)}\nonumber\\
&=-\sum_{i<j}\frac{g_2g_s^2\mr{Im}(g_{ij}^{Z}\Gamma_{ji}^h)v}{2\pi^2c_W}[\sum_{m=1}^{10}\overline{G}_{1,2,3}^{(m)}-\sum_{m=1}^{10}\overline{G}_{1,2,3}^{(m)}(i\leftrightarrow j)].
\end{align}
Here, the $\overline{G}_{1,2,3}^{(m)}(i\leftrightarrow j)$ stands for substitution of $m_i\leftrightarrow m_j$ in $\overline{G}_{1,2,3}^{(m)}$. By means of FeynCalc \cite{Shtabovenko:2020gxv}, the $\overline{G}_{1,2,3}^{(m)}(m=1,2,...,10)$ are simplified in terms of following ten loop functions:
\begin{align}
&B_0^i(s),\qquad\qquad\qquad D_0^{iiij}(s,t)/D_0^{iiij}(s,u)/D_0^{iijj}(t,u)\nonumber\\
&C_0^i(s)/C_0^{iij}(m_h^2,t)/C_0^{iij}(m_h^2,u)/C_0^{iij}(m_Z^2,t)/C_0^{iij}(m_Z^2,u)/C_0^{iji}(m_h^2,m_Z^2,s).
\end{align}
The shorthand notations are defined in App. \ref{app:loop}, which are basically scalar integrals $B_0/C_0/D_0$ with specific variables. In App. \ref{app:results}, we present the explicit expressions of $\overline{G}_{1,2,3}^{(m)}$. Note that the contributions in Eq. \eqref{eqn:SLQcontri:explicit:G123} are universal for the color triplet and electroweak multiplet scalars, and the explicit expressions of $\mr{Im}(g_{ij}^{Z}\Gamma_{ji}^h)$ are model dependent.

As pointed in Ref. \cite{Kniehl:2011aa}, the $h-G^0-Z$, $Z-\mr{LQ}-\mr{LQ}$, and $g-\mr{LQ}-\mr{LQ}$ couplings linearly depend on the momentum. If the total number of such vertex is odd, the amplitude will change sign under the reversed loop momentum flow. Thus, the contributions from diagonal interactions vanish when adding the clockwise and counter-clockwise diagrams together, which is obvious in Eq. \eqref{eqn:SLQcontri:explicit:G123}. 

For the contributions from off-diagonal interactions, the $g_{ij}^{Z}$ and $\Gamma_{ji}^h$ are model dependent. In Sec. \ref{sec:model:1LQ} and Sec. \ref{sec:model:2LQ}, we demonstrate $\mr{Im}(g_{ij}^{Z}\Gamma_{ji}^h)=0$ for $i\neq j$, which means that the contributions exactly vanish in the one-\sLQ{} and two-\sLQ{} models. For the three-\sLQ{} models, there can be non-zero contributions in the presence of three different \sLQ{} mass and $CP$ violation as expected in Eq. \eqref{eqn:model:general:CP3gen}. In Sec. \ref{sec:model:S1R2tS3}, we find that $\mr{Im}(g_{ij}^{Z}\Gamma_{ji}^h)$ does not vanish at $\mc{O}(v^2)$ in the $S_1+\widetilde{R}_2+S_3$ model, which implies there are enough degrees of freedom such that $CP$ violation can generate physical effects in this process. In Sec. \ref{sec:model:R2R2tS3}, we find that $\mr{Im}(g_{ij}^{Z}\Gamma_{ji}^h)$  vanishes at $\mc{O}(v^2)$ in the $R_2+\widetilde{R}_2+S_3$ model.

\section{Heavy internal mass expansion of the contributions}\label{sec:contri:expansion}
\subsection{General expansion results}

In the limit of $m_h^2,m_Z^2,\hat{s},\hat{t},\hat{u}\ll m_{\mr{LQ}}^2$, we can obtain the one-loop \sLQ{} contributions under heavy internal mass approximation. Adopting the expansions of the loop integrals in App. \ref{App:loop:expansion}, the coefficients $G_{1,2,3}$ defined in Eq. \eqref{eqn:SLQcontri:explicit:G123} can be expanded as
\begin{align}\label{eqn:SLQcontri:expansion:G123}
&G_1=-\sum_{i<j}\frac{g_2g_s^2\mr{Im}(g_{ij}^{Z}\Gamma_{ji}^h)v}{2\pi^2c_W}\cdot\Big[\frac{m_i^8-8m_i^6m_j^2+8m_i^2m_j^6-m_j^8+12m_i^4m_j^4\log\frac{m_i^2}{m_j^2}}{24m_i^2m_j^2(m_i^2-m_j^2)^4}+\mc{O}(\frac{1}{m_{\mr{LQ}}^6})\Big],\nonumber\\
&G_2=-\sum_{i<j}\frac{g_2g_s^2\mr{Im}(g_{ij}^{Z}\Gamma_{ji}^h)v}{2\pi^2c_W}\cdot\Big[\frac{3(m_i^4-m_j^4)-(m_i^4+4m_i^2m_j^2+m_j^4)\log\frac{m_i^2}{m_j^2}}{12(m_i^2-m_j^2)^4}+\mc{O}(\frac{1}{m_{\mr{LQ}}^6})\Big],\nonumber\\
&G_3=-\sum_{i<j}\frac{g_2g_s^2\mr{Im}(g_{ij}^{Z}\Gamma_{ji}^h)v}{2\pi^2c_W}\cdot\mc{O}(\frac{1}{m_{\mr{LQ}}^6}).
\end{align}
As we can see, the loop functions are at $\mc{O}(1/m_\mr{LQ}^4)$. Because the mixings are usually suppressed by the heavy \sLQ{} mass, there can be extra supressions from off-diagonal coupling product $\mr{Im}(g_{ij}^{Z}\Gamma_{ji}^h)$. The expansion results above have been checked by the LoopTools numerically \cite{Hahn:1998yk}. For simplicity, we define a mass splitting quantity $m_{ij}^2\equiv m_i^2-m_j^2$, which is usually subject to constraints from electroweak precision observables. Based on the expansion results in Eq. \eqref{eqn:SLQcontri:expansion:G123}, let us consider the following small and large \sLQ{} mass splitting scenarios.

\begin{itemize}[itemindent=-10pt]

\item Small \sLQ{} mass splitting scenario

If $m_{ij}^2\ll m_{i,j}^2$, the expansion results above can be further approximated as
\begin{align}\label{eqn:SLQcontri:expansion:G123:small}
&G_1=-\sum_{i<j}\frac{g_2g_s^2\mr{Im}(g_{ij}^{Z}\Gamma_{ji}^h)v}{120\pi^2c_W}\cdot\Big[\frac{m_{ij}^2}{m_j^6}+\mc{O}(\frac{1}{m_{\mr{LQ}}^6})\Big],\nonumber\\
&G_2=\sum_{i<j}\frac{g_2g_s^2\mr{Im}(g_{ij}^{Z}\Gamma_{ji}^h)v}{720\pi^2c_W}\cdot\Big[\frac{m_{ij}^2}{m_j^6}+\mc{O}(\frac{1}{m_{\mr{LQ}}^6})\Big].
\end{align}

\item Large \sLQ{} mass splitting scenario

If there is a large mass splitting between $m_i$ and $m_j$, things are a little complex. For a moderate large mass splitting, we should take the full form of Eq. \eqref{eqn:SLQcontri:expansion:G123}. For a extreme large mass splitting, the expansion results in Eq. \eqref{eqn:SLQcontri:expansion:G123} can be further approximated as
\begin{align}\label{eqn:SLQcontri:expansion:large}
&G_1=-\sum_{i<j}\frac{g_2g_s^2\mr{Im}(g_{ij}^{Z}\Gamma_{ji}^h)v}{2\pi^2c_W}\cdot\left\{ \begin{array}{ll}
-\frac{1}{24m_i^2m_j^2}+\mc{O}(\frac{1}{m_{\mr{LQ}}^6})& ,\quad\textrm{for}\quad m_i\ll m_j\\[2ex]
\frac{1}{24m_i^2m_j^2}+\mc{O}(\frac{1}{m_{\mr{LQ}}^6})& ,\quad\textrm{for}\quad m_i\gg m_j\\[2ex]\end{array} \right.,\nonumber\\
&G_2=-\sum_{i<j}\frac{g_2g_s^2\mr{Im}(g_{ij}^{Z}\Gamma_{ji}^h)v}{2\pi^2c_W}\cdot\left\{ \begin{array}{ll}
\frac{1}{12m_j^4}(\log\frac{m_j^2}{m_i^2}-3)+\mc{O}(\frac{1}{m_{\mr{LQ}}^6})& ,\quad\textrm{for}\quad m_i\ll m_j\\[2ex]
-\frac{1}{12m_i^4}(\log\frac{m_i^2}{m_j^2}-3)+\mc{O}(\frac{1}{m_{\mr{LQ}}^6})& ,\quad\textrm{for}\quad m_i\gg m_j\\[2ex]\end{array} \right..
\end{align}

\end{itemize}

\subsection{Expansion results in the $S_1+\widetilde{R}_2+S_3$ model}\label{sec:contri:expansion:S1R2tS3}

In the $S_1+\widetilde{R}_2+S_3$ model, there are three \sLQ{} mass scales $m_{S_1}, m_{\widetilde{R}_2^{1/3}}, m_{S_3^{1/3}}$, which are abbreviated as $m_{1}, m_{\widetilde{2}}, m_{3}$, respectively. For the heavy internal mass expansion results, we consider the following three scenarios: small mass splitting, two splitting mass scales, and three splitting mass scales.

\begin{itemize}[itemindent=-10pt]

\item Scenario with small \sLQ{} mass splitting

In this scenario, there are small splittings between the three mass scales $m_{1}, m_{\widetilde{2}}, m_{3}$. Inserting approximate $\mr{Im}(g_{ij}^{Z}\Gamma_{ji}^h)$ from Eq. \eqref{eqn:model:S1R2tS3:ImgZgammahapp} into Eq. \eqref{eqn:SLQcontri:expansion:G123:small}, we find that 
\begin{align}
&\frac{\mr{Im}(g_{12}^{Z}\Gamma_{21}^h)}{m_{\mr{LQ}}^4}\cdot\frac{m_{1\widetilde{2}}^2}{m_{\mr{LQ}}^2}+\frac{\mr{Im}(g_{13}^{Z}\Gamma_{31}^h)}{m_{\mr{LQ}}^4}\cdot\frac{m_{13}^2}{m_{\mr{LQ}}^2}+\frac{\mr{Im}(g_{23}^{Z}\Gamma_{32}^h)}{m_{\mr{LQ}}^4}\cdot\frac{m_{\widetilde{2}3}^2}{m_{\mr{LQ}}^2}\nonumber\\
&=\frac{v^2}{4m_{1\widetilde{2}}^2m_{\widetilde{2}3}^2}\mr{Im}(A_{1\widetilde{2}}^{\ast}A_{\widetilde{2}3}Y_{13}^{\ast})\frac{-m_{1\widetilde{2}}^2+m_{13}^2-m_{\widetilde{2}3}^2}{m_{\mr{LQ}}^6}=0.
\end{align}
This leads to $G_1=G_2=0$ at $\mr{Im}(g_{ij}^{Z}\Gamma_{ji}^h)\cdot\mc{O}(1/m_{\mr{LQ}}^4)$ level, which means that the \sLQ{} contributions vanish for small mass splitting scenario in the $S_1+\widetilde{R}_2+S_3$ model. This behaviour is expected becase the $G_{1,2,3}$ are exactly cancelled for the same loop particles from Eq. \eqref{eqn:SLQcontri:explicit:G123}.

\item Scenario with two splitting mass scales\\
In this scenario, there are two splitting mass scales, which covers the following six cases:
\begin{align}
&m_{1}\approx m_{\widetilde{2}}\ll m_{3},\quad m_{3}\ll m_{1}\approx m_{\widetilde{2}},\quad m_{1}\approx m_{3}\ll m_{\widetilde{2}},\nonumber\\
&m_{\widetilde{2}}\ll m_{1}\approx m_{3},\quad m_{\widetilde{2}}\approx m_{3} \ll m_{1},\quad m_{1}\ll m_{\widetilde{2}}\approx m_{3}.
\end{align}
Inserting the approximate $\mr{Im}(g_{ij}^{Z}\Gamma_{ji}^h)$ from Eq. \eqref{eqn:model:S1R2tS3:ImgZgammahapp} into Eq. \eqref{eqn:SLQcontri:expansion:large}, we find that 
\begin{align}
&G_{1,2}=-\frac{g_2g_s^2}{2\pi^2c_W}\frac{v^3}{4m_{1\widetilde{2}}^2m_{\widetilde{2}3}^2}\mr{Im}(A_{1\widetilde{2}}^{\ast}A_{\widetilde{2}3}Y_{13}^{\ast})\cdot\mc{O}(\frac{1}{m_{\mr{LQ}}^6}).
\end{align}
This behaviour is just expected by the $CP$ violated quantity presented in Eq. \eqref{eqn:model:general:CP3gen}. Due to the nearly degeneracy of two \sLQ{} masses, it can be also understood and treated as a two-field model in nature. Meanwhile, we have demonstrated that the \sLQ{} contributions vanish for the two-field models in Sec. \ref{sec:model:2LQ}.

\item Scenario with three splitting mass scales\\
In this scenario, the three mass scales are all split. Inserting the approximate $\mr{Im}(g_{ij}^{Z}\Gamma_{ji}^h)$ from Eq. \eqref{eqn:model:S1R2tS3:ImgZgammahapp} into Eq. \eqref{eqn:SLQcontri:expansion:large}, the $G_1$ can be expanded as
\begin{align}
&G_1\approx-\frac{g_2g_s^2v^3}{8\pi^2c_W}\mr{Im}(A_{1\widetilde{2}}^{\ast}A_{\widetilde{2}3}Y_{13}^{\ast})\cdot\left\{ \begin{array}{ll}
	\frac{1}{24m_{1}^2m_{\widetilde{2}}^4m_{3}^2}& ,\quad\textrm{for}\quad m_{1}\ll m_{\widetilde{2}}\ll m_{3}\\[2ex]
	\frac{1}{24m_{1}^2m_{\widetilde{2}}^4m_{3}^2}& ,\quad\textrm{for}\quad m_{1}\ll m_{3}\ll m_{\widetilde{2}}\\[2ex]
	\frac{1}{24m_{1}^4m_{\widetilde{2}}^2m_{3}^2}& ,\quad\textrm{for}\quad m_{\widetilde{2}}\ll m_{1}\ll m_{3}\\[2ex]
	-\frac{1}{24m_{1}^2m_{\widetilde{2}}^2m_{3}^4}& ,\quad\textrm{for}\quad m_{\widetilde{2}}\ll m_{3}\ll m_{1} \\[2ex]
	-\frac{1}{24m_{1}^2m_{\widetilde{2}}^4m_{3}^2}& ,\quad\textrm{for}\quad m_{3}\ll m_{1} \ll m_{\widetilde{2}}\\[2ex]
	-\frac{1}{24m_{1}^2m_{\widetilde{2}}^4m_{3}^2}& ,\quad\textrm{for}\quad m_{3}\ll m_{\widetilde{2}} \ll m_{1}\end{array} \right..
\end{align}
Inserting the approximate $\mr{Im}(g_{ij}^{Z}\Gamma_{ji}^h)$ from Eq. \eqref{eqn:model:S1R2tS3:ImgZgammahapp} into Eq. \eqref{eqn:SLQcontri:expansion:large}, the $G_2$ can be expanded as
\begin{align}
&G_2\approx-\frac{g_2g_s^2v^3}{8\pi^2c_W}\mr{Im}(A_{1\widetilde{2}}^{\ast}A_{\widetilde{2}3}Y_{13}^{\ast})\cdot\left\{ \begin{array}{ll}
	-\frac{1}{12m_{\widetilde{2}}^6m_{3}^2}(\log\frac{m_{\widetilde{2}}^2}{m_{1}^2}-3)& ,\quad\textrm{for}\quad m_{1}\ll m_{\widetilde{2}}\ll m_{3}\\[2ex]
	-\frac{1}{12m_{\widetilde{2}}^4m_{3}^4}(\log\frac{m_{3}^2}{m_{1}^2}-3)& ,\quad\textrm{for}\quad m_{1}\ll m_{3}\ll m_{\widetilde{2}}\\[2ex]
	-\frac{1}{12m_{1}^6m_{3}^2}(\log\frac{m_{1}^2}{m_{\widetilde{2}}^2}-3)& ,\quad\textrm{for}\quad m_{\widetilde{2}}\ll m_{1}\ll m_{3}\\[2ex]
	\frac{1}{12m_{1}^2m_{3}^6}(\log\frac{m_{3}^2}{m_{\widetilde{2}}^2}-3)& ,\quad\textrm{for}\quad m_{\widetilde{2}}\ll m_{3}\ll m_{1} \\[2ex]
	\frac{1}{12m_{1}^4m_{\widetilde{2}}^4}(\log\frac{m_{1}^2}{m_{3}^2}-3)& ,\quad\textrm{for}\quad m_{3}\ll m_{1} \ll m_{\widetilde{2}}\\[2ex]
	\frac{1}{12m_{1}^2m_{\widetilde{2}}^6}(\log\frac{m_{\widetilde{2}}^2}{m_{3}^2}-3)& ,\quad\textrm{for}\quad m_{3}\ll m_{\widetilde{2}} \ll m_{1}\end{array} \right..
\end{align}

Note that the $CP$ violation appears in the form of $\mr{Im}(A_{1\widetilde{2}}^{\ast}A_{\widetilde{2}3}Y_{13}^{\ast})$. The $A_{1\widetilde{2}}^{\ast}$ and $A_{\widetilde{2}3}$ are dimensional couplings in front of the cubic type $H\,\mr{LQ}^2$ interactions, which can be taken as the \sLQ{} mass scale. Therefore, $G_{1,2}$ are suppressed by the factor $1/m_{\mr{LQ}}^6$ eventually, which are further explained from viewpoint of SMEFT in the following Sec. \ref{sec:contri:EFT}.

In SM, the $F_1$ is expanded as \cite{Kniehl:1990zu}
\begin{align}
F_1=\frac{ig_2g_s^2}{16\pi^2vc_W}\cdot\frac{m_Z^2-t}{6m_t^2s}+\mc{O}(\frac{1}{m_t^4}).
\end{align}
Since $\sqrt{s}>m_Z+m_h$ and $\sqrt{2}m_t\approx v$, we can take $(m_Z^2-t)/(6m_t^2s)$ as $1/v^2$ roughly and this is enough to capture the magnitude. At leading order, the ratios between \sLQ{} and SM contributions are naively estimated as
\begin{align}
\frac{G_{1,2}}{F_1}\sim\frac{v^6\mr{Im}(A_{1\widetilde{2}}^{\ast}A_{\widetilde{2}3}Y_{13}^{\ast})}{m_{\mr{LQ}}^8}.
\end{align}
Note that the coupling $Y_{13}$ can be taken as $\mc{O}(1)$ and the lower limit of \sLQ{} mass is typically at 2 TeV. Then, we can obtain $G_{1,2}/F_1\sim v^6/m_{\mr{LQ}}^6\lesssim10^{-6}$. Even if the interference contributions can be extracted through $Z$ polarization and differential distributions, the $10^{-6}$ effects are too small. In fact, the magnitude estimation is quite general for the color triplet and electroweak multiplet scalar contributions. The contribution in Eq. \eqref{eqn:SLQcontri:explicit:G123} is a model independent formula for general $h-\mr{LQ}-\mr{LQ}$ and $Z-\mr{LQ}-\mr{LQ}$ interactions, which are valid for the color triplet scalars. Meanwhile, the off-diagonal coupling product $\mr{Im}(g_{ij}^{Z}\Gamma_{ji}^h)$ ($i\neq j$) is of order $\mc{O}(v^2/m_{\mr{LQ}}^2)$, which makes the loop expansion in Eq. \eqref{eqn:SLQcontri:expansion:G123} to be $vG_{1,2}\sim v^4/m_{\mr{LQ}}^6$.\\
Currently, the $V(W/Z)h$ associated production has been observed \cite{ATLAS:2018kot, CMS:2018nsn}, while the $gg\rightarrow Zh$ channel is still not observed yet. The cross section of $gg\rightarrow Zh$ is roughly $15\%$ of the $pp\rightarrow Zh$ process at the center-of-mass energy of $\sqrt{s}=14\mr{TeV}$ \cite{LHCHiggsCrossSectionWorkingGroup:2016ypw}. At near future high luminosity LHC (HL-LHC), the projection is performed by extrapolating the present measurements in the $h\rightarrow b\bar{b}$ channel. From simultaneous three-parameter fit of the processes $Wh$, $q\bar{q}\rightarrow Zh$, $gg\rightarrow Zh$, the expected uncertainty for $gg\rightarrow Zh$ production is $\sim40\%$ at $1\sigma$ confidence level \cite{Cepeda:2019klc}. Therefore, measurements of the $v^6/m_{\mr{LQ}}^6\lesssim10^{-6}$ \sLQ{} effects are quite challenging.
\end{itemize}

\subsection{Analysis in the effective field theory}\label{sec:contri:EFT}

In Sec. \ref{sec:contributions:general}, we construct the traditional $CP$-even and new $CP$-odd tensor structures from Ward identity and Bose symmetry in Eq. \eqref{eqn:contri:LQ:general:structureEven} and Eq. \eqref{eqn:contri:LQ:general:structureOdd}, respectively. Then, we compute the form factors in the \sLQ{} models. As matter of fact, they can also be analysed and confirmed within the SMEFT framework respecting $SU_C(3)\times SU_L(2)\times U_Y(1)$ gauge symmetry. Only high dimensional contact operators can contribute to the $gg\rightarrow Zh$ process directly, and the lowest order is at dimension-eight level. The $CP$-even ($CP$-odd) tensor structures are induced by the operators with (without) dual field strength tensor $\widetilde{G}^{\mu\nu,a}\equiv\epsilon^{\mu\nu\rho\sigma}G_{\rho\sigma}^a/2$. In view of the independent dimension-eight operators listed in papers \cite{Li:2020gnx, Murphy:2020rsh}, the three relevant operators are given as
\begin{align}
&G_{\mu\nu}^a\widetilde{G}^{\mu\nu,a}(D_{\rho}H)^{\dag}(D^{\rho}H),\quad G_{\mu\nu}^aG^{\mu\nu,a}(D_{\rho}H)^{\dag}(D^{\rho}H),\quad G_{\mu\rho}^aG^{\nu\rho,a}(D^{\mu}H)^{\dag}(D_{\nu}H).
\end{align}
After EWSB, we can check that none of the three operators above can produce the $ggZh$ contact interaction.

In Refs. \cite{Chala:2021cgt, Corbett:2024yoy}, the authors present nine redundant operators which are removed in the traditional basis. Here, we list the nine $G^2H^2D^2$-type operators, which can appear in a new basis. Containing dual field strength tensor, the four operators are given as
\begin{align}\label{eqn:contri:EFT:even}
&\mc{O}_{even,1a}=G_{\mu\nu}^a\widetilde{G}^{\mu\nu,a}H^{\dag}(D^2H),\qquad\qquad\qquad \mc{O}_{even,1b}=G_{\mu\nu}^a\widetilde{G}^{\mu\nu,a}(D^2H)^{\dag}H,\nonumber\\
&\mc{O}_{even,2a}=(D^{\mu}G_{\mu\nu}^a)\widetilde{G}^{\nu\rho,a}H^{\dag}(D_{\rho}H),\qquad\qquad \mc{O}_{even,2b}=(D^{\mu}G_{\mu\nu}^a)\widetilde{G}^{\nu\rho,a}(D_{\rho}H)^{\dag}H.
\end{align}
Without containing dual field strength tensor, the five operators are given as
\begin{align}\label{eqn:contri:EFT:odd}
&\mc{O}_{odd,1a}=G_{\mu\nu}^aG^{\mu\nu,a}H^{\dag}(D^2H),\qquad\qquad\qquad \mc{O}_{odd,1b}=G_{\mu\nu}^aG^{\mu\nu,a}(D^2H)^{\dag}H,\nonumber\\
&\mc{O}_{odd,2a}=(D^{\mu}G_{\mu\nu}^a)G^{\nu\rho,a}H^{\dag}(D_{\rho}H),\qquad\qquad \mc{O}_{odd,2b}=(D^{\mu}G_{\mu\nu}^a)G^{\nu\rho,a}(D_{\rho}H)^{\dag}H,\nonumber\\
&\mc{O}_{odd,3}=(D^{\mu}G_{\mu\rho}^a)(D_{\nu}G^{\nu\rho,a})H^{\dag}H.
\end{align}
Note that the operators $\mc{O}_{even,2a(2b)},\mc{O}_{odd,2a(2b)},\mc{O}_{odd,3}$ can not produce the $ggZh$ contact interaction. After EWSB, the $H^{\dag}(D^2H)$ can lead to the terms $h(\partial^{\mu}Z_{\mu})$ and $Z_{\mu}(\partial^{\mu}h)$. However, the $h(\partial^{\mu}Z_{\mu})$ part do not contribute for on-shell $Z$ boson. Therefore, the operators $\mc{O}_{even,1a(1b)}$ in Eq. \eqref{eqn:contri:EFT:even} will lead to the following mass eigenstate interaction:
\begin{align}\label{eqn:contri:EFT:evenEWSB}
&\mc{O}_{even,1}=G_{\mu\nu}^a\widetilde{G}^{\mu\nu,a}Z_{\rho}(\partial^{\rho}h).
\end{align}
The operator $\mc{O}_{even,1}$ in Eq. \eqref{eqn:contri:EFT:evenEWSB} implies the $F_1$ tensor structure $(k_1\cdot k_2)\epsilon^{\mu\nu\rho k_2}-k_2^{\mu}\epsilon^{\nu\rho k_1k_2}$, which is also checked by implementing the fr file to FeynRules \cite{Alloul:2013bka}. Moreover, the tensor structure can be confirmed by the heavy quark expansion of SM results in Ref. \cite{Kniehl:1990zu}, where the $F_{1}$ and $F_{2}$ parts satisfy
\begin{align}
&[\frac{m_Z^2-\hat{t}}{\hat{s}}(\frac{\hat{s}}{2}\epsilon^{\mu\nu\rho k_2}-k_2^{\mu}\epsilon^{\nu\rho k_1k_2})-(p_1^{\mu}+\frac{\hat{t}-m_Z^2}{\hat{s}}k_2^{\mu})\epsilon^{\nu\rho k_1k_2}+(k_1,\mu)\leftrightarrow(k_2,\nu)]\nonumber\\
&=(p_1\cdot k_1)\epsilon^{\mu\nu\rho k_2}-p_1^{\mu}\epsilon^{\nu\rho k_1k_2}-(p_1\cdot k_2)\epsilon^{\mu\nu\rho k_1}+p_1^{\nu}\epsilon^{\mu\rho k_1k_2}.
\end{align}
Applying the Schouten identity, the above result is just $p_1^{\rho}\epsilon^{\mu\nu k_1k_2}$, which causes no physical effects because of the transversality condition $\epsilon_{\rho}^{r_3,\ast}(p_1)p_1^{\rho}=0$. As a result, only one form factor $F_{1}$ can survive in Eq. \eqref{eqn:contri:LQ:general:structureEven} at dimension-eight level.

Similarly, the operators $\mc{O}_{odd,1a(1b)}$ in Eq. \eqref{eqn:contri:EFT:odd} will lead to the following mass eigenstate interaction:
\begin{align}\label{eqn:contri:EFT:oddEWSB1}
&\mc{O}_{odd,1}=G_{\mu\nu}^aG^{\mu\nu,a}Z_{\rho}(\partial^{\rho}h).
\end{align}
The operator $\mc{O}_{odd,1}$ in Eq. \eqref{eqn:contri:EFT:oddEWSB1} implies the $G_1$ tensor structure $[k_2^{\mu}k_1^{\nu}-(k_1\cdot k_2)g^{\mu\nu}]k_1^{\rho}$.

\begin{figure}[!h]
\centering
\includegraphics[scale=0.5]{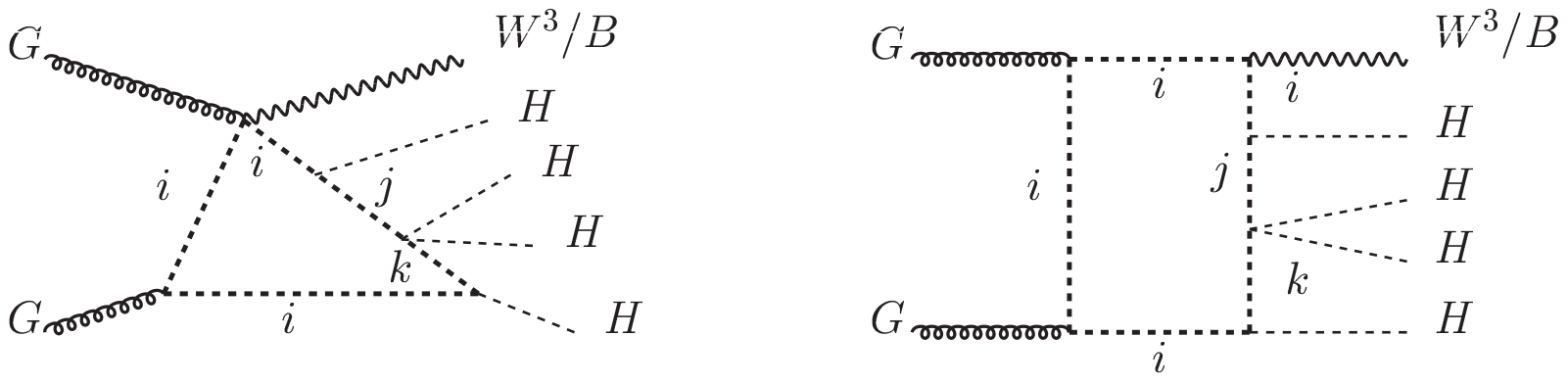}
\caption{Typical triangle (left) and box (right) diagrams contributing to the $ggZh$ contact effective operators. Here, the indices $i,j,k$ label three different fields. The diagrams are generated by JaxoDraw \cite{Binosi:2008ig}.}
\label{fig:contri:EFT:contact}
\end{figure}

As established previously, the presence of at least three \sLQ{} generations is necessary to generate non-zero contributions. In Fig. \ref{fig:contri:EFT:contact}, we show the diagrams that can be contracted into a $ggZh$ vertex from gauge eigenstate fields. Because $G^2$ is already of mass dimension four, the incorporation of $H^4$ and $W^3/B$ fields can not yield the dimension-eight operators and the leading operators are induced at dimension-ten level \footnote{Part of dimension-ten operators are shown in \cite{Chang:2022crb}, while systematic analyses lie beyond the scope of this work.}. That is the reason why the \sLQ{} contributions are negligible as confirmed by our heavy internal mass expansion of the \sLQ{} contributions in Eq. \eqref{eqn:SLQcontri:expansion:G123}. Considering $1/m_{\mr{LQ}}^2$ suppression associated with the off-diagonal coupling product $\mr{Im}(g_{ij}^{Z}\Gamma_{ji}^h)$, the two form factors $G_1$ and $G_2$ only survive at $\mc{O}(1/m_{\mr{LQ}}^6)$. The strong suppression arises from symmetry cancellations in the specific \sLQ{} models.
\section{Summary and conclusions}\label{sec:summary}
In this manuscript, we systematically investigate one-loop contributions to the $gg\rightarrow Zh$ process in the general \sLQ{} models quantitatively. Based on the Ward identity and Bose symmetry, we construct new $CP$ violated tensor structures. Starting from the most general \sLQ{} interactions, we derive the one-loop contributions in the $R_{\xi}$ gauge in Eq. \eqref{eqn:SLQcontri:explicit:G123}, which verify these new tensor structures. The gauge independence of amplitude is validated evidently by the vanishing color trace, vanishing diagonal $G^0-\mr{LQ}-\mr{LQ}$ interaction, and the identity $2B_1(k_1^2,m_i^2,m_i^2)+B_0(k_1^2,m_i^2,m_i^2)=0$. In light of the analytic contributions, we find that there should exist both the off-diagonal $h-\mr{LQ}-\mr{LQ}$ and $Z-\mr{LQ}-\mr{LQ}$ interactions to produce non-zero contributions. In the one-field and two-field models, the contributions vanish exactly; thus, the contributions are possible at least in the three-field models. Concretely, we exemplify the contributions in the $S_1+\widetilde{R}_2+S_3$ model, which require the presence of scalar leptoquark mass splittings and $CP$ violation in the form of $\mr{Im}(A_{1\widetilde{2}}^{\ast}A_{\widetilde{2}3}Y_{13}^{\ast})$. From the SMEFT perspective, the \sLQ{} contributions are indeed described by the dimension-ten operators. Numerically, the leading contributions are small because it is heavily suppressed by the \sLQ{} mass.

Although we deduce the results in the \sLQ{} conventions, it is general for the renormalizable color triplet and electroweak multiplet scalar models, which can also serve as a useful reference for studying the contributions in other new physics models. Given that the colored scalar contributions in the $gg\rightarrow Zh$ process are negligible, our study quantitatively reveals the screening effects of the color triplet scalars in this channel. If there is a anomaly observed in the $gg\rightarrow Zh$ channel in the future, we can exclude the models extended solely by color triplet scalars. The new $CP$-odd tensor structures are foundations of studies in other new physics models. Our preliminary attempts indicates that such tensor structures can also arise in the $CP$ violated fermion loops. Crucially, colored scalars only contribute to the $CP$-odd tensor structures, while fermion loops can generate both the $CP$-even and $CP$-odd tensor structures. For example, the $h\bar{t}i\gamma^5t$ interactions can lead to the contributions at dimension-eight level. The new fermion loops and modfied SM couplings will be studied exhaustively in the future works.

\begin{acknowledgments}
We would like to thank Ying-nan Mao and Bin Yan for helpful discussions. This work was supported in part by the Basic Research Program of Shanxi Province (Grant No. 202403021222062) and the startup research fund of Taiyuan University of Technology (Grant No. RY2400001554).
\end{acknowledgments}
\section*{Appendix}
\begin{appendices}

\section{Specific two scalar leptoquark extended models}\label{app:model:2LQ:specific}
In the following, we enumerate the five two-field models with mixing: $S_1+\widetilde{R}_2,~S_1+S_3,~\widetilde{S}_1+S_3,~R_2+\widetilde{R}_2,~\widetilde{R}_2+S_3$ \cite{Dorsner:2019itg, Crivellin:2021ejk}.
\begin{itemize}[itemindent=-10pt]
\item $S_1+\widetilde{R}_2$ model

For this model, there can be a mixing interaction $-A_{1\widetilde{2}}(\widetilde{R}_2^{\dag}H)S_1^{\dag}+\mathrm{h.c.}$. After EWSB, the related mass terms and scalar interactions can be parameterized as
\begin{align}
&-\left[\begin{array}{c}S_1\quad(\widetilde{R}_2^{-1/3})^{\dag}\end{array}\right]M_{0}^2\left[\begin{array}{c}(S_1)^{\dag}\\\widetilde{R}_2^{-1/3}\end{array}\right]
	-vh\left[\begin{array}{c}S_1\quad(\widetilde{R}_2^{-1/3})^{\dag}\end{array}\right]\Gamma_{0}^h\left[\begin{array}{c}(S_1)^{\dag}\\\widetilde{R}_2^{-1/3}\end{array}\right]\nonumber\\
	&-ivG^0\left[\begin{array}{c}S_1\quad(\widetilde{R}_2^{-1/3})^{\dag}\end{array}\right]\Gamma_{0}^{G^0}\left[\begin{array}{c}(S_1)^{\dag}\\\widetilde{R}_2^{-1/3}\end{array}\right].
\end{align}
In the above, the mass, $h$ coupling, and $G^0$ coupling matrices are given as
\begin{align}
&M_{0}^2=\left[\begin{array}{cc}\mu_{S_1}^2+\frac{1}{2}\lambda_{1S_1}v^2&\frac{1}{\sqrt{2}}A_{1\widetilde{2}}^{\ast}v\\\frac{1}{\sqrt{2}}A_{1\widetilde{2}}v&\mu_{\widetilde{R}_2}^2+\frac{1}{2}(\lambda_{1\widetilde{R}_2}-\frac{1}{2}\lambda_{2\widetilde{R}_2})v^2\end{array}\right],\nonumber\\
&\Gamma_{0}^h=\left[\begin{array}{cc}\lambda_{1S_1}&\frac{1}{\sqrt{2}v}A_{1\widetilde{2}}^{\ast}\\\frac{1}{\sqrt{2}v}A_{1\widetilde{2}}&\lambda_{1\widetilde{R}_2}-\frac{1}{2}\lambda_{2\widetilde{R}_2}\end{array}\right],\qquad
	\Gamma_{0}^{G^0}=\left[\begin{array}{cc}0&-\frac{1}{\sqrt{2}v}A_{1\widetilde{2}}^{\ast}\\\frac{1}{\sqrt{2}v}A_{1\widetilde{2}}&0\end{array}\right].
\end{align}

As for the $I_{3,\phi_i}$, they are given as
\begin{align}
I_{3,(S_1)^{\dag}}=0,\qquad I_{3,\widetilde{R}_2^{-1/3}}=-\frac{1}{2}.
\end{align}

\item $S_1+S_3$ model

For this model, there can be a mixing interaction $Y_{13}H^{\dag}(S_3^a\sigma_a)^{\dag}HS_1+\mathrm{h.c.}$. After EWSB, the related mass terms and scalar interactions can be parameterized as
\begin{align}
&-\left[\begin{array}{c}S_1\quad S_3^{1/3}\end{array}\right]M_{0}^2\left[\begin{array}{c}(S_1)^{\dag}\\(S_3^{1/3})^{\dag}\end{array}\right]
	-vh\left[\begin{array}{c}S_1\quad S_3^{1/3}\end{array}\right]\Gamma_{0}^h\left[\begin{array}{c}(S_1)^{\dag}\\(S_3^{1/3})^{\dag}\end{array}\right]\nonumber\\
	&-ivG^0\left[\begin{array}{c}S_1\quad S_3^{1/3}\end{array}\right]\Gamma_{0}^{G^0}\left[\begin{array}{c}(S_1)^{\dag}\\(S_3^{1/3})^{\dag}\end{array}\right].
\end{align}
In the above, the mass, $h$ coupling, and $G^0$ coupling matrices are given as
\begin{align}
&M_{0}^2=\left[\begin{array}{cc}\mu_{S_1}^2+\frac{1}{2}\lambda_{1S_1}v^2&\frac{1}{2}Y_{13}v^2\\\frac{1}{2}Y_{13}^{\ast}v^2&\mu_{S_3}^2+\frac{1}{2}\lambda_{1S_3}v^2\end{array}\right],\qquad \Gamma_{0}^h=\left[\begin{array}{cc}\lambda_{1S_1}&Y_{13}\\Y_{13}^{\ast}&\lambda_{1S_3}\end{array}\right],\qquad
	\Gamma_{0}^{G^0}=\left[\begin{array}{cc}0&0\\0&0\end{array}\right].
\end{align}

As for the $I_{3,\phi_i}$, they are given as
\begin{align}
I_{3,(S_1)^{\dag}}=I_{3,(S_3^{1/3})^{\dag}}=0.
\end{align}

\item $\widetilde{S}_1+S_3$ model

For this model, there can be a mixing interaction $Y_{\widetilde{1}3}H^T(i\sigma_2)(S_3^a\sigma_a)H\widetilde{S}_1^{\dag}+\mathrm{h.c.}$. After EWSB, the related mass terms and scalar interactions can be parameterized as
\begin{align}
&-\left[\begin{array}{c}(\widetilde{S}_1)^{\dag}\quad (S_3^{4/3})^{\dag}\end{array}\right]M_{0}^2\left[\begin{array}{c}\widetilde{S}_1\\S_3^{4/3}\end{array}\right]
	-vh\left[\begin{array}{c}(\widetilde{S}_1)^{\dag}\quad (S_3^{4/3})^{\dag}\end{array}\right]\Gamma_{0}^h\left[\begin{array}{c}\widetilde{S}_1\\S_3^{4/3}\end{array}\right]\nonumber\\
	&-ivG^0\left[\begin{array}{c}(\widetilde{S}_1)^{\dag}\quad (S_3^{4/3})^{\dag}\end{array}\right]\Gamma_{0}^{G^0}\left[\begin{array}{c}\widetilde{S}_1\\S_3^{4/3}\end{array}\right].
\end{align}
In the above, the mass, $h$ coupling, and $G^0$ coupling matrices are given as
\begin{align}
&M_{0}^2=\left[\begin{array}{cc}\mu_{\widetilde{S}_1}^2+\frac{1}{2}\lambda_{1\widetilde{S}_1}v^2&\frac{1}{\sqrt{2}}Y_{\widetilde{1}3}v^2\\\frac{1}{\sqrt{2}}Y_{\widetilde{1}3}^{\ast}v^2&\mu_{S_3}^2+\frac{1}{2}(\lambda_{1S_3}+\lambda_{2S_3})v^2\end{array}\right],\nonumber\\
&\Gamma_{0}^h=\left[\begin{array}{cc}\lambda_{1\widetilde{S}_1}&\sqrt{2}Y_{\widetilde{1}3}\\\sqrt{2}Y_{\widetilde{1}3}^{\ast}&\lambda_{1S_3}+\lambda_{2S_3}\end{array}\right],\qquad
	\Gamma_{0}^{G^0}=\left[\begin{array}{cc}0&\sqrt{2}Y_{\widetilde{1}3}\\-\sqrt{2}Y_{\widetilde{1}3}^{\ast}&0\end{array}\right].
\end{align}

As for the $I_{3,\phi_i}$, they are given as
\begin{align}
I_{3,\widetilde{S}_1}=0,\qquad I_{3,S_3^{4/3}}=1.
\end{align}

\item $R_2+\widetilde{R}_2$ model

For this model, there can be a mixing interaction $Y_{2\widetilde{2}}(R_2^{\dag}H)(H^Ti\sigma_2\widetilde{R}_2)+\mathrm{h.c.}$. After EWSB, the related mass terms and scalar interactions can be parameterized as
\begin{align}
&-\left[\begin{array}{c}(R_2^{2/3})^{\dag}\quad(\widetilde{R}_2^{2/3})^{\dag}\end{array}\right]M_{0}^2\left[\begin{array}{c}R_2^{2/3}\\\widetilde{R}_2^{2/3}\end{array}\right]
	-vh\left[\begin{array}{c}(R_2^{2/3})^{\dag}\quad(\widetilde{R}_2^{2/3})^{\dag}\end{array}\right]\Gamma_{0}^h\left[\begin{array}{c}R_2^{2/3}\\\widetilde{R}_2^{2/3}\end{array}\right]\nonumber\\
	&-ivG^0\left[\begin{array}{c}(R_2^{2/3})^{\dag}\quad(\widetilde{R}_2^{2/3})^{\dag}\end{array}\right]\Gamma_{0}^{G^0}\left[\begin{array}{c}R_2^{2/3}\\\widetilde{R}_2^{2/3}\end{array}\right].
\end{align}
In the above, the mass, $h$ coupling, and $G^0$ coupling matrices are given as
\begin{align}
&M_{0}^2=\left[\begin{array}{cc}\mu_{R_2}^2+\frac{1}{2}(\lambda_{1R_2}-\frac{1}{2}\lambda_{2R_2})v^2&\frac{1}{2}Y_{2\widetilde{2}}v^2\\\frac{1}{2}Y_{2\widetilde{2}}^{\ast}v^2&\mu_{\widetilde{R}_2}^2+\frac{1}{2}(\lambda_{1\widetilde{R}_2}+\frac{1}{2}\lambda_{2\widetilde{R}_2})v^2\end{array}\right],\nonumber\\
&\Gamma_{0}^h=\left[\begin{array}{cc}\lambda_{1R_2}-\frac{1}{2}\lambda_{2R_2}&Y_{2\widetilde{2}}\\[1ex]Y_{2\widetilde{2}}^{\ast}&\lambda_{1\widetilde{R}_2}+\frac{1}{2}\lambda_{2\widetilde{R}_2}\end{array}\right],\qquad
	\Gamma_{0}^{G^0}=\left[\begin{array}{cc}0&Y_{2\widetilde{2}}\\-Y_{2\widetilde{2}}^{\ast}&0\end{array}\right].
\end{align}

As for the $I_{3,\phi_i}$, they are given as
\begin{align}
I_{3,R_2^{2/3}}=-\frac{1}{2},\qquad I_{3,\widetilde{R}_2^{2/3}}=\frac{1}{2}.
\end{align}

\item $\widetilde{R}_2+S_3$ model

For this model, there can be a mixing interaction $A_{\widetilde{2}3}\widetilde{R}_2^{\dag}(S_3^a\sigma_a)^{\dag}H+\mathrm{h.c.}$. After EWSB, there are mixings between the $2/3$ and $-1/3$ electrically charged \sLQ{} individually.

\textcircled{1} For the \sLQ{} with electric charge $2/3$, the related mass terms and scalar interactions can be parameterized as
\begin{align}
&-\left[\begin{array}{c}(\widetilde{R}_2^{2/3})^{\dag}\quad S_3^{-2/3}\end{array}\right]M_{0}^2\left[\begin{array}{c}\widetilde{R}_2^{2/3}\\(S_3^{-2/3})^{\dag}\end{array}\right]
	-vh\left[\begin{array}{c}(\widetilde{R}_2^{2/3})^{\dag}\quad S_3^{-2/3}\end{array}\right]\Gamma_{0}^h\left[\begin{array}{c}\widetilde{R}_2^{2/3}\\(S_3^{-2/3})^{\dag}\end{array}\right]\nonumber\\
	&-ivG^0\left[\begin{array}{c}(\widetilde{R}_2^{2/3})^{\dag}\quad S_3^{-2/3}\end{array}\right]\Gamma_{0}^{G^0}\left[\begin{array}{c}\widetilde{R}_2^{2/3}\\(S_3^{-2/3})^{\dag}\end{array}\right].
\end{align}
In the above, the mass, $h$ coupling, and $G^0$ coupling matrices are given as
\begin{align}
&M_{0}^2=\left[\begin{array}{cc}\mu_{\widetilde{R}_2}^2+\frac{1}{2}(\lambda_{1\widetilde{R}_2}+\frac{1}{2}\lambda_{2\widetilde{R}_2})v^2&-A_{\widetilde{2}3}v\\-A_{\widetilde{2}3}^{\ast}v&\mu_{S_3}^2+\frac{1}{2}(\lambda_{1S_3}-\lambda_{2\widetilde{S}_3})v^2\end{array}\right],\nonumber\\
&\Gamma_{0}^h=\left[\begin{array}{cc}\lambda_{1\widetilde{R}_2}+\frac{1}{2}\lambda_{2\widetilde{R}_2}&-\frac{1}{v}A_{\widetilde{2}3}\\-\frac{1}{v}A_{\widetilde{2}3}^{\ast}&\lambda_{1S_3}-\lambda_{2\widetilde{S}_3}\end{array}\right],\qquad
	\Gamma_{0}^{G^0}=\left[\begin{array}{cc}0&-\frac{1}{v}A_{\widetilde{2}3}\\\frac{1}{v}A_{\widetilde{2}3}^{\ast}&0\end{array}\right].
\end{align} 

As for the $I_{3,\phi_i}$, they are given as
\begin{align}
I_{3,\widetilde{R}_2^{2/3}}=\frac{1}{2},\qquad I_{3,(S_3^{-2/3})^{\dag}}=1.
\end{align}

\textcircled{2} For the \sLQ{} with electric charge $-1/3$, the related mass terms and scalar interactions can be parameterized as
\begin{align}
&-\left[\begin{array}{c}(\widetilde{R}_2^{-1/3})^{\dag}\quad S_3^{1/3}\end{array}\right]M_{0}^2\left[\begin{array}{c}\widetilde{R}_2^{-1/3}\\(S_3^{1/3})^{\dag}\end{array}\right]
	-vh\left[\begin{array}{c}(\widetilde{R}_2^{-1/3})^{\dag}\quad S_3^{1/3}\end{array}\right]\Gamma_{0}^h\left[\begin{array}{c}\widetilde{R}_2^{-1/3}\\(S_3^{1/3})^{\dag}\end{array}\right]\nonumber\\
	&-ivG^0\left[\begin{array}{c}(\widetilde{R}_2^{-1/3})^{\dag}\quad S_3^{1/3}\end{array}\right]\Gamma_{0}^{G^0}\left[\begin{array}{c}\widetilde{R}_2^{-1/3}\\(S_3^{1/3})^{\dag}\end{array}\right].
\end{align}
In the above, the mass, $h$ coupling, and $G^0$ coupling matrices are given as
\begin{align}
&M_{0}^2=\left[\begin{array}{cc}\mu_{\widetilde{R}_2}^2+\frac{1}{2}(\lambda_{1\widetilde{R}_2}-\frac{1}{2}\lambda_{2\widetilde{R}_2})v^2&\frac{1}{\sqrt{2}}A_{\widetilde{2}3}v\\\frac{1}{\sqrt{2}}A_{\widetilde{2}3}^{\ast}v&\mu_{S_3}^2+\frac{1}{2}\lambda_{1S_3}v^2\end{array}\right],\nonumber\\
&\Gamma_{0}^h=\left[\begin{array}{cc}\lambda_{1\widetilde{R}_2}-\frac{1}{2}\lambda_{2\widetilde{R}_2}&\frac{1}{\sqrt{2}v}A_{\widetilde{2}3}\\\frac{1}{\sqrt{2}v}A_{\widetilde{2}3}^{\ast}&\lambda_{1S_3}\end{array}\right],\qquad
	\Gamma_{0}^{G^0}=\left[\begin{array}{cc}0&\frac{1}{\sqrt{2}v}A_{\widetilde{2}3}\\-\frac{1}{\sqrt{2}v}A_{\widetilde{2}3}^{\ast}&0\end{array}\right].
\end{align} 

As for the $I_{3,\phi_i}$, they are given as
\begin{align}
I_{3,\widetilde{R}_2^{-1/3}}=-\frac{1}{2},\qquad I_{3,(S_3^{1/3})^{\dag}}=0.
\end{align}

\end{itemize}

\section{Vanishing of the diagonal $G^0-\mr{LQ}-\mr{LQ}$ type interactions}\label{app:interactions:G0LQLQ}
\subsection{Two scalar leptoquark extended models}

Inserting Eq. \eqref{eqn:model:2LQ:U} into Eq. \eqref{eqn:model:general:gamma}, the $G^0$ coupling matrix in mass eigenstates is computed as
\begin{align}
\Gamma^{G^0}=\left[\begin{array}{cc}-\sin\theta\cos\theta[\Gamma_{0,12}^{G^0}e^{-i\delta}-(\Gamma_{0,12}^{G^0})^{\ast}e^{i\delta}]
	&e^{i\delta}[\Gamma_{0,12}^{G^0}\cos^2\theta e^{-i\delta}+(\Gamma_{0,12}^{G^0})^{\ast}\sin^2\theta e^{i\delta}]\\[4ex]
	-e^{-i\delta}[(\Gamma_{0,12}^{G^0})^{\ast}\cos^2\theta e^{i\delta}+\Gamma_{0,12}^{G^0}\sin^2\theta e^{-i\delta}]
	&\sin\theta\cos\theta[\Gamma_{0,12}^{G^0}e^{-i\delta}-(\Gamma_{0,12}^{G^0})^{\ast}e^{i\delta}]\end{array}\right].
\end{align}
Based on the expression of $M_{0,12}^2$ shown in Eq. \eqref{eqn:model:2LQ:M0sq}, we have the following relations:
\begin{align}
&\Gamma_{11}^{G^0}=-\sin\theta\cos\theta[\Gamma_{0,12}^{G^0}e^{-i\delta}-(\Gamma_{0,12}^{G^0})^{\ast}e^{i\delta}]=-2i\sin\theta\cos\theta\cdot\mr{Im}(\Gamma_{0,12}^{G^0}e^{-i\delta})\nonumber\\
&=-2i(m_{\phi_2}^2-m_{\phi_1}^2)\sin^2\theta\cos^2\theta\cdot\mr{Im}(\frac{\Gamma_{0,12}^{G^0}}{M_{0,12}^2}).
\end{align}

Again, the cubic type $H\,\mr{LQ}^2$ and quartic type $H^2\mr{LQ}^2$ interactions leads to the result $\Gamma_{11}^{G^0}=0$ due to real $\Gamma_{0,12}^{G^0}/M_{0,12}^2$ in the two-field models, which is also illustrated explicitly in App. \ref{app:model:2LQ:specific} with specific two scalar leptoquark extended model examples. Because of $\mr{tr}(\Gamma^{G^0})=0$, the $\Gamma_{22}^{G^0}=0$ is valid. Consequently, off-diagonal $G^0-\mr{LQ}-\mr{LQ}$ interactions are absent in the two-\sLQ{} models.

\subsection{Three scalar leptoquark $S_1+\widetilde{R}_2+S_3$ model}

Inserting expressions of $\Gamma_{0}^{G^0}$ given in Eq. \eqref{eqn:model:S1R2tS3:M0sqGam0} into Eq. \eqref{eqn:model:general:gammaCom}, the diagonal $G^0$ coupling matrix in mass eigenstates is computed as
\begin{align}\label{eqn:app:interaction:S1R2tS3:gammaii}
&(\Gamma^{G^0})_{ii}=\frac{1}{\sqrt{2}v}(-U_{1i}^{\ast}U_{2i}A_{1\widetilde{2}}^{\ast}+U_{2i}^{\ast}U_{1i}A_{1\widetilde{2}}+U_{2i}^{\ast}U_{3i}A_{\widetilde{2}3}-U_{3i}^{\ast}U_{2i}A_{\widetilde{2}3}^{\ast}).
\end{align}
Note that there is no sum over the index $i$.

Inserting expressions of $M_{0}^2$ given in Eq. \eqref{eqn:model:S1R2tS3:M0sqGam0} into $M_{0}^2=UM^2U^{\dag}$, we have the following relations:
\begin{align}\label{eqn:app:interaction:S1R2tS3:AA}
&\frac{1}{\sqrt{2}}A_{1\widetilde{2}}v=m_{S_1}^2U_{21}U_{11}^{\ast}+m_{\widetilde{R}_2^{1/3}}^2U_{22}U_{12}^{\ast}+m_{S_3^{1/3}}^2U_{23}U_{13}^{\ast},\nonumber\\
&\frac{1}{\sqrt{2}}A_{\widetilde{2}3}v=m_{S_1}^2U_{21}U_{31}^{\ast}+m_{\widetilde{R}_2^{1/3}}^2U_{22}U_{32}^{\ast}+m_{S_3^{1/3}}^2U_{23}U_{33}^{\ast}.
\end{align}

Inserting expressions of $A_{1\widetilde{2}}$ and $A_{\widetilde{2}3}$ given in Eq. \eqref{eqn:app:interaction:S1R2tS3:AA} into Eq. \eqref{eqn:app:interaction:S1R2tS3:gammaii}, the $(\Gamma^{G^0})_{ii}$ is computed as
\begin{align}
&v^2(\Gamma^{G^0})_{ii}=m_{S_1}^2(-U_{1i}^{\ast}U_{2i}U_{11}U_{21}^{\ast}+U_{2i}^{\ast}U_{1i}U_{21}U_{11}^{\ast}+U_{2i}^{\ast}U_{3i}U_{21}U_{31}^{\ast}-U_{3i}^{\ast}U_{2i}U_{31}U_{21}^{\ast})\nonumber\\
	&+m_{\widetilde{R}_2^{1/3}}^2(-U_{1i}^{\ast}U_{2i}U_{12}U_{22}^{\ast}+U_{2i}^{\ast}U_{1i}U_{22}U_{12}^{\ast}+U_{2i}^{\ast}U_{3i}U_{22}U_{32}^{\ast}-U_{3i}^{\ast}U_{2i}U_{32}U_{22}^{\ast})\nonumber\\
	&+m_{S_3^{1/3}}^2(-U_{1i}^{\ast}U_{2i}U_{13}U_{23}^{\ast}+U_{2i}^{\ast}U_{1i}U_{23}U_{13}^{\ast}+U_{2i}^{\ast}U_{3i}U_{23}U_{33}^{\ast}-U_{3i}^{\ast}U_{2i}U_{33}U_{23}^{\ast}).
\end{align}
Especially, the $(\Gamma^{G^0})_{11}$ is computed as
\begin{align}
&v^2(\Gamma^{G^0})_{11}=m_{\widetilde{R}_2^{1/3}}^2[U_{21}^{\ast}U_{22}(U_{11}U_{12}^{\ast}+U_{31}U_{32}^{\ast})-U_{21}U_{22}^{\ast}(U_{11}^{\ast}U_{12}+U_{31}^{\ast}U_{32})]\nonumber\\
	&+m_{S_3^{1/3}}^2[U_{21}^{\ast}U_{23}(U_{11}U_{13}^{\ast}+U_{31}U_{33}^{\ast})-U_{21}U_{23}^{\ast}(U_{11}^{\ast}U_{13}+U_{31}^{\ast}U_{33})].
\end{align}
The unitarity of $U$ matrix leads to the following identities:
\begin{align}
&U_{11}U_{12}^{\ast}+U_{31}U_{32}^{\ast}=-U_{21}U_{22}^{\ast},\qquad
U_{11}U_{13}^{\ast}+U_{31}U_{33}^{\ast}=-U_{21}U_{23}^{\ast}.
\end{align}
Thus, we have $(\Gamma^{G^0})_{11}=0$. Similarly, the $(\Gamma^{G^0})_{22}=0$ also holds on. Because of $\mr{tr}(\Gamma^{G^0})=\mr{tr}(\Gamma_0^{G^0})=0$, the $\Gamma_{33}^{G^0}=0$ is valid. Consequently, off-diagonal $G^0-\mr{LQ}-\mr{LQ}$ interactions are absent in the $S_1+\widetilde{R}_2+S_3$ model.
\subsection{Three scalar leptoquark $R_2+\widetilde{R}_2+S_3$ model}
Inserting expressions of $\Gamma_{0}^{G^0}$ given in Eq. \eqref{eqn:model:R2R2tS3:M0sqGam0} into Eq. \eqref{eqn:model:general:gammaCom}, the diagonal $G^0$ coupling matrix in mass eigenstates is computed as
\begin{align}\label{eqn:app:interaction:R2R2tS3:gammaii}
&(\Gamma^{G^0})_{ii}=U_{1i}^{\ast}U_{2i}Y_{2\widetilde{2}}-U_{2i}^{\ast}U_{1i}Y_{2\widetilde{2}}^{\ast}-\frac{1}{v}U_{2i}^{\ast}U_{3i}A_{\widetilde{2}3}+\frac{1}{v}U_{3i}^{\ast}U_{2i}A_{\widetilde{2}3}^{\ast}.
\end{align}
Note that there is no sum over the index $i$.

Inserting expressions of $M_{0}^2$ given in Eq. \eqref{eqn:model:R2R2tS3:M0sqGam0} into $M_{0}^2=UM^2U^{\dag}$, we have the following relations:
\begin{align}\label{eqn:app:interaction:R2R2tS3:YA}
&\frac{1}{2}Y_{2\widetilde{2}}v^2=m_{R_2^{2/3}}^2U_{11}U_{21}^{\ast}+m_{\widetilde{R}_2^{2/3}}^2U_{12}U_{22}^{\ast}+m_{S_3^{2/3}}^2U_{13}U_{23}^{\ast},\nonumber\\
&-A_{\widetilde{2}3}v=m_{R_2^{2/3}}^2U_{21}U_{31}^{\ast}+m_{\widetilde{R}_2^{2/3}}^2U_{22}U_{32}^{\ast}+m_{S_3^{2/3}}^2U_{23}U_{33}^{\ast},\nonumber\\
&0=m_{R_2^{2/3}}^2U_{11}U_{31}^{\ast}+m_{\widetilde{R}_2^{2/3}}^2U_{12}U_{32}^{\ast}+m_{S_3^{2/3}}^2U_{13}U_{33}^{\ast}.
\end{align}

Inserting expressions of $Y_{2\widetilde{2}}$ and $A_{\widetilde{2}3}$ given in Eq. \eqref{eqn:app:interaction:R2R2tS3:YA} into Eq. \eqref{eqn:app:interaction:R2R2tS3:gammaii}, the $(\Gamma^{G^0})_{ii}$ is computed as
\begin{align}
&v^2(\Gamma^{G^0})_{ii}=m_{R_2^{2/3}}^2(2U_{1i}^{\ast}U_{2i}U_{11}U_{21}^{\ast}-2U_{2i}^{\ast}U_{1i}U_{21}U_{11}^{\ast}+U_{2i}^{\ast}U_{3i}U_{21}U_{31}^{\ast}-U_{3i}^{\ast}U_{2i}U_{31}U_{21}^{\ast})\nonumber\\
	&+m_{\widetilde{R}_2^{2/3}}^2(2U_{1i}^{\ast}U_{2i}U_{12}U_{22}^{\ast}-2U_{2i}^{\ast}U_{1i}U_{22}U_{12}^{\ast}+U_{2i}^{\ast}U_{3i}U_{22}U_{32}^{\ast}-U_{3i}^{\ast}U_{2i}U_{32}U_{22}^{\ast})\nonumber\\
	&+m_{S_3^{2/3}}^2(2U_{1i}^{\ast}U_{2i}U_{13}U_{23}^{\ast}-2U_{2i}^{\ast}U_{1i}U_{23}U_{13}^{\ast}+U_{2i}^{\ast}U_{3i}U_{23}U_{33}^{\ast}-U_{3i}^{\ast}U_{2i}U_{33}U_{23}^{\ast}).
\end{align}
Especially, the $(\Gamma^{G^0})_{11}$ is computed as
\begin{align}\label{eqn:app:interaction:R2R2tS3:gamma11}
&v^2(\Gamma^{G^0})_{11}=m_{\widetilde{R}_2^{2/3}}^2(2U_{11}^{\ast}U_{21}U_{12}U_{22}^{\ast}-2U_{21}^{\ast}U_{11}U_{22}U_{12}^{\ast}+U_{21}^{\ast}U_{31}U_{22}U_{32}^{\ast}-U_{31}^{\ast}U_{21}U_{32}U_{22}^{\ast})\nonumber\\
	&+m_{S_3^{2/3}}^2(2U_{11}^{\ast}U_{21}U_{13}U_{23}^{\ast}-2U_{21}^{\ast}U_{11}U_{23}U_{13}^{\ast}+U_{21}^{\ast}U_{31}U_{23}U_{33}^{\ast}-U_{31}^{\ast}U_{21}U_{33}U_{23}^{\ast}).
\end{align}
The unitarity of $U$ matrix leads to the following identities:
\begin{align}
&U_{11}^{\ast}U_{12}+U_{31}^{\ast}U_{32}=-U_{21}^{\ast}U_{22},\qquad
U_{11}^{\ast}U_{13}+U_{31}^{\ast}U_{33}=-U_{21}^{\ast}U_{23}.
\end{align}
Adopting the unitary relations above, the $(\Gamma^{G^0})_{11}$ in Eq. \eqref{eqn:app:interaction:R2R2tS3:gamma11} can be converted as
\begin{align}
&v^2(\Gamma^{G^0})_{11}=3m_{\widetilde{R}_2^{2/3}}^2(-U_{11}^{\ast}U_{12}U_{31}U_{32}^{\ast}+U_{12}^{\ast}U_{11}U_{32}U_{31}^{\ast})+3m_{S_3^{2/3}}^2(-U_{11}^{\ast}U_{13}U_{31}U_{33}^{\ast}+U_{13}^{\ast}U_{11}U_{33}U_{31}^{\ast})\nonumber\\
&=3U_{11}U_{31}^{\ast}(m_{\widetilde{R}_2^{2/3}}^2U_{12}^{\ast}U_{32}+m_{S_3^{2/3}}^2U_{13}^{\ast}U_{33})-3U_{11}^{\ast}U_{31}(m_{\widetilde{R}_2^{2/3}}^2U_{12}U_{32}^{\ast}+m_{S_3^{2/3}}^2U_{13}U_{33}^{\ast}).
\end{align}
Considering the third line in Eq. \eqref{eqn:app:interaction:R2R2tS3:YA}, we can verify $(\Gamma^{G^0})_{11}=0$. Similarly, the $(\Gamma^{G^0})_{22}=0$ also holds on. Because of $\mr{tr}(\Gamma^{G^0})=\mr{tr}(\Gamma_0^{G^0})=0$, the $\Gamma_{33}^{G^0}=0$ is valid. Consequently, off-diagonal $G^0-\mr{LQ}-\mr{LQ}$ interactions are absent in the $R_2+\widetilde{R}_2+S_3$ model.

\section{Definitions and expansions of the loop integrals}\label{app:loop}
\subsection{Definitions and shorthand notations of the loop integrals}

For simplicity, we adopt the following abbreviations for the scalar integrals $B_0/C_0/D_0$:\footnote{Similar notations apply for the exchange of $i\leftrightarrow j$.}
\begin{align}
&B_0^i(s)\equiv B_0(s,m_i^2,m_i^2),\nonumber\\
&C_0^i(s)\equiv C_0(0,0,s,m_i^2,m_i^2,m_i^2),\hspace{10ex} C_0^{iji}(m_h^2,m_Z^2,s)\equiv C_0(m_h^2,m_Z^2,s,m_i^2,m_j^2,m_i^2),\nonumber\\
&C_0^{iij}(m_h^2,t)\equiv C_0(0,m_h^2,t,m_i^2,m_i^2,m_j^2),\qquad C_0^{iij}(m_h^2,u)\equiv C_0(0,m_h^2,u,m_i^2,m_i^2,m_j^2),\nonumber\\
&C_0^{iij}(m_Z^2,t)\equiv C_0(0,m_Z^2,t,m_i^2,m_i^2,m_j^2),\qquad C_0^{iij}(m_Z^2,u)\equiv C_0(0,m_Z^2,u,m_i^2,m_i^2,m_j^2),\nonumber\\
&D_0^{iiij}(s,t)\equiv D_0(0,0,m_h^2,m_Z^2,s,t,m_i^2,m_i^2,m_i^2,m_j^2),\nonumber\\
&D_0^{iiij}(s,u)\equiv D_0(0,0,m_h^2,m_Z^2,s,u,m_i^2,m_i^2,m_i^2,m_j^2),\nonumber\\
&D_0^{iijj}(t,u)\equiv D_0(0,m_h^2,0,m_Z^2,t,u,m_i^2,m_i^2,m_j^2,m_j^2).
\end{align}
The tensor integrals are defined in Ref. \cite{Denner:1991kt}, which can be reduced into scalar integrals with the help of FeynCalc \cite{Shtabovenko:2020gxv}.
\subsection{Heavy internal mass expansion of the loop integrals}\label{App:loop:expansion}
In this appendix, we list the related heavy internal mass expansions of the scalar loop integrals $B_0/C_0/D_0$ which are checked by the Package-X \cite{Patel:2016fam}, and part of the results are obtained in our paper \cite{He:2020fqj}. According to the explicit results of form factors $G_{1,2,3}$ in App. \ref{app:results}, the highest power of \sLQ{} mass in front of the loop integrals are $1\cdot B_0$, $m_{\mr{LQ}}^4\cdot C_0$, and $m_{\mr{LQ}}^6\cdot D_0$. Therefore, the loop integrals should be expanded at least up to $1/m_{\mr{LQ}}^4$ for $B_0$, $1/m_{\mr{LQ}}^8$ for $C_0$, and $1/m_{\mr{LQ}}^{10}$ for $D_0$, respectively.

\subsubsection{Related heavy internal mass expansion of $B_0$ function}
The $B_0$ function is defined as
\begin{align}
&B_0(k^2,m_0^2,m_1^2)\equiv\frac{(2\pi\mu)^{4-D}}{i\pi^2}\int d^Dq\frac{1}{(q^2-m_0^2)[(q+k)^2-m_1^2]}\nonumber\\
&=\Delta_{\epsilon}-\int_0^1dx\log\frac{xm_0^2+(1-x)m_1^2-x(1-x)k^2}{\mu^2}+\mc{O}(\epsilon),
\end{align}
where $\Delta_{\epsilon}\equiv1/\epsilon-\gamma_E+\log4\pi$ with $\gamma_E$ being the Euler's constant. Here, $D=4-2\epsilon$ is the spacetime dimension, and $\mu$ is the renormalization scale.

$\bullet$ When the two internal masses are equal, the $B_0$ function can be expanded as
\begin{align}
&B_0(k^2,m_i^2,m_i^2)=\Delta_{\epsilon}-\log\frac{m_i^2}{\mu^2}+\frac{k^2}{6m_i^2}+\frac{k^4}{60m_i^4}+\frac{k^6}{420m_i^6}+\mathcal{O}(\frac{k^8}{m_i^8}).
\end{align}
\subsubsection{Related heavy internal mass expansion of $C_0$ function}
The $C_0$ function is defined as
\begin{align}
&C_0(k_1^2,k_{12}^2,k_2^2,m_0^2,m_1^2,m_2^2)\equiv\frac{(2\pi\mu)^{4-D}}{i\pi^2}\int d^Dq\frac{1}{(q^2-m_0^2)[(q+k_1)^2-m_1^2][(q+k_2)^2-m_2^2]}\nonumber\\
=&-\int_0^1\int_0^1\int_0^1dxdydz\frac{\delta(x+y+z-1)}{xm_0^2+ym_1^2+zm_2^2-xyk_1^2-xzk_2^2-yzk_{12}^2},
\end{align}
where $k_{12}\equiv k_1-k_2$.

$\bullet$ When the three internal masses are all equal, the $C_0$ function can be expanded as
\begin{align}\label{eqn:Apploop:expansion:C0:mi}
&C_0(k_1^2,k_{12}^2,k_2^2,m_i^2,m_i^2,m_i^2)\nonumber\\
=&-\frac{1}{2m_i^2}-\frac{k_1^2+k_2^2+k_{12}^2}{24m_i^4}-\frac{(k_1^2+k_2^2)^2+(k_1^2+k_{12}^2)^2+(k_2^2+k_{12}^2)^2}{360m_i^6}\nonumber\\
&-\frac{3(k_1^2+k_2^2+k_{12}^2)(k_1^4+k_2^4+k_{12}^4)+4k_1^2k_2^2k_{12}^2}{3360m_i^8}+\mathcal{O}(\frac{k^8}{m_i^{10}}).
\end{align}

$\bullet$ When the first two internal masses are equal, the $C_0$ function can be expanded as
\begin{align}
&C_0(k_1^2,k_{12}^2,k_2^2,m_i^2,m_i^2,m_j^2)\approx\frac{1+\log r_{ij}^2-r_{ij}^2}{m_j^2(1-r_{ij}^2)^2}\nonumber\\
	&-\frac{2+6r_{ij}^2\log r_{ij}^2+3r_{ij}^2-6r_{ij}^4+r_{ij}^6}{12m_j^4r_{ij}^2(1-r_{ij}^2)^4}k_1^2+\frac{5+2(1+2r_{ij}^2)\log r_{ij}^2-4r_{ij}^2-r_{ij}^4}{4m_j^4(1-r_{ij}^2)^4}(k_2^2+k_{12}^2)\nonumber\\
	&-\frac{3-30r_{ij}^2-20r_{ij}^4(1+3\log r_{ij}^2)+60r_{ij}^6-15r_{ij}^8+2r_{ij}^{10}}{180m_j^6r_{ij}^4(1-r_{ij}^2)^6}k_1^4\nonumber\\
&+\frac{10+9r_{ij}^2+3(1+6r_{ij}^2+3r_{ij}^4)\log r_{ij}^2-18r_{ij}^4-r_{ij}^6}{9m_j^6(1-r_{ij}^2)^6}(k_{12}^4+k_2^2k_{12}^2+k_2^4)\nonumber\\
&-\frac{3+44r_{ij}^2+12r_{ij}^2(2+3r_{ij}^2)\log r_{ij}^2-36r_{ij}^4-12r_{ij}^6+r_{ij}^8}{36m_j^6r_{ij}^2(1-r_{ij}^2)^6}k_1^2(k_2^2+k_{12}^2)\nonumber\\
	&+\frac{(1-r_{ij}^2)(47+239r_{ij}^2+131r_{ij}^4+3r_{ij}^6)+12(1+12r_{ij}^2+18r_{ij}^4+4r_{ij}^6)\log r_{ij}^2}{48m_j^8(1-r_{ij}^2)^8}(k_2^2+k_{12}^2)(k_2^4+k_{12}^4)\nonumber\\
	&+\frac{(1-r_{ij}^2)(-4-159r_{ij}^2-239r_{ij}^4-19r_{ij}^6+r_{ij}^8)-60r_{ij}^2(1+4r_{ij}^2+2r_{ij}^4)\log r_{ij}^2}{240m_j^8r_{ij}^2(1-r_{ij}^2)^8}k_1^2(3k_2^4+3k_{12}^4+4k_2^2k_{12}^2)\nonumber\\
	&+\frac{(1-r_{ij}^2)(-2+34r_{ij}^2+319r_{ij}^4+79r_{ij}^6-11r_{ij}^8+r_{ij}^{10})+60r_{ij}^4(3+4r_{ij}^2)\log r_{ij}^2}{240m_j^8r_{ij}^4(1-r_{ij}^2)^8}k_1^4(k_2^2+k_{12}^2)\nonumber\\
	&+\frac{(1-r_{ij}^2)(-4+38r_{ij}^2-214r_{ij}^4-319r_{ij}^6+101r_{ij}^8-25r_{ij}^{10}+3r_{ij}^{12})-420r_{ij}^6\log r_{ij}^2}{1680m_j^8r_{ij}^6(1-r_{ij}^2)^8}k_1^6.
\end{align}
The $r_{ij}$ is defined as $r_{ij}\equiv m_i/m_j$. In the above, the expansions agree with Eq. \eqref{eqn:Apploop:expansion:C0:mi} in the limit of $m_i=m_j$ (or $r_{ij}\rightarrow1$). If the first and third internal masses are equal, the $C_0$ function can be correlated with the case where the first two internal masses are equal, through the relation of
\begin{align}
&C_0(k_1^2,k_{12}^2,k_2^2,m_i^2,m_j^2,m_i^2)=C_0(k_2^2,k_{12}^2,k_1^2,m_i^2,m_i^2,m_j^2).
\end{align}

\subsubsection{Related heavy internal mass expansion of $D_0$ function}
The $D_0$ function is defined as
\begin{align}
&D_0(k_1^2,k_{12}^2,k_{23}^2,k_3^2,k_2^2,k_{13}^2,m_0^2,m_1^2,m_2^2,m_3^2)\nonumber\\
\equiv&\frac{(2\pi\mu)^{4-D}}{i\pi^2}\int d^Dq\frac{1}{(q^2-m_0^2)[(q+k_1)^2-m_1^2][(q+k_2)^2-m_2^2][(q+k_3)^2-m_3^2]}\nonumber\\
=&\int_0^1\int_0^1\int_0^1\int_0^1dxdydzdw\frac{\delta(x+y+z+w-1)}{[xm_0^2+ym_1^2+zm_2^2+wm_3^2-xyk_1^2-xzk_2^2-xwk_3^2-yzk_{12}^2-ywk_{13}^2-zwk_{23}^2]^2},
\end{align}
where we have $k_{12}\equiv k_1-k_2$, $k_{23}\equiv k_2-k_3$, and $k_{13}\equiv k_1-k_3$.

$\bullet$ When the four internal masses are all equal, the $D_0$ function can be expanded as
\begin{align}
&D_0(k_1^2,k_{12}^2,k_{23}^2,k_3^2,k_2^2,k_{13}^2,m_i^2,m_i^2,m_i^2,m_i^2)\nonumber\\
=&\frac{1}{6m_i^4}\big\{1+\frac{k_1^2+k_2^2+k_3^2+k_{12}^2+k_{13}^2+k_{23}^2}{10m_i^2}+\frac{1}{280m_i^4}[2(k_1^2+k_2^2+k_3^2+k_{12}^2+k_{13}^2+k_{23}^2)^2\nonumber\\
	&+(k_1^2-k_{23}^2)^2+(k_2^2-k_{13}^2)^2+(k_3^2-k_{12}^2)^2+k_1^4+k_2^4+k_3^4+k_{12}^4+k_{13}^4+k_{23}^4]\nonumber\\
	&+\frac{1}{1260m_i^6}[(k_1^2+k_2^2)^3+(k_1^2+k_3^2)^3+(k_2^2+k_3^2)^3+(k_{12}^2+k_{13}^2)^3+(k_{12}^2+k_{23}^2)^3+(k_{13}^2+k_{23}^2)^3\nonumber\\
	&+(k_1^2+k_2^2)(k_{13}^2+k_{23}^2)^2+(k_1^2+k_3^2)(k_{12}^2+k_{23}^2)^2+(k_2^2+k_3^2)(k_{12}^2+k_{13}^2)^2\nonumber\\
	&+(k_{13}^2+k_{23}^2)(k_1^2+k_2^2)^2+(k_{12}^2+k_{23}^2)(k_1^2+k_3^2)^2+(k_{12}^2+k_{13}^2)(k_2^2+k_3^2)^2\nonumber\\
	&+3(k_1^2+k_{23}^2)(k_2^2+k_{13}^2)(k_3^2+k_{12}^2)+k_1^6+k_2^6+k_3^6+k_{12}^6+k_{13}^6+k_{23}^6+2k_{12}^2(k_1^2+k_2^2)(k_1^2+k_2^2+k_{12}^2)\nonumber\\
	&+2k_{13}^2(k_1^2+k_3^2)(k_1^2+k_3^2+k_{13}^2)+2k_{23}^2(k_2^2+k_3^2)(k_2^2+k_3^2+k_{23}^2)-k_1^2k_{23}^2(k_1^2+k_{23}^2)-k_2^2k_{13}^2(k_2^2+k_{13}^2)\nonumber\\
	&-k_3^2k_{12}^2(k_3^2+k_{12}^2)-3k_1^2k_2^2k_{12}^2-3k_1^2k_3^2k_{13}^2-3k_2^2k_3^2k_{23}^2+k_{12}^2k_{13}^2k_{23}^2]+\mathcal{O}(\frac{k^8}{m_i^8})\big\}.
\end{align}

$\bullet$ When the first three internal masses are equal, the $D_0$ function can be expanded as
\begin{align}
&D_0(k_1^2,k_{12}^2,k_{23}^2,k_3^2,k_2^2,k_{13}^2,m_i^2,m_i^2,m_i^2,m_j^2)\nonumber\\
\approx&\frac{1}{m_j^4}\cdot\frac{1+2r_{ij}^2\log r_{ij}^2-r_{ij}^4}{2r_{ij}^2(1-r_{ij}^2)^3}+\frac{(k_3^2+k_{13}^2+k_{23}^2)}{m_j^6}\cdot\frac{1+6r_{ij}^2(1+r_{ij}^2)\log r_{ij}^2+9r_{ij}^2-9r_{ij}^4-r_{ij}^6}{6r_{ij}^2(1-r_{ij}^2)^5}\nonumber\\
&+\frac{(k_1^2+k_2^2+k_{12}^2)}{m_j^6}\cdot\frac{1-8r_{ij}^2-12r_{ij}^4\log r_{ij}^2+8r_{ij}^6-r_{ij}^8}{24r_{ij}^4(1-r_{ij}^2)^5}.
\end{align}
In our \textit{Mathematica} code, we expand the $D_0^{iiij}(s,t)$ function up to $\mc{O}(1/m_{j}^{10})$. Here, we only present the results up to $\mc{O}(1/m_{j}^{6})$ because the $1/m_{j}^{8}$ and $1/m_{j}^{10}$ terms are quite lengthy. If the last three internal masses are equal, the $D_0$ function can be correlated with the case where the first three internal masses are equal, through the relation of
\begin{align}
&D_0(k_1^2,k_{12}^2,k_{23}^2,k_3^2,k_2^2,k_{13}^2,m_i^2,m_i^2,m_i^2,m_j^2)=D_0(k_{13}^2,k_{12}^2,k_2^2,k_3^2,k_{23}^2,k_1^2,m_j^2,m_i^2,m_i^2,m_i^2).
\end{align}

$\bullet$ When the first two internal masses are equal, the $D_0$ function can be expanded as
\begin{align}
&D_0(k_1^2,k_{12}^2,k_{23}^2,k_3^2,k_2^2,k_{13}^2,m_i^2,m_i^2,m_j^2,m_j^2)\nonumber\\
\approx&-\frac{1}{m_j^4}\cdot\frac{2+(1+r_{ij}^2)\log r_{ij}^2-2r_{ij}^2}{(1-r_{ij}^2)^3}-\frac{k_2^2+k_3^2+k_{12}^2+k_{13}^2}{m_j^6}\cdot\frac{3+(1+4r_{ij}^2+r_{ij}^4)\log r_{ij}^2-3r_{ij}^4}{2(1-r_{ij}^2)^5}\nonumber\\
&+\frac{k_1^2}{m_j^6}\cdot\frac{1+6r_{ij}^2(1+r_{ij}^2)\log r_{ij}^2+9r_{ij}^2-9r_{ij}^4-r_{ij}^6}{6r_{ij}^2(1-r_{ij}^2)^5}+\frac{k_{23}^2}{m_j^6}\cdot\frac{1+6r_{ij}^2(1+r_{ij}^2)\log r_{ij}^2+9r_{ij}^2-9r_{ij}^4-r_{ij}^6}{6(1-r_{ij}^2)^5}.
\end{align}
In our \textit{Mathematica} code, we also expand the $D_0^{iijj}(t,u)$ function up to $\mc{O}(1/m_{j}^{10})$. If the second and third internal masses are equal, the $D_0$ function can be correlated with the case where the first two internal masses are equal, through the relation of
\begin{align}
&D_0(k_1^2,k_{12}^2,k_{23}^2,k_3^2,k_2^2,k_{13}^2,m_i^2,m_i^2,m_j^2,m_j^2)=D_0(k_3^2,k_{23}^2,k_{12}^2,k_1^2,k_2^2,k_{13}^2,m_i^2,m_j^2,m_j^2,m_i^2).
\end{align}

\section{Computation details of the amplitudes}\label{app:amp}
\subsection{Non-vanishing contributions of the amplitudes}\label{app:amp:contri}
$\bullet$ For the clockwise triangle-like diagram (a) in Fig. \ref{fig:SLQcontri:Feynman}, the amplitude is calculated as
\begin{align}
&i\mc{M}_{a,iij}^{clock}=\frac{g_2g_s^2g_{ij}^{gZ}\Gamma_{ji}^hv}{c_W}\epsilon_{\mu}^a(k_1)\epsilon_{\nu}^a(k_2)\epsilon_{\rho}^{\ast}(p_1)g^{\nu\rho}\int\frac{d^Dq}{(2\pi)^D}\frac{(2q+k_1)^{\mu}}{(q^2-m_i^2)[(q+k_1)^2-m_i^2][(q+p_1-k_2)^2-m_j^2]}\nonumber\\
&=\frac{ig_2g_s^2g_{ij}^{gZ}\Gamma_{ji}^hv}{8\pi^2c_W}\epsilon_{\mu}^a(k_1)\epsilon_{\nu}^a(k_2)\epsilon_{\rho}^{\ast}(p_1)g^{\nu\rho}(p_1-k_2)^{\mu}C_2(0,m_h^2,u,m_i^2,m_i^2,m_j^2).
\end{align}
For the counter-clockwise triangle-like diagram (a) in Fig. \ref{fig:SLQcontri:Feynman}, the amplitude is calculated as
\begin{align}
&i\mc{M}_{a,iij}^{counter}=\frac{g_2g_s^2g_{ji}^{gZ}\Gamma_{ij}^hv}{c_W}\epsilon_{\mu}^a(k_1)\epsilon_{\nu}^a(k_2)\epsilon_{\rho}^{\ast}(p_1)g^{\nu\rho}\int\frac{d^Dq}{(2\pi)^D}\frac{(2q-k_1)^{\mu}}{(q^2-m_i^2)[(q-k_1)^2-m_i^2][(q+k_2-p_1)^2-m_j^2]}\nonumber\\
&=\frac{ig_2g_s^2g_{ji}^{gZ}\Gamma_{ij}^hv}{8\pi^2c_W}\epsilon_{\mu}^a(k_1)\epsilon_{\nu}^a(k_2)\epsilon_{\rho}^{\ast}(p_1)g^{\nu\rho}(k_2-p_1)^{\mu}C_2(0,m_h^2,u,m_i^2,m_i^2,m_j^2).
\end{align}
Due to the physical polarization condition, we can drop the terms such as $\epsilon_{\mu}^a(k_1)k_1^{\mu}, \epsilon_{\nu}^a(k_2)k_2^{\nu}, \epsilon_{\rho}^{\ast}(p_1)p_1^{\rho}$. Utilizing the Hermitian conditions $g^{gZ}=(g^{gZ})^{\dag}$ and $\Gamma^h=(\Gamma^h)^{\dag}$, we have $g_{ji}^{gZ}\Gamma_{ij}^h=(g_{ij}^{gZ})^{\ast}(\Gamma_{ji}^h)^{\ast}$. Then, the total amplitude for the triangle-like diagram (a) in Fig. \ref{fig:SLQcontri:Feynman} is calculated as
\begin{align}
&i\mc{M}_{a,iij}=i\mc{M}_{a,iij}^{clock}+i\mc{M}_{a,iij}^{counter}\nonumber\\
&=-\frac{g_2g_s^2\mr{Im}(g_{ij}^{gZ}\Gamma_{ji}^h)v}{4\pi^2c_W}\epsilon_{\mu}^a(k_1)\epsilon_{\nu}^a(k_2)\epsilon_{\rho}^{\ast}(p_1)g^{\nu\rho}(p_1-k_2)^{\mu}C_2(0,m_h^2,u,m_i^2,m_i^2,m_j^2).
\end{align}

$\bullet$ For the triangle-like diagram (b) in Fig. \ref{fig:SLQcontri:Feynman}, the total amplitude can be derived through the replacement of $(\mu,\nu)\leftrightarrow(\nu,\mu)$ and $(k_1,k_2)\leftrightarrow(k_2,k_1)$ in $\mc{M}_{a}$. Then, it is calculated as
\begin{align}
&i\mc{M}_{b,iij}=-\frac{g_2g_s^2\mr{Im}(g_{ij}^{gZ}\Gamma_{ji}^h)v}{4\pi^2c_W}\epsilon_{\mu}^a(k_1)\epsilon_{\nu}^a(k_2)\epsilon_{\rho}^{\ast}(p_1)g^{\mu\rho}(p_1-k_1)^{\nu}C_2(0,m_h^2,t,m_i^2,m_i^2,m_j^2).
\end{align}

$\bullet$ For the clockwise triangle-like diagram (c) in Fig. \ref{fig:SLQcontri:Feynman}, the amplitude is calculated as
\begin{align}
&i\mc{M}_{c,iij}^{clock}=\frac{g_2g_s^2g_{ji}^{Z}\Gamma_{ij}^hv}{c_W}\epsilon_{\mu}^a(k_1)\epsilon_{\nu}^a(k_2)\epsilon_{\rho}^{\ast}(p_1)g^{\mu\nu}\int\frac{d^Dq}{(2\pi)^D}\frac{(2q-p_1)^{\rho}}{(q^2-m_i^2)[(q-k_1-k_2)^2-m_i^2][(q-p_1)^2-m_j^2]}\nonumber\\
&=-\frac{ig_2g_s^2g_{ji}^{Z}\Gamma_{ij}^hv}{8\pi^2c_W}\epsilon_{\mu}^a(k_1)\epsilon_{\nu}^a(k_2)\epsilon_{\rho}^{\ast}(p_1)g^{\mu\nu}p_2^{\rho}C_1(s,m_h^2,m_Z^2,m_i^2,m_i^2,m_j^2).
\end{align}
For the counter-clockwise triangle-like diagram (c) in Fig. \ref{fig:SLQcontri:Feynman}, the amplitude is calculated as
\begin{align}
&i\mc{M}_{c,iij}^{counter}=\frac{g_2g_s^2g_{ij}^{Z}\Gamma_{ji}^hv}{c_W}\epsilon_{\mu}^a(k_1)\epsilon_{\nu}^a(k_2)\epsilon_{\rho}^{\ast}(p_1)g^{\mu\nu}\int\frac{d^Dq}{(2\pi)^D}\frac{(2q+p_1)^{\rho}}{(q^2-m_i^2)[(q+k_1+k_2)^2-m_i^2][(q+p_1)^2-m_j^2]}\nonumber\\
&=\frac{ig_2g_s^2g_{ij}^{Z}\Gamma_{ji}^hv}{8\pi^2c_W}\epsilon_{\mu}^a(k_1)\epsilon_{\nu}^a(k_2)\epsilon_{\rho}^{\ast}(p_1)g^{\mu\nu}p_2^{\rho}C_1(s,m_h^2,m_Z^2,m_i^2,m_i^2,m_j^2).
\end{align}
Then, the total amplitude for the triangle-like diagram (c) in Fig. \ref{fig:SLQcontri:Feynman} is calculated as
\begin{align}
&i\mc{M}_{c,iij}=i\mc{M}_{c,iij}^{clock}+i\mc{M}_{c,iij}^{counter}\nonumber\\
&=-\frac{g_2g_s^2\mr{Im}(g_{ij}^{Z}\Gamma_{ji}^h)v}{4\pi^2c_W}\epsilon_{\mu}^a(k_1)\epsilon_{\nu}^a(k_2)\epsilon_{\rho}^{\ast}(p_1)g^{\mu\nu}p_2^{\rho}C_1(s,m_h^2,m_Z^2,m_i^2,m_i^2,m_j^2).
\end{align}

$\bullet$ For the clockwise and counter-clockwise box diagram (d) in Fig. \ref{fig:SLQcontri:Feynman}, the amplitude is calculated as
\begin{align}
&i\mc{M}_{d,iiij}^{clock}=-\frac{g_2g_s^2g_{ji}^{Z}\Gamma_{ij}^hv}{2c_W}\epsilon_{\mu}^a(k_1)\epsilon_{\nu}^a(k_2)\epsilon_{\rho}^{\ast}(p_1)\nonumber\\
	&\int\frac{d^Dq}{(2\pi)^D}\frac{(2q+k_1)^{\mu}(2q-k_2)^{\nu}(2q+2k_1-p_1)^{\rho}}{(q^2-m_i^2)[(q+k_1)^2-m_i^2][(q-k_2)^2-m_i^2][(q+k_1-p_1)^2-m_j^2]},\nonumber\\
&i\mc{M}_{d,iiij}^{counter}=-\frac{g_2g_s^2g_{ij}^{Z}\Gamma_{ji}^hv}{2c_W}\epsilon_{\mu}^a(k_1)\epsilon_{\nu}^a(k_2)\epsilon_{\rho}^{\ast}(p_1)\nonumber\\
	&\int\frac{d^Dq}{(2\pi)^D}\frac{(2q-k_1)^{\mu}(2q+k_2)^{\nu}(2q-2k_1+p_1)^{\rho}}{(q^2-m_i^2)[(q-k_1)^2-m_i^2][(q+k_2)^2-m_i^2][(q-k_1+p_1)^2-m_j^2]}.
\end{align}
Then, the total amplitude for the box diagram (d) in Fig. \ref{fig:SLQcontri:Feynman} is calculated as
\begin{align}
&i\mc{M}_{d,iiij}=i\mc{M}_{d,iiij}^{clock}+i\mc{M}_{d,iiij}^{counter}\nonumber\\
&=\frac{ig_2g_s^2\mr{Im}(g_{ij}^{Z}\Gamma_{ji}^h)v}{c_W}\epsilon_{\mu}^a(k_1)\epsilon_{\nu}^a(k_2)\epsilon_{\rho}^{\ast}(p_1)\nonumber\\
	&\int\frac{d^Dq}{(2\pi)^D}\frac{(2q+k_1)^{\mu}(2q-k_2)^{\nu}(2q+2k_1-p_1)^{\rho}}{(q^2-m_i^2)[(q+k_1)^2-m_i^2][(q-k_2)^2-m_i^2][(q+k_1-p_1)^2-m_j^2]}\nonumber\\
&=-\frac{g_2g_s^2\mr{Im}(g_{ij}^{Z}\Gamma_{ji}^h)v}{2\pi^2c_W}\epsilon_{\mu}^a(k_1)\epsilon_{\nu}^a(k_2)\epsilon_{\rho}^{\ast}(p_1)\big\{g^{\mu\nu}[(D_{00}+D_{001}+D_{003})k_1^{\rho}-D_{002}k_2^{\rho}]\nonumber\\
	&+g^{\mu\rho}[(D_{001}+D_{003})k_1^{\nu}-D_{003}p_1^{\nu}]+g^{\nu\rho}(-D_{002}k_2^{\mu}-D_{003}p_1^{\mu})\nonumber\\
	&+k_2^{\mu}k_1^{\nu}[-(D_{12}+D_{23}+D_{112}+D_{233}+2D_{123})k_1^{\rho}+(D_{122}+D_{223})k_2^{\rho}]\nonumber\\
	&-p_1^{\mu}k_1^{\nu}k_1^{\rho}(D_{13}+D_{33}+D_{113}+D_{333}+2D_{133})-p_1^{\nu}k_2^{\mu}k_2^{\rho}D_{223}\nonumber\\
	&+p_1^{\mu}p_1^{\nu}[(D_{33}+D_{133}+D_{333})k_1^{\rho}-D_{233}k_2^{\rho}]\nonumber\\
	&+p_1^{\mu}k_1^{\nu}k_2^{\rho}(D_{233}+D_{123})+p_1^{\nu}k_2^{\mu}k_1^{\rho}(D_{23}+D_{233}+D_{123})\big\}.
\end{align}
Here, the variables of $D$ functions are $(0,s,m_h^2,t,0,m_Z^2,m_i^2,m_i^2,m_i^2,m_j^2)$.

$\bullet$ For the box diagram (e) in Fig. \ref{fig:SLQcontri:Feynman}, the total amplitude can be derived through the replacement of $(\mu,\nu)\leftrightarrow(\nu,\mu)$ and $(k_1,k_2)\leftrightarrow(k_2,k_1)$ in $\mc{M}_{d}$. Then, it is calculated as
\begin{align}
&i\mc{M}_{e,iiij}=-\frac{g_2g_s^2\mr{Im}(g_{ij}^{Z}\Gamma_{ji}^h)v}{2\pi^2c_W}\epsilon_{\mu}^a(k_1)\epsilon_{\nu}^a(k_2)\epsilon_{\rho}^{\ast}(p_1)\big\{g^{\mu\nu}[(D_{00}+D_{001}+D_{003})k_2^{\rho}-D_{002}k_1^{\rho}]\nonumber\\
	&+g^{\nu\rho}[(D_{001}+D_{003})k_2^{\mu}-D_{003}p_1^{\mu}]+g^{\mu\rho}(-D_{002}k_1^{\nu}-D_{003}p_1^{\nu})\nonumber\\
	&+k_2^{\mu}k_1^{\nu}[-(D_{12}+D_{23}+D_{112}+D_{233}+2D_{123})k_2^{\rho}+(D_{122}+D_{223})k_1^{\rho}]\nonumber\\
	&-p_1^{\nu}k_2^{\mu}k_2^{\rho}(D_{13}+D_{33}+D_{113}+D_{333}+2D_{133})-p_1^{\mu}k_1^{\nu}k_1^{\rho}D_{223}\nonumber\\
	&+p_1^{\mu}p_1^{\nu}[(D_{33}+D_{133}+D_{333})k_2^{\rho}-D_{233}k_1^{\rho}]\nonumber\\
	&+p_1^{\nu}k_2^{\mu}k_1^{\rho}(D_{233}+D_{123})+p_1^{\mu}k_1^{\nu}k_2^{\rho}(D_{23}+D_{233}+D_{123})\big\}.
\end{align}
Here, the variables of $D$ functions are $(0,s,m_h^2,u,0,m_Z^2,m_i^2,m_i^2,m_i^2,m_j^2)$.

$\bullet$ For the clockwise and counter-clockwise box diagram (f) in Fig. \ref{fig:SLQcontri:Feynman}, the amplitude is calculated as
\begin{align}
&i\mc{M}_{f,jjii}^{clock}=-\frac{g_2g_s^2g_{ij}^{Z}\Gamma_{ji}^hv}{2c_W}\epsilon_{\mu}^a(k_1)\epsilon_{\nu}^a(k_2)\epsilon_{\rho}^{\ast}(p_1)\nonumber\\
	&\int\frac{d^Dq}{(2\pi)^D}\frac{(2q+k_1-2p_1)^{\mu}(2q-k_2)^{\nu}(2q-p_1)^{\rho}}{(q^2-m_j^2)[(q-k_2)^2-m_j^2][(q-p_1)^2-m_i^2][(q+k_1-p_1)^2-m_i^2]},\nonumber\\
&i\mc{M}_{f,jjii}^{counter}=-\frac{g_2g_s^2g_{ji}^{Z}\Gamma_{ij}^hv}{2c_W}\epsilon_{\mu}^a(k_1)\epsilon_{\nu}^a(k_2)\epsilon_{\rho}^{\ast}(p_1)\nonumber\\
	&\int\frac{d^Dq}{(2\pi)^D}\frac{(2q-k_1+2p_1)^{\mu}(2q+k_2)^{\nu}(2q+p_1)^{\rho}}{(q^2-m_j^2)[(q+k_2)^2-m_j^2][(q+p_1)^2-m_i^2][(q-k_1+p_1)^2-m_i^2]}.
\end{align}
Then, the total amplitude for the box diagram (f) in Fig. \ref{fig:SLQcontri:Feynman} is calculated as
\begin{align}
&i\mc{M}_{f,jjii}=i\mc{M}_{f,jjii}^{clock}+i\mc{M}_{f,jjii}^{counter}\nonumber\\
&=-\frac{ig_2g_s^2\mr{Im}(g_{ij}^{Z}\Gamma_{ji}^h)v}{c_W}\epsilon_{\mu}^a(k_1)\epsilon_{\nu}^a(k_2)\epsilon_{\rho}^{\ast}(p_1)\nonumber\\
	&\int\frac{d^Dq}{(2\pi)^D}\frac{(2q+k_1-2p_1)^{\mu}(2q-k_2)^{\nu}(2q-p_1)^{\rho}}{(q^2-m_j^2)[(q-k_2)^2-m_j^2][(q-p_1)^2-m_i^2][(q+k_1-p_1)^2-m_i^2]}\nonumber\\
&=\frac{g_2g_s^2\mr{Im}(g_{ij}^{Z}\Gamma_{ji}^h)v}{2\pi^2c_W}\epsilon_{\mu}^a(k_1)\epsilon_{\nu}^a(k_2)\epsilon_{\rho}^{\ast}(p_1)\big\{g^{\mu\nu}(D_{003}k_1^{\rho}-D_{001}k_2^{\rho})\nonumber\\
	&+g^{\mu\rho}[D_{003}k_1^{\nu}-(D_{002}+D_{003})p_1^{\nu}]+g^{\nu\rho}[-D_{001}k_2^{\mu}-(D_{00}+D_{002}+D_{003})p_1^{\mu}]\nonumber\\
	&+k_2^{\mu}k_1^{\nu}(-D_{133}k_1^{\rho}+D_{113}k_2^{\rho})-p_1^{\mu}k_1^{\nu}k_1^{\rho}(D_{33}+D_{333}+D_{233})-p_1^{\nu}k_2^{\mu}k_2^{\rho}(D_{112}+D_{113})\nonumber\\
	&+p_1^{\mu}p_1^{\nu}[(D_{33}+D_{23}+D_{333}+D_{223}+2D_{233})k_1^{\rho}-(D_{12}+D_{13}+D_{122}+D_{133}+2D_{123})k_2^{\rho}]\nonumber\\
	&+p_1^{\mu}k_1^{\nu}k_2^{\rho}(D_{13}+D_{133}+D_{123})+p_1^{\nu}k_2^{\mu}k_1^{\rho}(D_{133}+D_{123})\big\}.
\end{align}
Here, the variables of $D$ functions are $(0,u,0,t,m_Z^2,m_h^2,m_j^2,m_j^2,m_i^2,m_i^2)$.
\subsection{Vanishing contributions of the amplitudes}\label{app:amp:vanish}
In this section, we illustrate how the contributions vanish explicitly in $R_{\xi}$ gauge.

For the clockwise triangle diagram (a) in Fig. \ref{fig:SLQcontri:Feynman:vanishInt}, the amplitude is calculated as
\begin{align}
&i\mc{M}_{fig3a}^{clock}=-\frac{g_2^2g_s^2m_Z(I_3-Qs_W^2)}{2c_W^2}\frac{\epsilon_{\mu}^a(k_1)\epsilon_{\nu}^a(k_2)\epsilon_{\rho}^{\ast}(p_1)}{(k_1+k_2)^2-m_Z^2}\int\frac{d^Dq}{(2\pi)^D}\frac{1}{(q^2-m_i^2)[(q+k_1)^2-m_i^2][(q-k_2)^2-m_i^2]}\nonumber\\
	&\cdot(2q+k_1)^{\mu}(2q-k_2)^{\nu}[(2q+k_1-k_2)^{\rho}-(1-\xi)\frac{(2q)\cdot(k_1+k_2)(k_1+k_2)^{\rho}}{(k_1+k_2)^2-\xi m_Z^2}].
\end{align}
For the counter-clockwise triangle diagram (a) in Fig. \ref{fig:SLQcontri:Feynman:vanishInt}, the amplitude is calculated as
\begin{align}
&i\mc{M}_{fig3a}^{counter}=-\frac{g_2^2g_s^2m_Z(I_3-Qs_W^2)}{2c_W^2}\frac{\epsilon_{\mu}^a(k_1)\epsilon_{\nu}^a(k_2)\epsilon_{\rho}^{\ast}(p_1)}{(k_1+k_2)^2-m_Z^2}\int\frac{d^Dq}{(2\pi)^D}\frac{1}{(q^2-m_i^2)[(q-k_1)^2-m_i^2][(q+k_2)^2-m_i^2]}\nonumber\\
	&\cdot(2q-k_1)^{\mu}(2q+k_2)^{\nu}[(2q+k_2-k_1)^{\rho}-(1-\xi)\frac{(2q)\cdot(k_1+k_2)(k_1+k_2)^{\rho}}{(k_1+k_2)^2-\xi m_Z^2}].
\end{align}
We can check that $\mc{M}_{fig3a}^{clock}+\mc{M}_{fig3a}^{counter}=0$.

For the bubble-like diagram (b) in Fig. \ref{fig:SLQcontri:Feynman:vanishInt}, the amplitude is calculated as
\begin{align}
&i\mc{M}_{fig3b}=-\frac{g_2^2g_s^2g_{ii}^{gZ}m_Z}{c_W^2}\frac{\epsilon_{\mu}^a(k_1)\epsilon_{\nu}^a(k_2)\epsilon_{\rho}^{\ast}(p_1)}{(k_1+k_2)^2-m_Z^2}[g^{\nu\rho}-(1-\xi)\frac{(k_1+k_2)^{\nu}(k_1+k_2)^{\rho}}{(k_1+k_2)^2-\xi m_Z^2}]\nonumber\\
	&\cdot\int\frac{d^Dq}{(2\pi)^D}\frac{(2q+k_1)^{\mu}}{(q^2-m_i^2)[(q+k_1)^2-m_i^2]}\nonumber\\
&\sim \epsilon_{\mu}^a(k_1)k_1^{\mu}\cdot[2B_1(k_1^2,m_i^2,m_i^2)+B_0(k_1^2,m_i^2,m_i^2)]=0.
\end{align}
The identity above holds on because of the transversality condition of the gluon. Furthermore, the equality $2B_1(k_1^2,m_i^2,m_i^2)+B_0(k_1^2,m_i^2,m_i^2)=0$ is also true indeed.

For the bubble-like diagram (c) in Fig. \ref{fig:SLQcontri:Feynman:vanishInt}, the amplitude is identified as zero through the replacement of $(\mu,\nu)\leftrightarrow(\nu,\mu)$ and $(k_1,k_2)\leftrightarrow(k_2,k_1)$ in $\mc{M}_{fig3b}$.

For the bubble-like diagram (d) in Fig. \ref{fig:SLQcontri:Feynman:vanishInt}, the amplitude is calculated as
\begin{align}
&i\mc{M}_{fig3d}=\frac{g_2^2g_s^2g_{ii}^{Z}m_Z}{c_W^2}\frac{\epsilon_{\mu}^a(k_1)\epsilon_{\nu}^a(k_2)\epsilon_{\rho}^{\ast}(p_1)g^{\mu\nu}}{(k_1+k_2)^2-m_Z^2}\int\frac{d^Dq}{(2\pi)^D}\frac{(2q+k_1+k_2)^{\rho}-(1-\xi)\frac{(k_1+k_2)\cdot(2q+k_1+k_2)(k_1+k_2)^{\rho}}{(k_1+k_2)^2-\xi m_Z^2}}{(q^2-m_i^2)[(q+k_1+k_2)^2-m_i^2]}\nonumber\\
&\sim (k_1+k_2)^{\rho}[1-(1-\xi)\frac{(k_1+k_2)^2}{(k_1+k_2)^2-\xi m_Z^2}]\cdot[2B_1\big((k_1+k_2)^2,m_i^2,m_i^2)\big)+B_0\big((k_1+k_2)^2,m_i^2,m_i^2)\big)]\nonumber\\
&=0.
\end{align}

\section{Explicit results of the form factors $G_{1,2,3}$}\label{app:results}
In Eq. \eqref{eqn:SLQcontri:explicit:G123}, the form factor $G_{1,2,3}$ is given as
\begin{align}
&G_{1,2,3}=-\sum_{i,j}\frac{g_2g_s^2\mr{Im}(g_{ij}^{Z}\Gamma_{ji}^h)v}{2\pi^2c_W}\sum_{m=1}^{10}\overline{G}_{1,2,3}^{(m)}.
\end{align}
In the following, the expressions of $\overline{G}_{1,2,3}^{(m)}$ are enumerated explicitly. First of all, let us define the $\lambda$ function as
\begin{align}
\lambda(x,y,z)=x^2+y^2+z^2-2xy-2xz-2yz.
\end{align}
Hence, we have
\begin{align}
&\lambda(m_h^2,m_Z^2,s)=m_h^4+m_Z^4+s^2-2m_h^2m_Z^2-2s(m_h^2+m_Z^2) =(t+u)^2-4m_h^2m_Z^2.
\end{align}
\subsection{Ten explicit expressions of $\overline{G}_{1}^{(m)}$}\label{app:results:G1}

\begin{align}
&\overline{G}_1^{(1)}=[\frac{m_Z^2(u-m_Z^2)}{(tu-m_h^2m_Z^2)^2}+\frac{m_Z^2(s+m_Z^2-m_h^2)}{\lambda(m_h^2,m_Z^2,s)(tu-m_h^2m_Z^2)}]B_0^i(s).
\end{align}

\begin{align}
&\overline{G}_1^{(2)}=\big\{\frac{m_Z^2s(2m_Z^2-t-2u)}{2(tu-m_h^2m_Z^2)^2}+\frac{m_Z^2s(m_Z^2-u)(2m_h^2m_Z^2-(t+u)^2)}{2(tu-m_h^2m_Z^2)^3}\nonumber\\
	&+m_Z^2s[\frac{(u-m_Z^2)(t+u)}{(tu-m_h^2m_Z^2)^3}-\frac{1}{2(tu-m_h^2m_Z^2)^2}](m_i^2-m_j^2)\nonumber\\
	&+(m_Z^2-u)[-\frac{m_Z^2}{(tu-m_h^2m_Z^2)^2}+\frac{1}{s(tu-m_h^2m_Z^2)}]m_i^2+\frac{m_Z^2s(u-m_Z^2)(m_i^2-m_j^2)^2}{(tu-m_h^2m_Z^2)^3}\big\}C_0^i(s).
\end{align}

\begin{align}
&\overline{G}_1^{(3)}=\big\{\frac{(m_Z^2-u)}{2s^3}+\frac{m_Z^4t(t-m_h^2)}{2(tu-m_h^2m_Z^2)^3}\nonumber\\
	&+[\frac{m_Z^4(t-m_h^2)}{(tu-m_h^2m_Z^2)^3}+\frac{m_Z^2(t-m_h^2)}{2s(tu-m_h^2m_Z^2)^2}+\frac{(u-m_Z^2)}{2s^2(tu-m_h^2m_Z^2)}](m_i^2-m_j^2)\big\}(m_h^2-t)C_0^{iij}(m_h^2,t).
\end{align}

\begin{align}
&\overline{G}_1^{(4)}=[\frac{(u-m_Z^2)}{2s^3}+\frac{m_Z^2(m_Z^2-u)u^2}{2(tu-m_h^2m_Z^2)^3}+\frac{m_Z^2(m_Z^2-u)u}{(tu-m_h^2m_Z^2)^3}(m_i^2-m_j^2)](m_h^2-u)C_0^{iij}(m_h^2,u).
\end{align}

\begin{align}
&\overline{G}_1^{(5)}=\big\{\frac{(u-m_Z^2)}{2s^3}+\frac{m_Z^4t(t-m_h^2)}{2(tu-m_h^2m_Z^2)^3}+[\frac{m_Z^4(t-m_h^2)}{(tu-m_h^2m_Z^2)^3}-\frac{m_Z^2}{2(tu-m_h^2m_Z^2)^2}](m_i^2-m_j^2)\big\}(m_Z^2-t)C_0^{iij}(m_Z^2,t).
\end{align}

\begin{align}
&\overline{G}_1^{(6)}=\big\{\frac{(m_Z^2-t)}{2s^3(tu-m_h^2m_Z^2)}+\frac{m_Z^2u(u-m_h^2)}{2s(tu-m_h^2m_Z^2)^3}+\frac{m_Z^2(m_h^2-u)}{2s^2(tu-m_h^2m_Z^2)^2}\nonumber\\
	&+[\frac{m_Z^2(u-m_h^2)}{s(tu-m_h^2m_Z^2)^3}+\frac{(t-m_Z^2)}{2s^2(tu-m_h^2m_Z^2)^2}](m_i^2-m_j^2)\big\}(m_Z^2-u)^3C_0^{iij}(m_Z^2,u).
\end{align}

\begin{align}
&\overline{G}_1^{(7)}=\big\{\frac{m_Z^4(s+m_h^2-m_Z^2)}{\lambda(m_h^2,m_Z^2,s)(tu-m_h^2m_Z^2)}+\frac{m_Z^2(m_Z^2-u)(t+u)\lambda(m_h^2,m_Z^2,s)}{2(tu-m_h^2m_Z^2)^3}+\frac{m_Z^2[(t+2u-2m_Z^2)(t+u)-2m_h^2m_Z^2]}{2(tu-m_h^2m_Z^2)^2}\nonumber\\
	&+[\frac{m_Z^2(m_Z^2-u)\lambda(m_h^2,m_Z^2,s)}{(tu-m_h^2m_Z^2)^3}+\frac{m_Z^2(t+3u-4m_Z^2)}{2(tu-m_h^2m_Z^2)^2}+\frac{m_Z^2(m_h^2-m_Z^2-s)}{\lambda(m_h^2,m_Z^2,s)(tu-m_h^2m_Z^2)}](m_i^2-m_j^2)\big\}C_0^{iji}(m_h^2,m_Z^2,s).
\end{align}

\begin{align}
&\overline{G}_1^{(8)}=\big\{\frac{m_Z^2s(m_Z^2-u)}{2(tu-m_h^2m_Z^2)^3}(m_i^2-m_j^2)^3+[\frac{3m_Z^4s(t-m_h^2)}{2(tu-m_h^2m_Z^2)^3}-\frac{m_Z^2s}{(tu-m_h^2m_Z^2)^2}](m_i^2-m_j^2)^2\nonumber\\
	&+[\frac{3m_Z^2(m_Z^2-u)}{2(tu-m_h^2m_Z^2)^2}+\frac{u-m_Z^2}{2s(tu-m_h^2m_Z^2)}](m_i^4-m_i^2m_j^2)+[\frac{3m_Z^4st(t-m_h^2)}{2(tu-m_h^2m_Z^2)^3}-\frac{m_Z^2st}{2(tu-m_h^2m_Z^2)^2}](m_i^2-m_j^2)\nonumber\\
	&+[-\frac{1}{2s}+\frac{3m_Z^4(t-m_h^2)}{2(tu-m_h^2m_Z^2)^2}+\frac{m_Z^2(u-m_Z^2)}{2s(tu-m_h^2m_Z^2)}]m_i^2+\frac{m_Z^4st^2(t-m_h^2)}{2(tu-m_h^2m_Z^2)^3}\big\}D_0^{iiij}(s,t).
\end{align}

\begin{align}
&\overline{G}_1^{(9)}=\big\{\frac{m_Z^2s(m_Z^2-u)}{2(tu-m_h^2m_Z^2)^3}(m_i^2-m_j^2)^3+\frac{3m_Z^2su(m_Z^2-u)}{2(tu-m_h^2m_Z^2)^3}(m_i^2-m_j^2)^2\nonumber\\
	&+[\frac{3m_Z^2(m_Z^2-u)}{2(tu-m_h^2m_Z^2)^2}+\frac{u-m_Z^2}{2s(tu-m_h^2m_Z^2)}](m_i^4-m_i^2m_j^2)+\frac{3m_Z^2su^2(m_Z^2-u)}{2(tu-m_h^2m_Z^2)^3}(m_i^2-m_j^2)\nonumber\\
	&+[\frac{3m_Z^2u(m_Z^2-u)}{2(tu-m_h^2m_Z^2)^2}+\frac{u(u-m_Z^2)}{2s(tu-m_h^2m_Z^2)}]m_i^2+\frac{m_Z^2su^3(m_Z^2-u)}{2(tu-m_h^2m_Z^2)^3}\big\}D_0^{iiij}(s,u).
\end{align}

\begin{align}
&\overline{G}_1^{(10)}=\big\{\frac{m_Z^2s(m_Z^2-u)}{2(tu-m_h^2m_Z^2)^3}(m_i^2-m_j^2)^3+\frac{m_Z^2(m_Z^2-u)}{2(tu-m_h^2m_Z^2)^2}(m_i^2-m_j^2)^2\nonumber\\
	&+[\frac{3m_Z^2(m_Z^2-u)}{2(tu-m_h^2m_Z^2)^2}+\frac{(u-m_Z^2)}{2s(tu-m_h^2m_Z^2)}](m_i^2m_j^2-m_j^4)+\frac{(m_Z^2-u)}{2s^2}m_i^2\nonumber\\
	&+[\frac{(m_Z^2-u)}{s^2}+\frac{m_Z^2(u-m_Z^2)}{2s(tu-m_h^2m_Z^2)}]m_j^2+\frac{(m_Z^2-u)(tu-m_h^2m_Z^2)}{2s^3}\big\}D_0^{iijj}(t,u).
\end{align}
\subsection{Ten explicit expressions of $\overline{G}_{2}^{(m)}$}\label{app:results:G2}

\begin{align}
&\overline{G}_2^{(1)}=[\frac{m_Z^2s}{(tu-m_h^2m_Z^2)^2}+\frac{1}{2(tu-m_h^2m_Z^2)}-\frac{1}{\lambda(m_h^2,m_Z^2,s)}-\frac{(m_Z^2-t)^2+m_Z^2s}{\lambda(m_h^2,m_Z^2,s)(tu-m_h^2m_Z^2)}]B_0^i(s).
\end{align}

\begin{align}
&\overline{G}_2^{(2)}=\big\{\frac{s(2m_Z^2t+u^2-t^2)}{4(tu-m_h^2m_Z^2)^2}+\frac{m_Z^2s^2(t^2+u^2)}{2(tu-m_h^2m_Z^2)^3}+[\frac{m_Z^2s^2(t+u)}{(tu-m_h^2m_Z^2)^3}+\frac{s(m_Z^2+u)}{2(tu-m_h^2m_Z^2)^2}](m_i^2-m_j^2)\nonumber\\
	&+\frac{m_Z^2s}{(tu-m_h^2m_Z^2)^2}m_i^2+[\frac{m_Z^2s^2}{(tu-m_h^2m_Z^2)^3}+\frac{s}{2(tu-m_h^2m_Z^2)^2}](m_i^2-m_j^2)^2\big\}C_0^i(s).
\end{align}

\begin{align}
&\overline{G}_2^{(3)}=\big\{\frac{m_Z^2t(m_h^2-t)(m_Z^2-t)}{2(tu-m_h^2m_Z^2)^3}+\frac{t(m_h^2-t)(t-m_Z^2)}{4s(tu-m_h^2m_Z^2)^2}+\frac{(m_h^2-t)(m_Z^2-t)}{4s^2(tu-m_h^2m_Z^2)}\nonumber\\
	&+\frac{m_Z^2(m_h^2-t)(m_Z^2-t)}{(tu-m_h^2m_Z^2)^3}(m_i^2-m_j^2)\big\}(m_h^2-t)C_0^{iij}(m_h^2,t).
\end{align}

\begin{align}
&\overline{G}_2^{(4)}=\big\{\frac{1}{4s^2}-\frac{u^2}{4(tu-m_h^2m_Z^2)^2}-\frac{m_Z^2su^2}{2(tu-m_h^2m_Z^2)^3}\nonumber\\
	&+[-\frac{m_Z^2su}{(tu-m_h^2m_Z^2)^3}-\frac{u}{2(tu-m_h^2m_Z^2)^2}](m_i^2-m_j^2)\big\}(m_h^2-u)C_0^{iij}(m_h^2,u).
\end{align}

\begin{align}
&\overline{G}_2^{(5)}=\big\{\frac{1}{4s^2}+\frac{t(t-2m_Z^2)}{4(tu-m_h^2m_Z^2)^2}-\frac{m_Z^2st^2}{2(tu-m_h^2m_Z^2)^3}\nonumber\\
	&+[-\frac{m_Z^2st}{(tu-m_h^2m_Z^2)^3}-\frac{m_Z^2}{2(tu-m_h^2m_Z^2)^2}](m_i^2-m_j^2)\big\}(m_Z^2-t)C_0^{iij}(m_Z^2,t).
\end{align}

\begin{align}
&\overline{G}_2^{(6)}=\big\{\frac{(t-m_Z^2)}{4s^2(tu-m_h^2m_Z^2)}+\frac{m_Z^2u(m_h^2-u)}{2(tu-m_h^2m_Z^2)^3}+\frac{(m_Z^2u+su-m_h^2m_Z^2)}{4s(tu-m_h^2m_Z^2)^2}\nonumber\\
	&+[\frac{m_Z^2(m_h^2-u)}{(tu-m_h^2m_Z^2)^3}+\frac{1}{2(tu-m_h^2m_Z^2)^2}](m_i^2-m_j^2)\big\}(m_Z^2-u)^2C_0^{iij}(m_Z^2,u).
\end{align}

\begin{align}
&\overline{G}_2^{(7)}=\big\{-\frac{m_Z^2s(t+u)\lambda(m_h^2,m_Z^2,s)}{2(tu-m_h^2m_Z^2)^3}+\frac{4m_h^2m_Z^2(m_Z^2-t)+(t+u)(2m_h^2m_Z^2-2m_Z^2t+4m_Z^2s+t^2-u^2)}{4(tu-m_h^2m_Z^2)^2}\nonumber\\
	&+\frac{m_Z^2(m_Z^2-t)(m_Z^2-s-m_h^2)}{\lambda(m_h^2,m_Z^2,s)(tu-m_h^2m_Z^2)}+[-\frac{m_Z^2s\lambda(m_h^2,m_Z^2,s)}{(tu-m_h^2m_Z^2)^3}+\frac{4m_Z^2s+m_Z^2(m_h^2-t)+m_h^2(m_Z^2-u)+su}{2(tu-m_h^2m_Z^2)^2}\nonumber\\
	&+\frac{(m_Z^2-t)(m_Z^2+s-m_h^2)}{\lambda(m_h^2,m_Z^2,s)(tu-m_h^2m_Z^2)}](m_i^2-m_j^2)\big\}C_0^{iji}(m_h^2,m_Z^2,s).
\end{align}

\begin{align}
&\overline{G}_2^{(8)}=\big\{[-\frac{m_Z^2s^2}{2(tu-m_h^2m_Z^2)^3}-\frac{s}{4(tu-m_h^2m_Z^2)^2}](m_i^2-m_j^2)^3\nonumber\\
	&+[-\frac{3m_Z^2s^2t}{2(tu-m_h^2m_Z^2)^3}-\frac{s(2m_Z^2+t)}{4(tu-m_h^2m_Z^2)^2}](m_i^2-m_j^2)^2+[-\frac{3m_Z^2s}{2(tu-m_h^2m_Z^2)^2}-\frac{1}{2(tu-m_h^2m_Z^2)}](m_i^4-m_i^2m_j^2)\nonumber\\
	&+[-\frac{3m_Z^2s^2t^2}{2(tu-m_h^2m_Z^2)^3}+\frac{st(t-4m_Z^2)}{4(tu-m_h^2m_Z^2)^2}](m_i^2-m_j^2)\nonumber\\
	&+[-\frac{3m_Z^2st}{2(tu-m_h^2m_Z^2)^2}+\frac{(t-2m_Z^2)}{2(tu-m_h^2m_Z^2)}]m_i^2-\frac{m_Z^2s^2t^3}{2(tu-m_h^2m_Z^2)^3}+\frac{st^2(t-2m_Z^2)}{4(tu-m_h^2m_Z^2)^2}\big\}D_0^{iiij}(s,t).
\end{align}

\begin{align}
&\overline{G}_2^{(9)}=\big\{[-\frac{m_Z^2s^2}{2(tu-m_h^2m_Z^2)^3}-\frac{s}{4(tu-m_h^2m_Z^2)^2}](m_i^2-m_j^2)^3\nonumber\\
	&+[-\frac{3m_Z^2s^2u}{2(tu-m_h^2m_Z^2)^3}-\frac{3su}{4(tu-m_h^2m_Z^2)^2}](m_i^2-m_j^2)^2+[-\frac{3m_Z^2s}{2(tu-m_h^2m_Z^2)^2}-\frac{1}{2(tu-m_h^2m_Z^2)}](m_i^4-m_i^2m_j^2)\nonumber\\
	&+[-\frac{3m_Z^2s^2u^2}{2(tu-m_h^2m_Z^2)^3}-\frac{3su^2}{4(tu-m_h^2m_Z^2)^2}](m_i^2-m_j^2)+[-\frac{3m_Z^2su}{2(tu-m_h^2m_Z^2)^2}-\frac{u}{2(tu-m_h^2m_Z^2)}]m_i^2\nonumber\\
	&-\frac{m_Z^2s^2u^3}{2(tu-m_h^2m_Z^2)^3}-\frac{su^3}{4(tu-m_h^2m_Z^2)^2}\big\}D_0^{iiij}(s,u).
\end{align}

\begin{align}
&\overline{G}_2^{(10)}=\big\{[-\frac{m_Z^2s^2}{2(tu-m_h^2m_Z^2)^3}-\frac{s}{4(tu-m_h^2m_Z^2)^2}](m_i^2-m_j^2)^3\nonumber\\
	&+[\frac{m_Z^2s}{(tu-m_h^2m_Z^2)^2}+\frac{1}{4(tu-m_h^2m_Z^2)}](m_i^2-m_j^2)^2+[-\frac{3m_Z^2s}{2(tu-m_h^2m_Z^2)^2}-\frac{1}{2(tu-m_h^2m_Z^2)}](m_i^4-m_i^2m_j^2)\nonumber\\
	&-\frac{1}{4s}m_i^2+[\frac{m_Z^2}{2(tu-m_h^2m_Z^2)}-\frac{1}{4s}]m_j^2-\frac{(tu-m_h^2m_Z^2)}{4s^2}\big\}D_0^{iijj}(t,u).
\end{align}
\subsection{Ten explicit expressions of $\overline{G}_{3}^{(m)}$}\label{app:results:G3}

\begin{align}
&\overline{G}_3^{(1)}=[\frac{s(t-m_Z^2)}{(tu-m_h^2m_Z^2)^2}+\frac{s(s+m_Z^2-m_h^2)}{\lambda(m_h^2,m_Z^2,s)(tu-m_h^2m_Z^2)}]B_0^i(s).
\end{align}

\begin{align}
&\overline{G}_3^{(2)}=\big\{-\frac{s^2u}{2(tu-m_h^2m_Z^2)^2}+\frac{s^2(t-m_Z^2)(t^2+u^2)}{2(tu-m_h^2m_Z^2)^3}+\frac{s(t-m_Z^2)}{(tu-m_h^2m_Z^2)^2}m_i^2\nonumber\\
	&+[-\frac{s^2}{2(tu-m_h^2m_Z^2)^2}+\frac{s^2(t+u)(t-m_Z^2)}{(tu-m_h^2m_Z^2)^3}](m_i^2-m_j^2)+\frac{s^2(t-m_Z^2)}{(tu-m_h^2m_Z^2)^3}(m_i^2-m_j^2)^2\big\}C_0^i(s).
\end{align}

\begin{align}
&\overline{G}_3^{(3)}=\frac{st(m_h^2-t)(m_Z^2-t)}{(tu-m_h^2m_Z^2)^3}(m_i^2-m_j^2+\frac{t}{2})C_0^{iij}(m_h^2,t).
\end{align}

\begin{align}
&\overline{G}_3^{(4)}=\big\{-\frac{m_Z^2su}{2(tu-m_h^2m_Z^2)^3}+[-\frac{m_Z^2s}{(tu-m_h^2m_Z^2)^3}-\frac{1}{2(tu-m_h^2m_Z^2)^2}](m_i^2-m_j^2)\big\}(m_h^2-u)^2C_0^{iij}(m_h^2,u).
\end{align}

\begin{align}
&\overline{G}_3^{(5)}=\big\{\frac{st^2}{2(tu-m_h^2m_Z^2)^3}+[\frac{st}{(tu-m_h^2m_Z^2)^3}+\frac{1}{2(tu-m_h^2m_Z^2)^2}](m_i^2-m_j^2)\big\}(m_Z^2-t)^2C_0^{iij}(m_Z^2,t).
\end{align}

\begin{align}
&\overline{G}_3^{(6)}=\big\{\frac{m_Z^2u(u-m_h^2)}{2(tu-m_h^2m_Z^2)^3}+[\frac{m_Z^2(u-m_h^2)}{(tu-m_h^2m_Z^2)^3}-\frac{1}{2(tu-m_h^2m_Z^2)^2}](m_i^2-m_j^2)\big\}s(m_Z^2-u)C_0^{iij}(m_Z^2,u).
\end{align}

\begin{align}
&\overline{G}_3^{(7)}=\big\{\frac{m_Z^2s(s+m_h^2-m_Z^2)}{\lambda(m_h^2,m_Z^2,s)(tu-m_h^2m_Z^2)}+\frac{s(m_Z^2-t)(t+u)\lambda(m_h^2,m_Z^2,s)}{2(tu-m_h^2m_Z^2)^3}+\frac{s[(2t+u-2m_Z^2)(t+u)-2m_h^2m_Z^2]}{2(tu-m_h^2m_Z^2)^2}\nonumber\\
	&+[\frac{s(m_Z^2-t)\lambda(m_h^2,m_Z^2,s)}{(tu-m_h^2m_Z^2)^3}+\frac{s(3t+u-4m_Z^2)}{2(tu-m_h^2m_Z^2)^2}+\frac{s(m_h^2-m_Z^2-s)}{\lambda(m_h^2,m_Z^2,s)(tu-m_h^2m_Z^2)}](m_i^2-m_j^2)\big\}C_0^{iji}(m_h^2,m_Z^2,s).
\end{align}

\begin{align}
&\overline{G}_3^{(8)}=\big\{\frac{s^2(m_Z^2-t)}{2(tu-m_h^2m_Z^2)^3}(m_i^2-m_j^2)^3+\frac{3s^2t(m_Z^2-t)}{2(tu-m_h^2m_Z^2)^3}(m_i^2-m_j^2)^2+\frac{3s(m_Z^2-t)}{2(tu-m_h^2m_Z^2)^2}(m_i^4-m_i^2m_j^2)\nonumber\\
	&+\frac{3s^2t^2(m_Z^2-t)}{2(tu-m_h^2m_Z^2)^3}(m_i^2-m_j^2)+\frac{3st(m_Z^2-t)}{2(tu-m_h^2m_Z^2)^2}m_i^2+\frac{s^2t^3(m_Z^2-t)}{2(tu-m_h^2m_Z^2)^3}\big\}D_0^{iiij}(s,t).
\end{align}

\begin{align}
&\overline{G}_3^{(9)}=\big\{\frac{s^2(m_Z^2-t)}{2(tu-m_h^2m_Z^2)^3}(m_i^2-m_j^2)^3+[\frac{3m_Z^2s^2(u-m_h^2)}{2(tu-m_h^2m_Z^2)^3}-\frac{s^2}{(tu-m_h^2m_Z^2)^2}](m_i^2-m_j^2)^2\nonumber\\
	&+\frac{3s(m_Z^2-t)}{2(tu-m_h^2m_Z^2)^2}(m_i^4-m_i^2m_j^2)+s(m_h^2-u)(m_Z^2-u)[\frac{3m_Z^2(m_h^2-u)}{2(tu-m_h^2m_Z^2)^3}+\frac{1}{2(tu-m_h^2m_Z^2)^2}]m_i^2\nonumber\\
	&+s^2u[\frac{3m_Z^2(m_h^2-u)}{2(tu-m_h^2m_Z^2)^3}+\frac{1}{2(tu-m_h^2m_Z^2)^2}]m_j^2+\frac{m_Z^2s^2u^2(u-m_h^2)}{2(tu-m_h^2m_Z^2)^3}\big\}D_0^{iiij}(s,u).
\end{align}

\begin{align}
&\overline{G}_3^{(10)}=\big\{\frac{s^2(m_Z^2-t)}{2(tu-m_h^2m_Z^2)^3}(m_i^2-m_j^2)^3+\frac{s(m_Z^2-t)}{2(tu-m_h^2m_Z^2)^2}(2m_i^4-m_i^2m_j^2-m_j^4)\nonumber\\
	&+\frac{(m_Z^2-t)}{2(tu-m_h^2m_Z^2)}m_i^2\big\}D_0^{iijj}(t,u).
\end{align}

\end{appendices}
\end{sloppypar}
\bibliography{gg2Zh-LQs-arXiv_v2}
\end{document}